\documentclass[a4paper,11pt]{article}
\pdfoutput=1 
\usepackage{jheppub}

\usepackage[utf8]{inputenc}

\usepackage{amssymb,amsmath,amsfonts}
\usepackage{mathtools}
\usepackage{mathrsfs}
\usepackage{bbm}
\usepackage{slashed}
\usepackage{nicefrac}
\usepackage{graphicx}
\usepackage{caption}
\usepackage{subcaption}
\usepackage{esint}
\usepackage{slashbox}

\usepackage{graphicx}
\usepackage[dvipsnames]{xcolor}
\usepackage{array}

\usepackage{hyperref}
\usepackage{xparse}
\usepackage{xspace}
\usepackage{soul}

\usepackage{tikz}
\usetikzlibrary{decorations.pathmorphing}
\usetikzlibrary{automata,positioning}

\usepackage{cancel}
\usepackage[normalem]{ulem}

\usepackage{xifthen}
\usepackage{dsfont}
\usepackage[titletoc]{appendix}
\usepackage{booktabs}
\usepackage{units}
\usepackage{braket}

\newcommand{\gettitle}{}
\hypersetup{linkcolor=black
	colorlinks,
	linkcolor={red!75!black},
	citecolor={blue!75!black},
	urlcolor={blue!75!black},
	pdftitle={\gettitle},
	pdfauthor={Caucal, Mehtar-Tani},
	pdfkeywords={Perturbative QCD} {Jet quenching},
	bookmarksopen=true,
	bookmarksopenlevel=2,
	bookmarksnumbered=true
}
\makeatletter
\newcommand\makebig[2]{%
  \@xp\newcommand\@xp*\csname#1\endcsname{\bBigg@{#2}}%
  \@xp\newcommand\@xp*\csname#1l\endcsname{\@xp\mathopen\csname#1\endcsname}%
  \@xp\newcommand\@xp*\csname#1r\endcsname{\@xp\mathclose\csname#1\endcsname}%
}
\makeatother
\makebig{biggg} {1.3}

\def\del{\partial}

\newcommand{\eqn}[1]{Eq.~\eqref{#1}}

\long\def\comment#1{ }

\newcommand{\nn}{\nonumber\\ }

\def\be{\begin{eqnarray*}}
\def\ee{\end{eqnarray*}}
\def\beq{\begin{eqnarray}}
\def\eeq{\end{eqnarray}}
\newcommand{\bea}{\beq \begin{aligned}}
\newcommand{\eea}{\end{aligned}\eeq}

\def\k{{\boldsymbol k}_\perp}

\def\q{{\boldsymbol q}}

\def\x{{\boldsymbol x}_\perp}

\def\0{{\boldsymbol 0}}

\def\x{{\boldsymbol x}}

\newcommand{\kt}{\k}
\newcommand{\xt}{\x}

\def\rme{{\rm e}}

\def\and{ \quad\text{and}\quad}

\def\abar{{\rm \bar\alpha}}

\def\cP{{\cal P}}

\def\cO{{\cal O}}

\def\rmd{{\rm d}}

\def\der{\text{d}}
\def\dif{\text{d}}

\def\abar{\bar{\alpha}_s}

\def\qhat{\hat{q}}
\def\kt{\boldsymbol{k}_\perp}
\def\xt{\boldsymbol{x}_\perp}

\def\der{\mathrm{d}}

\begin{document}

\title{Universality aspects of quantum corrections to transverse momentum broadening in QCD media }

\author[a]{Paul Caucal,}
\emailAdd{pcaucal@bnl.gov}
\affiliation[a]{Physics Department, Brookhaven National Laboratory, Upton, NY 11973, USA}
\author[a,b]{Yacine Mehtar-Tani,}
\emailAdd{mehtartani@bnl.gov}
\affiliation[b]{RIKEN BNL Research Center, Brookhaven National Laboratory, Upton, NY 11973, USA}

\abstract{
We study non-linear quantum corrections to transverse momentum broadening (TMB) of a fast parton propagating in dense QCD matter in the leading logarithmic approximation. These non-local corrections yield an anomalous super-diffusive behavior characterized by a heavy tailed distribution which is associated with Lévy random walks. Using a formal analogy with the physics of traveling waves, we show that at late times the transverse momentum distribution tends to a universal scaling regime. We derive analytic solutions in terms of an asymptotic expansion around the scaling limit for both fixed and running coupling. We note that our analytic approach yields a good agreement with the exact numerical solutions down to realistic values of medium length.  Finally, we discuss the interplay between system size and energy dependence of the diffusion coefficient $\hat q$ and its connection with the gluon distribution function that is manifest at large transverse momentum transfer. 
}

\date{\today}
\maketitle
\flushbottom

\section{Introduction}\label{sec:intro}
The QCD jets that form in high energy scattering processes involving heavy nuclei such as heavy ion collisions (HIC) and electron-ion collisions undergo substantial final state effects caused by multiple interactions with hot or cold nuclear matter. An important experimental signature of the latter is the strong suppression of jets and high-$p_t$ hadrons, commonly referred to as ``jet quenching'' \cite{Blaizot:2015lma,Qin:2015srf}, that was observed in HIC at RHIC in the early 2000's \cite{PHENIX:2001hpc,STAR:2002ggv} and more recently at the LHC \cite{CMS:2011iwn,ATLAS:2014ipv,ALICE:2013dpt}. In order to gain more insight on the transport properties of the quark gluon plasma (QGP) that is created in the aftermath of the collision, extensive studies of jet events and their substructure are currently carried out \cite{Cunqueiro:2021wls,Andrews:2018jcm}.

One of the key observable is the dijet  acoplanarity that measures the amount of azimuthal decorrelation that a dijet system suffers because of final state interactions \cite{STAR:2002svs,PHENIX:2008osq,ATLAS:2010isq}. It results in particular in the broadening of the jet transverse momentum w.r.t. their initial direction of propagation. Although this effect remains illusive at the LHC due to the large contribution of multi-jet events that are formally resummed in the so-called Sudakov from factor \cite{Mueller:2016gko}, there are hints of momentum broadening at RHIC energies \cite{Mueller:2016gko,Chen:2016vem} (see also \cite{Ringer:2019rfk} for a similar discussion using substructure techniques).  Transverse momentum broadening plays also an important indirect role in the process of radiative energy loss which is the main cause of the phenomenon of jet quenching.

Up until recently, transverse momentum broadening (TMB) was mostly studied at tree level. While multiple scatterings are resummed to all order in the limit where the in medium mean free path $\ell_{\rm mfp}$ is much smaller than the medium length $L$, the scattering rate is usually computed at leading order, i.e. $\alpha_s^2$, in the limit of large momentum transfer $q_\perp \gg T$, where $T$ is the plasma temperature.  However, higher order corrections turned out to be enhanced by potentially large double logarithms of the medium length $\alpha_s\ln^2(L)$  \cite{Liou:2013qya} and therefore must not be neglected. These corrections were shown to be related to a renormalization of the jet quenching parameter $\hat q\equiv \rmd \langle k_\perp^2\rangle_{\rm typ}/\rmd t$
that measures the typical transverse momentum squared per unit time accumulate by the fast parton in the plasma \cite{Blaizot:2014bha,Iancu:2014kga}. On the other hand, they involve gluon fluctuations that may stretch beyond the in-medium correlation length all the way up to the medium length reflecting non-locality of interactions beyond leading order. As a consequence, the diffusion coefficient is time dependent and results in an anomalous diffusion in transverse momentum space in contrast with normal diffusion at tree-level \cite{Caucal:2021lgf}. 

A systematic control of these logarithmically enhanced quantum corrections is not only crucial  for precision phenomenology at RHIC and LHC, as well as at the EIC, but also for probing this novel quantum transport phenomenon in experiment. Some progress has been made in this direction. The single logarithmic corrections enhanced by $\alpha_s\ln(L)$ have been computed in the seminal paper by Liou, Mueller and Wu \cite{Liou:2013qya}. More recently, Zakharov pointed out that relaxing the soft gluon approximation, usually performed in the computation of the radiative corrections, may lead to non-negligible NLO corrections at RHIC or LHC kinematics \cite{Zakharov:2018rst}. Furthermore, the interplay between the dense and dilute limit has been studied in \cite{Blaizot:2019muz}. E.~Iancu proposed a non-linear evolution equation \textit{\`{a} la} JIMWLK to account for the dominant radiative corrections to all orders \cite{Iancu:2014kga}. However, due to its complexity, there is, so far, no known solution (neither analytic nor numeric) to this evolution equation. Nevertheless, in the double logarithmic approximation (DLA), which aims at resumming the subset of these corrections whose general term takes the form $\alpha_s^n\ln^{2n}(L)$, the equation simplifies and can be studied analytically or numerically both at fixed \cite{Liou:2013qya,Blaizot:2014bha,Iancu:2018trm} and running coupling \cite{Iancu:2014sha}. 

Another way to think about this resummation program is in terms of Wilsonian renormalization: the divergence of the quenching parameter (defined at a time scale $\tau_0\sim 1/T$) due to an additional soft and collinear gluon emission with lifetime between $\tau_0$ and $\tau_0+d\tau$ is absorbed into a redefinition of the quenching parameter 
at a time scale $\tau=\tau_0+d\tau$ \cite{Blaizot:2014bha}. This renormalization of $\qhat$ is likely process-independent. For instance it has been shown that the leading, double logarithmic, radiative corrections to the medium-induced gluon spectrum of an off-shell quark can also be included by using the leading order BDMPS-Z spectrum \cite{Baier:1996kr,Zakharov:1996fv,Zakharov:1997uu} with the same effective (renormalized) $\qhat$  as the one involved in the radiative corrections to TMB \cite{Blaizot:2014bha,Iancu:2014kga}. Recently, the authors of \cite{Arnold:2021mow,Arnold:2021pin} demonstrated that this universality also holds at single logarithmic accuracy.

In this work, we perform an analytic and numerical study of the large system size dependence of the jet quenching parameter and the medium saturation momentum that is related to the typical transverse momentum broadening. We derive novel analytic results that are grounded on the double logarithmic approximation for the quantum evolution of $\qhat$. The latter appears to suppress the sensitivity to the initial, tree-level value of the quenching parameter, a property that we shall refer to as ``universality". We carried out the computation of all universal terms in the large $L$ expansion of both the jet quenching parameter and the saturation scale.

The double logarithmic evolution equation is considered both in the fixed coupling approximation and with running coupling correction. In the former case, we detail the mathematical results discussed in our short letter \cite{Caucal:2021lgf}. More precisely, thanks to a deep connection between the fixed coupling DLA equation and the physics of the propagation of traveling waves into unstable states \cite{2003,ARONSON197833,dee1983propagating,bramson1986microscopic,PhysRevA.39.6367,collet2014instabilities} --- typically governed by the Fisher-Kolmogoroff-Petrovsky-Piscounoff (FKPP)  equation \cite{fisher1937,10003528013} ---, we were able to compute all the universal sub-asymptotic corrections borrowing analytic techniques developed in that context. While previous studies were limited to its linearized version, in our approach, we successfully obtained the asymptotic expansion for the full non-linear version of the evolution equation. As we shall show, the non-linearities yield a sizable corrections, especially when $L$ is not asymptotically large. 

These techniques are also successfully applied to the evolution equation for the quenching parameter with running coupling. We provide in particular a proof of the numerically motivated ansatz proposed by Iancu and Triantafyllopoulos in Ref.~\cite{Iancu:2014sha}. In that case too, we are able to push further the accuracy of the asymptotic expansion for large system sizes so that our formulas can be used even for realistic values of the in-medium jet path length.

We emphasize that the aforementioned system size dependence of the renormalized $\hat q$  is different from the energy (or ``$x$") dependence considered in theoretical studies of $\qhat$ in the higher twist approach \cite{Casalderrey-Solana:2007xns,Kang:2014ela,Bianchi:2017wpt}. For high energy jets, with $E\gtrsim \qhat L^2$, the $\ln^2(L)$ enhancement is the dominant contribution from radiative corrections, as we argue in Secs.~\ref{sub:xdep-qhat}-\ref{sub:Edep-Qs}.

The physical interpretation of our results has been largely discussed in \cite{Caucal:2021lgf}. We briefly summarize here our main findings, a more detailed discussion is provided in section~\ref{sec:physics}. For asymptotically large system size, the transverse momentum broadening distribution reaches a universal distribution which satisfies geometric scaling. Namely, it is a function of $k_\perp/Q_s$ only. This asymptotic distribution has interesting properties. First, its typical width that defines the saturation scale or equivalently the typical transverse momentum transfer (measured, for instance, by the median) scales like $L^{1/2+\sqrt{\abar}}$ and thus, grows with the system size faster than $L^{1/2}$. This is characteristic of a super-diffusive regime. Furthermore, the large $k_\perp$ tail of the asymptotic distribution exhibits a harder power spectrum $1/k_\perp^{4-2\sqrt{\abar}}$ when compared to the Rutherford type, i.e.\ $1/k_\perp^4$. These two properties --- super-diffusion and heavy-tailed distribution--- are reminiscent of L\'{e}vy distributions which describe the probability density for the position of a particle undergoing a L\'{e}vy flight. The self-similarity of gluon fluctuations and the non-local nature of the jet-medium interactions are two essential characteristics of L\'{e}vy flights, shared by other systems in various physical areas like optics \cite{PhysRevA.53.3409,PhysRevLett.79.2221}, turbulence and polymer transport theory \cite{shlesinger1993strange,1995LNP...450.....S}.

The article is organized as follows. In the first section we set up the formalism. We introduce in particular the notations and the evolution equations that we shall study in the rest of the paper. For readers who are only interested in the final results, the second section summarizes our main analytic findings regarding the quenching parameter and the saturation momentum as a function of the system size. In sections \ref{sec:fc-analysis} and \ref{sec:rc-analysis}, we detail our analytic approach, based on the similarity with the problem of traveling waves, to compute the asymptotic and pre-asymptotic behavior of the quenching parameter and the saturation momentum for both the fixed and running coupling evolution. Section \ref{sec:numerics} presents an analytic approach to obtaining formulas approximating the exact solution down to realistic values of the system size. We briefly discuss qualitative aspects of the TMB distribution in section \ref{sec:physics}, and in particular the connection with statistical physics through the L\'{e}vy flight physical picture. Finally, in section \ref{sec:Edep-qhat}, we comment on the relation between the operator definition of gluon distributions in DIS and quenching parameter and we discuss the energy dependence of $\qhat$ at smaller jet energy. We highlight possible applications of our formula to phenomenology of heavy-ion collisions and small-$x$ physics in our summary, Sec.~\ref{sec:conclusion}.

\section{Transverse momentum broadening in DLA }\label{sec:qhat-evol}

Consider the eikonal propagation of a high energy parton in a dense QCD medium. By eikonal propagation, we mean that the incoming parton moves along, say, the positive light cone direction\footnote{We use light cone coordinates defined as $x^+=(x^0+x^3)/\sqrt{2}$ and $x^-=(x^0-x^3)/\sqrt{2}$.} with a large longitudinal component of its momentum, $P^+\equiv E$. More precisely, the momentum transferred from the medium is assumed to be small compared to the parton energy, that is, $|\k|\ll E$.

\begin{figure}[t] 
  \centering
  \begin{subfigure}[t]{0.48\textwidth}
     \includegraphics[page=1,width=\textwidth]{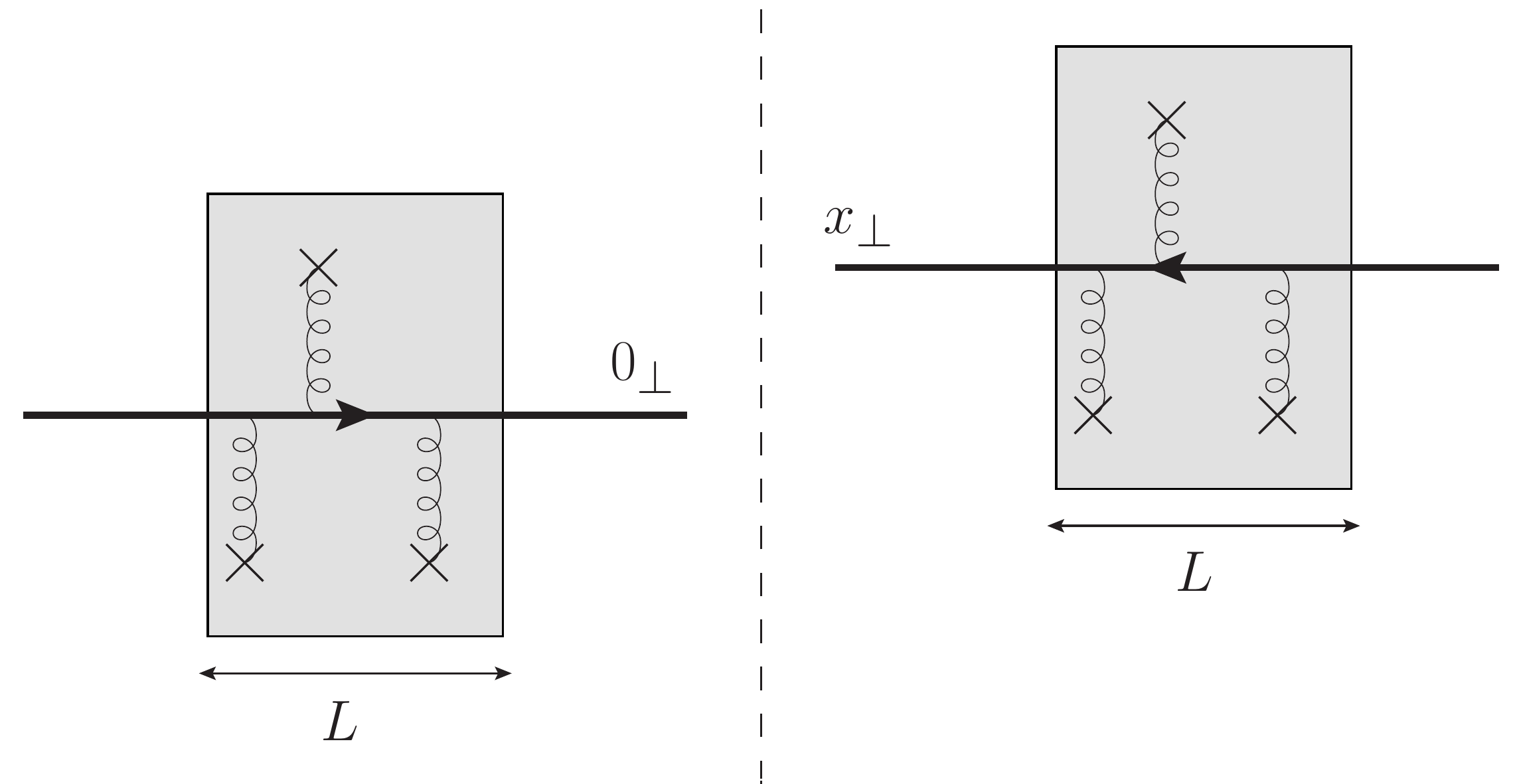} 
    \caption{\small }
    \label{fig:carton-LO}
    \end{subfigure}
  \begin{subfigure}[t]{0.48\textwidth}
     \includegraphics[page=1,width=\textwidth]{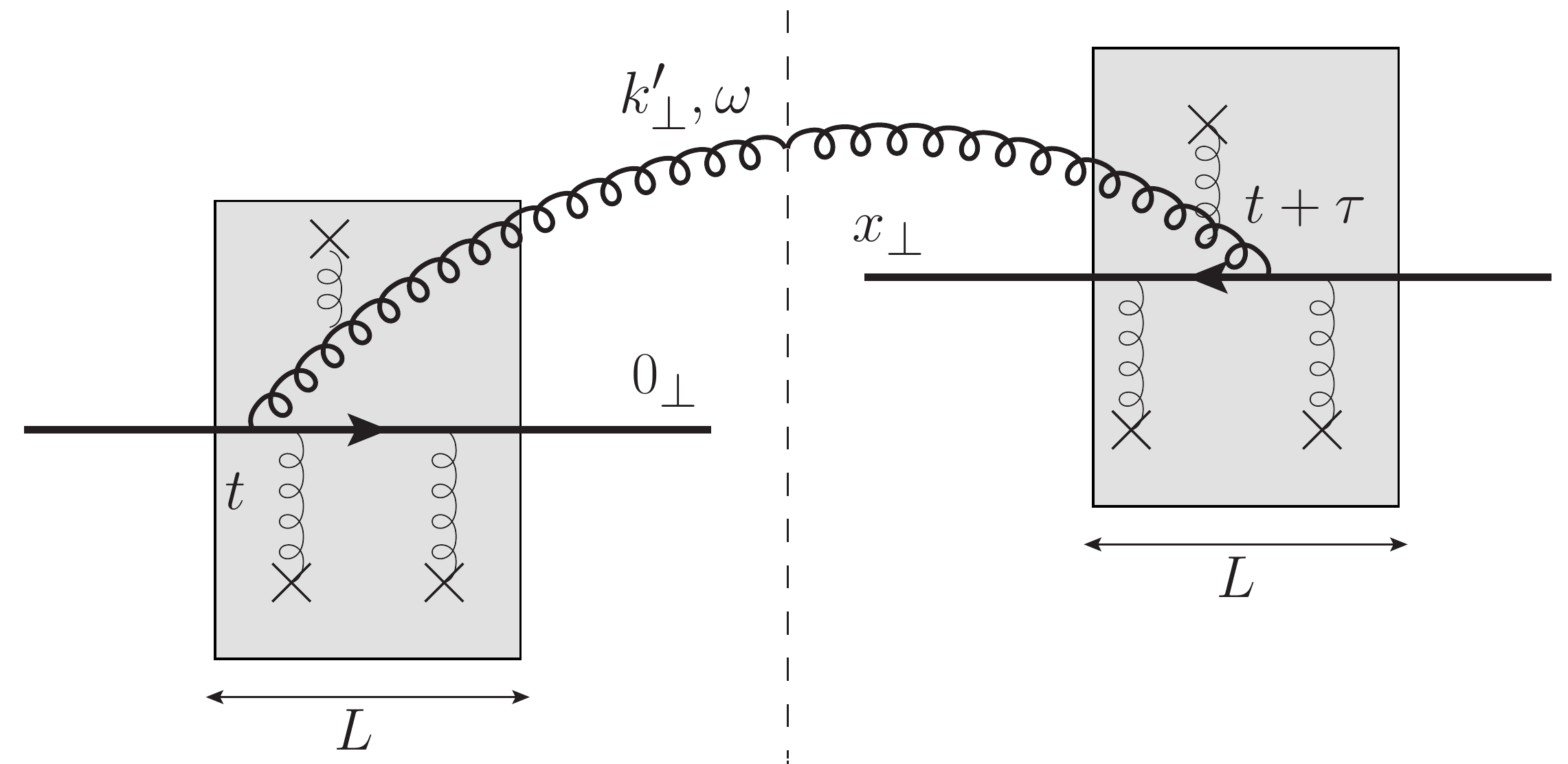} 
    \caption{\small}
    \label{fig:cartoon-NLO}
    \end{subfigure}    
        \caption{(Left) An illustration of a tree-level diagram \cite{BINOSI200476} contributing to Transverse momentum broadening (TMB), where vertical gluons depict multiple scatterings between the incoming quark and the medium scattering centers. The dotted vertical line represents the cut between the amplitude and its complex conjugate (c.c.). (Right) Typical NLO correction to the tree-level distribution due to the radiation of a real gluon of light-cone energy $\omega$ and transverse momentum $k'_\perp$. The transverse vector $x_\perp$ is the difference in transverse position of the quark in the amplitude and its  c.c. It is the Fourier conjugate of the measured transverse momentum broadening.}\label{fig:feynman-diagrams}
\end{figure}

In this approximation, the transverse momentum broadening distribution of the high energy parton can be related to the forward scattering amplitude $S(\xt)$ of an effective dipole in color representation $R=A,F$ with transverse size $\xt$(see e.g. \cite{DEramo:2010wup,Kovchegov:2012mbw,Blaizot:2013vha,Benzke:2012sz}) by a Fourier transform as follows
\beq\label{eq:kt-dist-definition}
\mathcal{P}(\kt)\equiv \frac{\rmd N}{\rmd^2\kt }  = \int \rmd^2 \xt  \, \rme^{-i \kt\cdot \xt}\, S(\xt) \,.
\eeq
where 
\beq
 S(\xt) \equiv \frac{1}{N_c}\mathrm{Tr} \langle U(\xt) U^\dag (\0) \rangle\,, \label{eq:Sxt-def}
\eeq
and 
\beq 
U(\xt) \equiv \cP \exp\left[ig \int_{-\infty}^{+\infty}\rmd x^+ t^a A_a^-(x^+,\xt) \right]\,
\eeq
is a path ordered Wilson line along $x^+$ and $t^a$ are the $SU(3)$ generators in the fundamental representation if the fast parton is a quark. A Feynman graph representation of this tree-level calculation is shown in Fig.\,\ref{fig:carton-LO}.
The medium background field $A^-$ has support in $[0,L]$ where $L$ is the length of the medium and $\langle ...\rangle$ stands for an ensemble average over the medium color charge configurations. 

Note that \eqn{eq:kt-dist-definition} may be encountered in a variety of high energy scattering processes. For instance, in gluon saturation physics at small-$x$ probed in inclusive DIS or forward particle production in high energy collisions such in pA collisions for instance, $A^-$ is the classical field produced by a boosted nuclear/proton target. It also arises in jet production in heavy ion collisions. In that case, $A^-$ is the classical field generated by the color charges of the QGP.

\subsection{Leading order: independent multiple scattering approximation}
In the independent multiple scattering approximation the background field correlations are Gaussian \cite{Blaizot:2013vha}:
\beq
g^2 \langle A^{a, -}(x^+,\q)A^{b,-}(y^+,\q')\rangle  =\delta^{ab} (2\pi)^2\delta^{(2)}(\q-\q')\,\delta(x^+-y^+) \, C(\q) \,.
\eeq
Here, the leading order collision rate $C$ encodes the Coulomb tail 
\beq
C(\q) \simeq \frac{g^4 \,n}{\q^4} \,, \label{eq:collrate}
\eeq
where $n$ stands for the density of scattering centers. The expression \eqref{eq:collrate} for the collision rate is valid in the perturbative regime for $q_\perp \gg \mu$. Throughout this paper, $\mu$ denotes a generic non-perturbative scale. For a weakly coupled quark-gluon plasma in thermal equilibrium at temperature $T$, $\mu$ is of order the Debye mass $m_D\sim gT$. On the other hand, in the case of a boosted nucleus, $\mu$ is a scale of order the inverse nucleon size. The precise form of the collision rate in the non-perturbative domain will not be discussed in the present article. We will mainly focus on the TMB distribution in the perturbative regime where the effect of leading quantum corrections is to strongly suppress non-perturbative physics, resulting in a universal behavior as shall be discussed in detail. 
 
Under the independent multiple scattering approximation, $S(\xt)$ exponentiates as 
\beq\label{eq:dipole-amp}
 S(\xt) = \exp\left[ - \frac{1}{4} \qhat_R^{(0)}(1/\x_\perp^2) \, L \, \xt^2 \right]\,,
\eeq
where the leading order diffusion coefficient $\qhat^{(0)}$ reads 
\beq 
\qhat_R^{(0)}(Q^2) = \int_{\mu^2}^{Q^2} \frac{\rmd^2 \q}{(2\pi)^2} \q^2 C(\q) \approx  4\pi\alpha_s^2 C_R\,n \, \ln \frac{Q^2}{\mu^2 }\,,\label{eq:qhat-tree}
\eeq
up to powers of $ \mu/Q$ suppressed terms. The color representation $R=A,F$ of the hard parton enters through the Casimir $C_R$. Thereafter, we shall write the tree-level quenching parameter in a more generic way as $\qhat^{(0)}=\qhat_0\ln(Q^2/\mu^2)$, with the constant $\qhat_0$ and $\mu$ being model dependent. For instance, for a weakly coupled QGP the bare quenching parameter $\qhat_0$ and the infrared transverse scale $\mu^2$ can be obtained from the hard thermal loop value of the collision rate \cite{Aurenche:2002pd}. They read respectively $\qhat_0=\alpha_sC_Rm_D^2T$ and $\mu=m_D\rme^{-1+\gamma_E}/2$ \cite{Barata:2020rdn}, with $T$ the plasma temperature, $m_D$ the Debye mass.

\eqn{eq:dipole-amp} encompasses two regimes separated by an emergent scale, the saturation scale $Q_s^2(L)$ (in analogy with gluon saturation at small-$x$): the regime $k_\perp \gg Q_s$ is characterized by rare hard events that correspond to $S\ll 1$. The TMB distribution falls like a power law, $\mathcal{P}(\kt)\propto 1/\kt^4$ that is characteristic of Rutherford-like scattering. When  $k_\perp \ll Q_s$ the unitarity bound is saturated, i.e., $S\simeq 1$ and all scattering centers contribute equally. For $\kt\sim Q_s$, the $\kt$ distribution is typically Gaussian, $\mathcal{P}(\kt)\propto \exp(-\kt^2/Q_s^2)$. 
As is customary in small-$x$ physics \cite{Kowalski:2003hm,Lappi:2011ju} or in Moli\`ere theory of multiple scattering \cite{Moliere+1948+78+97,PhysRev.89.1256,Barata:2020rdn}, the saturation scale is mathematically defined by the relation $S(\xt^2=1/Q_s^2(L))\equiv \mathrm{e}^{-1/4}$, or equivalently, 
\beq
Q_s^2 \equiv \hat q_R^{(0)}(Q_s^2)L\,.\label{eq:def-Qs}
\eeq
Eq.\,\eqref{eq:qhat-tree} can be solved for moderate values of $L$ iteratively and one finds approximately $Q_s^2\sim \qhat_0L\ln(\qhat_0L/\mu^2)$ at tree level for the saturation scale. Notice that the saturation scale $Q_s$ depends implicitly on the color representation of the high energy parton via the Casimir dependence of the quenching parameter.

\subsection{One-loop corrections and the double logarithmic resummation}
\label{sub:DLAresum}

\paragraph*{TMB at one loop.} Let us now discuss the transverse momentum distribution at one loop order. At order $\alpha_s$, one must include two contributions, one real and one virtual gluon attached to the incoming parton. A typical Feynman diagram representation of a real NLO correction that contributes to $\qhat$ is presented in Fig.\,\ref{fig:cartoon-NLO}. The limit in which the additional gluon is soft was addressed in detail in \cite{Blaizot:2014bha} (see also the appendix A of \cite{Liou:2013qya} for a calculation in Zakharov's formalims). Leaving technical details aside for the sake of clarity,  we may write schematically: 
\begin{equation}
\mathcal{P}(\kt)=\mathcal{P}^{(0)}(\kt)+\alpha_s\mathcal{P}^{(1)}(\kt)+\mathcal{O}(\alpha_s^2)\,,\label{eq:TMB-1loop}
\end{equation}
where $\mathcal{P}^{(0)}(\kt)$ is the tree-level TMB distribution given by Eqs.\,\eqref{eq:kt-dist-definition}-\eqref{eq:dipole-amp}. Even though Eq.\,\eqref{eq:TMB-1loop} looks like a standard perturbative expansion, one should keep in mind that each term actually resums to all orders powers of $\alpha_s^2n L $ due to multiple scatterings.
One can show that the leading contribution from $\alpha_s\mathcal{P}^{(1)}(\kt)$ comes from radiative corrections which are quasi-local, in the sense that they can occur everywhere inside the medium over the path length $L$ of the high energy parton. The average $\kt^2$ associated with such contributions reads
\begin{equation}
\left\langle\kt^2\right\rangle_{\rm 1-loop}\sim\frac{\alpha_sN_c}{\pi}L\times\int\frac{\der\omega}{\omega}\int^{\sqrt{\omega/\qhat_0}}\frac{\der\tau}{\tau}\qhat_0\,,\label{eq:mean-kt-1}
\end{equation}
where $\omega$ and $\tau$ are respectively the energy and the lifetime of the gluon fluctuation. This lifetime is bounded from above by the typical formation time of a medium-induced emission triggered by multiple soft scatterings. The latter bound ensures that the radiative process is triggered by a single scattering with medium constituents. The other boundaries of the double integral will be specified below. This double integral has a typical double logarithmic structure, $\langle \kt^2\rangle\sim \alpha_s \ln^2$, and therefore, the smallness of $\alpha_s$ can be compensated by the large logarithm squared, spoiling the convergence of the series \eqref{eq:TMB-1loop}. Note also that $\langle \kt^2\rangle$ is enhanced by the system size $L$, as a result of the quasi-locality of the double logarithmic quantum corrections \cite{Blaizot:2014bha}.

In this paper, we are mainly interested in the resummation of double logarithmic contributions of the form \eqref{eq:mean-kt-1} to all orders in the series \eqref{eq:TMB-1loop}. This resummation can be performed at the level of the quenching parameter itself, or in other words, the double logarithmic corrections \textit{exponentiate} owing to the fact that the logarithmic  $\tau$ integral is dominated by the regime $\tau \ll L$ (for more details see discussion in \cite{Blaizot:2014bha}):
\begin{equation}
\mathcal{P}(\kt)=\int \rmd^2 \xt  \, \rme^{-i \kt\cdot \xt}\, \exp\left[ - \frac{1}{4} \left(\qhat^{(0)}+\alpha_s\qhat^{(1)}+...\right) \, L \, \xt^2 \right]\,,
\end{equation}
with $\qhat^{(1)}\sim\qhat_0\ln^2$, $\qhat^{(n)}\sim\qhat_0\ln^{2n}$, etc.
A complete NLO computation of the TMB distribution would require both the resummation of the double and single logarithms via this exponentiation property, and a proper matching with the fixed order result \eqref{eq:TMB-1loop}. Such calculation is beyond the scope of this paper. We refer the interested reader to Refs.~\cite{Liou:2013qya,Arnold:2021mow} where some aspects of the single logarithmic contribution to momentum broadening and radiative energy loss are discussed.  

\paragraph*{Double logarithmic phase space for $\qhat$.} We now detail the double logarithmic phase space for the quenching parameter $\qhat$. In order to specify these boundaries, it is more convenient to perform the change of variable $\omega\to \kt'^2=2\omega/\tau$. From the logarithmic corrections, the quenching parameter acquires a $\tau$ and $\kt^2=1/\xt^2$ dependence. At one loop, the double logarithmic contribution reads
\begin{equation}
\qhat^{(1)}(\tau,\kt^2)=\frac{\alpha_sN_c}{\pi}\int_{\tau_0}^\tau\frac{\der\tau'}{\tau'}\int_{Q_s^2(\tau')}^{\kt^2}\frac{\der\kt'^2}{\kt'^2}\qhat^{(0)}\,.
\end{equation}
Here $\tau_0$ is a cut-off time scale of the order of the mean free path, which reflects the uncertainty of this calculation due to the non-perturbative physics of the plasma. $Q_s(\tau)$ is the saturation scale at the lifetime $\tau$, defined through the relation \eqref{eq:def-Qs} or in terms of the function $\qhat(\tau,\kt^2)$:
\begin{equation}
Q_s^2(\tau)\equiv\qhat(\tau,Q_s^2(\tau))\tau\,.
\end{equation}
 It is instructive to estimate this integral using a constant tree-level $\qhat$ value, namely $\qhat(\tau,\kt^2)=\qhat_0$ leading to $Q_s^2(\tau)\simeq\qhat_0\tau$. The consequences of this approximation on the evolution of the quenching parameter will be discuss in details in this paper. The single hard scattering condition $\tau'\leqslant \sqrt{\omega/\qhat}$ becomes $\kt'^2\geqslant Q_s^2(\tau')$ with our new variables. Note that the color factor is $N_c$ since we are dealing with the scattering of a gluon at the one loop order. 
 
 The main difference with the standard double log encountered in DGLAP is that the collinear and soft logs talk to each other through the saturation line that plays the role of physical cutoff for the collinear singularity as a result of coherence effects from multiple scattering. 

If one is interested in $k_\perp \gg Q_s(L)$ we would have (neglecting the logarithmic dependence of the $\hat q^{(0)}\equiv \qhat_0$ at leading order for simplicity)
\beq
\hat q^{(1)}(L,\kt^2\geqslant Q_s^2(L))= \abar  \int_{\tau_0}^L \frac{\rmd \tau' }{ \tau' }  \int_{Q_s^2(\tau')}^{k_\perp^2} \frac{\der \kt'^2 }{\kt'^2 } \, \qhat_0= \abar \qhat_0\left( \ln \frac{\kt^2}{\mu^2}\ln \frac{L}{\tau_0}-\frac{1}{2}\ln^2 \frac{L}{\tau_0}\right)\,,\nn
 \eeq
with $\mu^2=\qhat_0\tau_0$ and $\abar = \alpha_sN_c/\pi$, where the microscopic scale $\tau_0$ is related to the in-medium mean-free-path which for a thermal plasma scales as $(g^2T)^{-1}$ at weak coupling. 

\begin{figure}[t] 
  \centering

     \includegraphics[width=\textwidth]{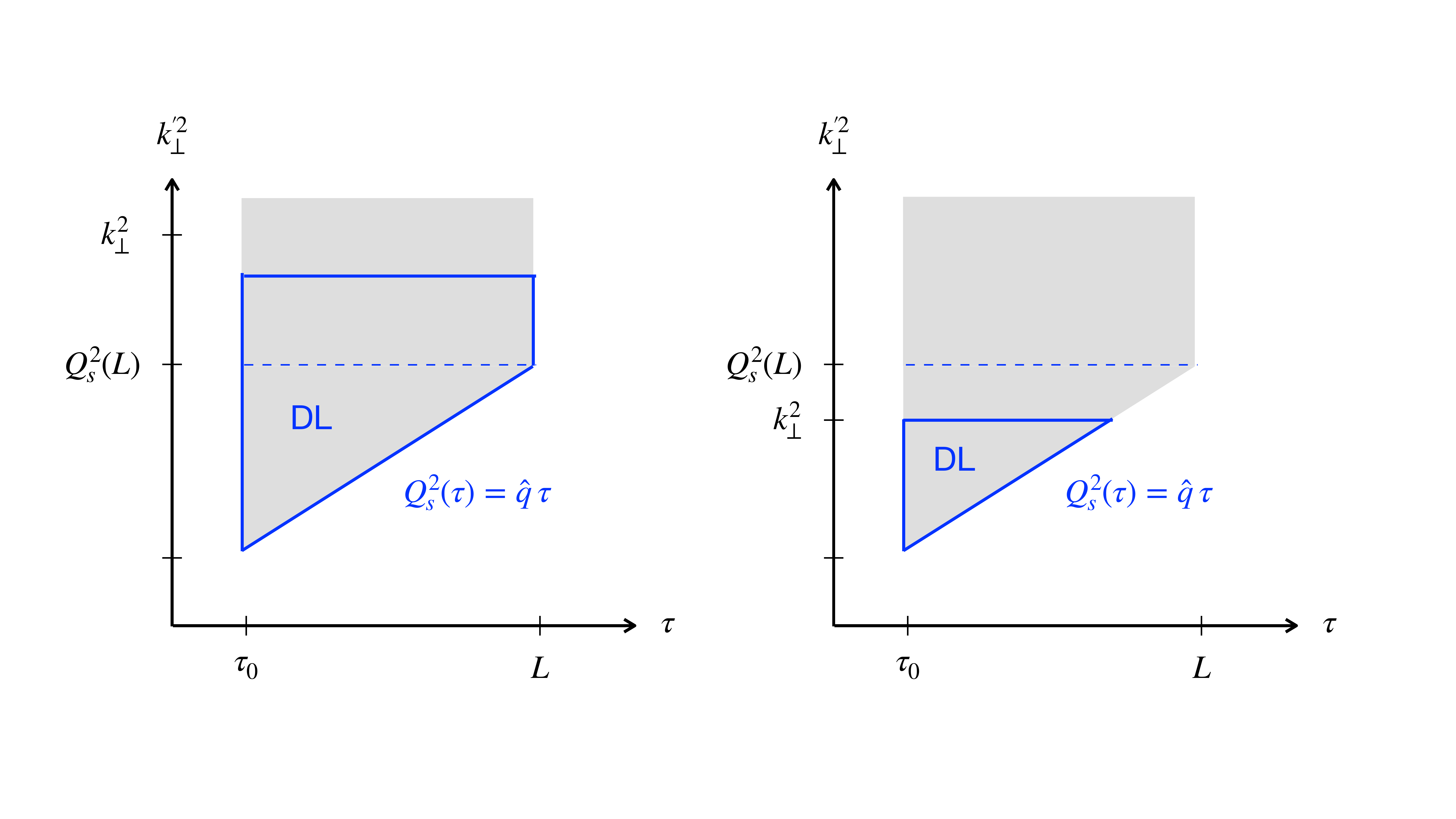}

    \caption{The double logarithmic phase space for gluons fluctuations with lifetime $\tau$ and transverse momentum $k_\perp'$ in the dilute (left) and dense (right) regimes. The choice between these two regimes depends on the final momentum $k_\perp$ of the double logarithmic cascade. For the TMB distribution, we set $k_\perp=1/x_\perp$ where $x_\perp$ is the initial effective dipole size. In this sketch, we have simplified the saturation line $Q_s(\tau)$ by its tree-level form $\qhat_0\tau$. }\label{fig:qhat-phase-space}
\end{figure}

This double integration corresponds to the area of the right trapezoid depicted in Figure~\ref{fig:qhat-phase-space} (left panel). When the upper limit of the $\kt'$ integration falls below $Q_s^2$, that is, $\kt^2 < Q_s(L)^2$, one is left with the area of a triangle and one obtains the following double log  (cf. Figure~\ref{fig:qhat-phase-space} (right panel)) 
\beq
 \hat q^{(1)}(\kt^2 < Q_s^2)=  \frac{\abar }{2}\hat q^{(0)}  \ln^2 \frac{\kt^2}{\mu^2}\,.
\eeq
To obtain the corrections to the typical value of transverse momentum broadening we must evaluate $\hat q $ at $\kt^2 = Q_s^2(L)\simeq \hat q^{(0)} L $ which yields the Liou-Mueller-Wu result \cite{Liou:2013qya}
\beq
\langle k_\perp^2\rangle_{\rm 1-loop,DL} \simeq \left( \hat q^{(0)}+ \hat q^{(1)}\right) L =\qhat_0 L \left(1 + \frac{\abar}{2}\ln^2 \frac{L}{\tau_0}\right)\,.
\eeq

\begin{figure}[t] 
  \centering
     \includegraphics[width=10cm]{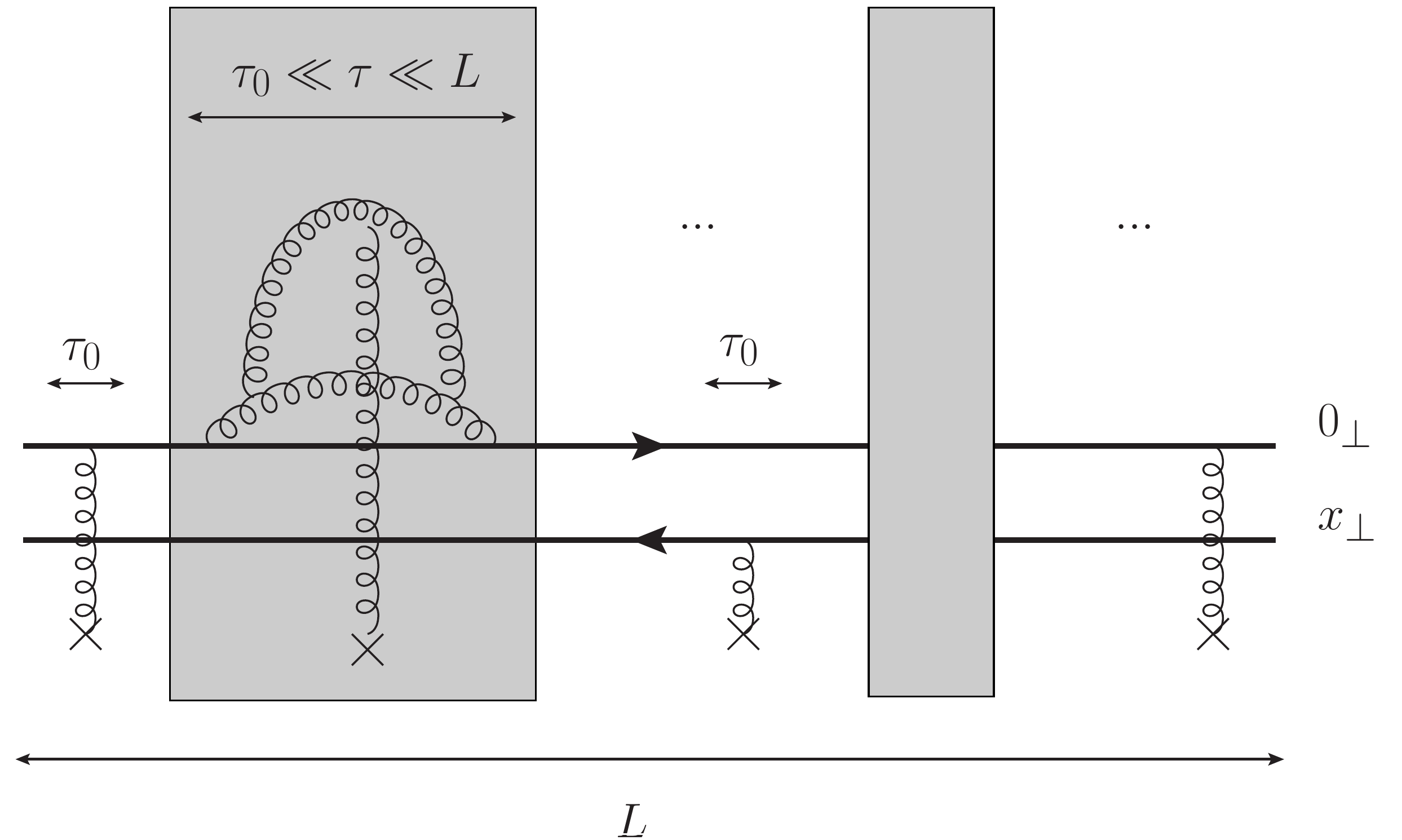} 
    \caption{An illustration of multiple radiative corrections considered in this paper. Each block represents a tower of gluon fluctuations triggered by a single scattering with a medium constituent, with strongly decreasing lifetime and transverse momentum along the cascade (so the transverse size of a gluon increases from the parent to its daughter). The exponentiation resums several such blocks over the path length $L$ of the incoming hard effective dipole with transverse size $x_\perp\sim 1/k_\perp$.}\label{fig:cartoon-resum}
\end{figure}

\paragraph*{Resummation.} In the double logarithmic accuracy (DLA) these radiative corrections can be resummed to all orders via an evolution equation ordered in 
$\tau$~\cite{Liou:2013qya,Blaizot:2014bha,Iancu:2014kga}: 
\begin{align}
\qhat(\tau,\kt^2)&=\qhat^{(0)}(\tau_0,\kt^2)+\int_{\tau_0}^{\tau}\frac{\dif\tau'}{\tau'}\int_{Q^2_{ s}(\tau')}^{\kt^2}\frac{\dif \kt'^2}{\kt'^2} \ \abar(\kt'^2) \ \qhat(\tau',\kt'^2)\,,\label{eq:qhat-DL}\\
Q^2_{s}(\tau)&=\qhat(\tau,Q_{ s}^2(\tau))\tau\,,\label{eq:Qsat}
\end{align}
where $\hat q^{(0)} (\tau_0,\kt)$ corresponds to the tree-level initial condition. The strong coupling constant appears inside the integral over $\kt'$ to account for its running with the transverse scale. It is convenient to re-express these two equations in terms of the logarithmic variables 
\beq \label{eq:log-var}
Y=\ln\frac{\tau}{\tau_0} \quad \text{and} \quad  \rho=\ln\frac{\kt^2}{\qhat_0\tau_0}\,.
\eeq
Thus, 
\begin{align}
\qhat(Y,\rho)&=\qhat^{(0)}(0,\rho)+\int_0^Y\der Y'\int_{\rho_s(Y')}^\rho\der\rho'\,\abar(\rho')\qhat(Y',\rho')\,,\\
\qhat(Y,\rho_s(Y))&=\qhat_0 \rme^{\rho_s(Y)-Y}\,.
\end{align}
This non-linear evolution equation resums  the double logarithms $\alpha_sY\rho$ to all orders. Also, it is valid in the large $N_c$ limit which is reflected in the overall $N_c$ factor absorbed in the constant $\abar$. In principle, since the definition of $Q_s$ is ``flavor" dependent, there should be a coupling between the evolution of the quenching parameter $\qhat_F$ in the fundamental representation and the adjoint one $\qhat_A$. In this paper, we do not consider the effect of such a coupling (which is beyond DLA) and focus on the evolution of $\qhat_A$ only. At this accuracy, the fundamental $\qhat$ can be obtained from $\qhat_A$ using $\qhat_F=C_F/C_A\qhat_A$. A graphical representation of this evolution equation is displayed in Figure~\,\ref{fig:cartoon-resum}.

For the running coupling evolution, we use the one loop beta function to determine the $\rho$ dependence of $\abar$:
\begin{equation}
\abar(\rho) = \frac{b_0}{\rho+\rho_0}\,,
\end{equation}
with 
\beq 
\frac{1}{b_0}= \frac{11}{12}-\frac{N_fT_R}{3N_c}\,,
\eeq
where $N_f=5$ is the number of quark flavors. 

\paragraph*{Dense and dilute regime.} In determining the transverse momentum distribution we have to distinguish between the dense regime $\rho<\rho_s(Y)$ and the dilute one $\rho \geqslant \rho_s(Y)$.  In the latter, recall that the quenching parameter is function of two independent variables  $Y=\ln (L/\tau_0)$ and $\rho$. In the dense regime, however, there is a subtlety in the choice of variables. Given a general solution $\qhat(Y,\rho)$, the variable $Y$ is no longer an independent function of $\rho$. Again, this variable is related to the upper limit of the $\tau$ integral in the DL phase-space as illustrated in Figure~\ref{fig:qhat-phase-space}. Therefore, the logarithmic variable $Y$ must be fixed such that the quenching parameter $\qhat$ which appears inside the forward scattering amplitude is only a function of $\rho$. The relevant time scale $\tau_s(\kt^2)$, or in logarithmic variables $Y_s(\rho)$, at which the quenching parameter $\qhat$ must be evaluated is the largest time allowed by the saturation condition:
\begin{equation}
Y_s(\rho)=Y\quad \Leftrightarrow\quad \rho = \rho_s(Y)\,.
\end{equation}
The function $Y_s(\rho)$ is then the inverse function of $\rho_s(Y)$. Again, in the dilute regime, the value of $Y$ is fixed by the typical path length $L$ of the hard parton inside the dense medium \cite{Blaizot:2019muz}, 
\begin{equation}
Y=\ln(L/\tau_0)\,.
\end{equation}
This is illustrated in  Figure~\ref{fig:qhat-phase-space} where we see that for $k^2_\perp <Q^2_s(L) $ we have $\tau < k_\perp^2/\hat q(\tau_s) \sim k_\perp^2/\hat q_0$, where $\tau_s\equiv \tau_s(k_\perp^2)$ is defined by $\tau_s= k_\perp^2/\hat q(\tau_s)  $, whereas for $k^2_\perp <Q^2_s(L) $ we simply have $\tau < L$. 

In summary, the function of $\kt^2$ (or $\rho$) to be used in the forward dipole amplitude is:
\begin{align}
\qhat(L,1/\xt^2)=\begin{cases}
              \qhat_>(Y,\rho)=\qhat(Y,\rho) & \textrm{ if } \rho\geqslant \rho_s(Y)\label{eq:qhat-geoscal}\\
                  \qhat_<(Y,\rho)=\qhat(Y_s(\rho),\rho) & \textrm{ if }\rho<\rho_s(Y)\,.                
               \end{cases}
\end{align}
As we shall see in Sec.\,\ref{sec:numerics}, this function is continuous and derivable in $\rho=\rho_s(Y)$, but the second derivative is not in general continuous. 

\paragraph*{Linearization.} Exact analytic solutions of the non-linear system \eqref{eq:qhat-DL}-\eqref{eq:Qsat} are in general difficult to obtain. However, there are analytic solutions to the  fixed coupling, linearized problem that consists in approximating $Q_s^2(\tau)\simeq \qhat_0\tau$ in the lower bound of the $\kt'$ integral in Eq.\,\eqref{eq:qhat-DL} \cite{Iancu:2014sha,Mueller:2016xoc}. Under these approximations, the integro-differential equation for $\qhat$ decouples from the implicit equation satisfied by $Q_s$. The saturation scale is defined instead by
\begin{equation}
Q_s^2(L)=\qhat(L,\qhat_0 L)L\,, \qquad \rho_s(Y)=Y+\ln\left(\frac{\qhat(Y,Y)}{\qhat_0}\right)\,.\label{eq:def-Qs-linear}
\end{equation} 
In terms of the logarithmic variables $Y$ and $\rho$, $\qhat(Y,\rho)$ satisfies
\beq\label{eq:evol-log-var}
 \hat q(Y,\rho) = \hat q_0(\rho)+\abar \int_0^Y \rmd Y' \int^\rho_{Y'} \rmd \rho' \hat q(Y',\rho')\,.
\eeq
The analytic solutions for $\qhat(Y,\rho)$ can be obtained by iterations \cite{Liou:2013qya,Iancu:2014sha,Mueller:2016xoc}. For constant tree-level initial conditions, $\qhat^{(0)}(\rho)=\qhat_0$, the solution reads
\begin{equation}
  \qhat(Y,\rho)=\qhat_0\left[\textrm{I}_0\left(2\sqrt{\abar Y\rho}\right)-\frac{Y}{\rho}\textrm{I}_2\left(2\sqrt{\abar Y\rho}\right)\right]\,,\label{eq:analytic-sol-1}
\end{equation}
while for leading twist tree-level initial conditions $\qhat^{(0)}(\rho)=\qhat_0\rho$, one obtains
\begin{equation}
\qhat(Y,\rho)=\qhat_0\frac{\rho}{\sqrt{\abar Y \rho}}\left[\textrm{I}_1\left(2\sqrt{\abar Y\rho}\right)-\frac{Y^2}{\rho^2}\textrm{I}_3\left(2\sqrt{\abar Y\rho}\right)\right]\,,\label{eq:analytic-sol-2}
\end{equation}
where the $\mathrm{I}_n(x)$'s are the modified Bessel functions of the first kind. Note that since we consider the linearized evolution equation, the solution with initial condition $\qhat^{(0)}(\rho)=\qhat_0\rho+{\rm const.}$ is simply the linear superposition of the two solutions above.
In the asymptotic limit, when $\abar Y \rho \gg 1$ the quenching parameter behaves roughly  like 
\beq\label{eq:asymp-exp}
\qhat(Y,\rho) \sim \rme^{2\sqrt{\abar Y\rho}} \,.
\eeq
A detailed discussion of the asymptotics of these solutions will be presented in Section~\ref{sec:fc-linear}.

\section{Asymptotics of TMB: highlights of the main results for $E\gg \hat q L^2$}\label{sec:results}

This section is meant for the reader who is less interested in the formal derivations detailed in the next two sections, but rather in the analytic formulas we have obtained that can be used for heavy-ion or small-$x$ phenomenology. 

We will focus on high energy limit: $E\gg \omega_c\equiv  \hat q L^2$ and postpone the discussion of the opposite regime in Section~\ref{sub:Edep-Qs}.

In the following we summarize our analytic results for the asymptotic expansion of the quenching parameter and the saturation scale in both fixed coupling and running coupling scenarios. We display our results for $\qhat(L,\kt^2=1/\xt^2)$ as a function of $\rho=\ln(\kt^2/(\qhat_0\tau_0))$ and $Y=\ln(L/\tau_0)$. Similarly, the asymptotic expansion of the saturation scale is written in terms of $\rho_s(Y)=\ln(Q_s^2(L)/\mu)$ so that 
\beq
Q_s^2(L) = \mu^2 \,\rme^{\rho_s(Y)} \,.
\eeq

\subsection*{Fixed coupling}

For the fixed coupling evolution, the saturation scale is given by
\begin{equation}
\rho_s(Y)=c\,Y-\frac{3c}{1+c}\ln(Y)+\kappa-\frac{6c\sqrt{2\pi(c-1)}}{(1+c)^2}\frac{1}{\sqrt{Y}}+\mathcal{O}\left(\frac{1}{Y}\right)\,,\label{eq:rhos-fc-summary}
\end{equation}
where  $c=1+2\sqrt{\abar+\abar^2}+2\abar$ is the celerity of the front 
and $\kappa$ is a non-universal integration constant to be determined numerically. 

Defining
\beq x=\rho-\rho_s(Y), \qquad \rm{and} \qquad \beta=\frac{c-1}{2c}\,,
\eeq
 we have
\beq
&&\frac{\qhat(Y,x)L}{Q_s^2(L)}= \nn
&&\begin{dcases}
             \exp\left(\beta  x-\frac{\beta  x^2}{4cY}\right)\left[1+\beta  x-\frac{3x}{c(1+c)Y}\left(1+\frac{\beta (c+4)x}{6}\right)+\mathcal{O}\left(\frac{1}{Y^2}\right)\right] & \textrm{ if } x\ge 0\label{eq:qhat-fc-summary}\nn
            \exp\left(2\beta x-\frac{3}{c(1+c)}\frac{x}{Y}+\mathcal{O}\left(\frac{1}{Y^2}\right)\right) & \textrm{ if } x<0\,.         
               \end{dcases}\nn
\eeq

\subsection*{Running coupling}
\begin{figure}[t] 
  \centering
     \includegraphics[page=1,width=10cm]{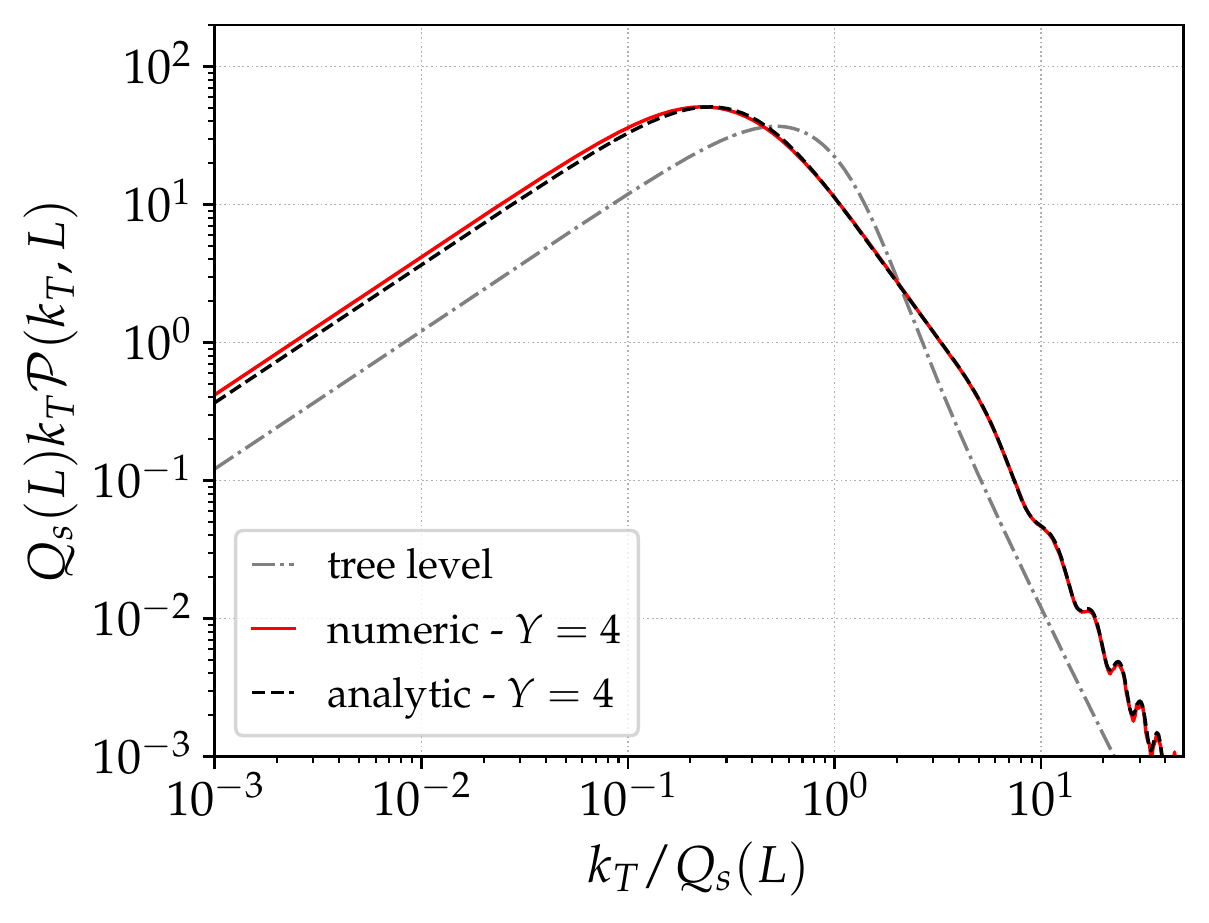} 
        \caption{The numerical and analytic TMB distribution after running coupling evolution obtained from the asymptotic study performed in this paper. The value of $Y=\ln(L/\tau_0)$ is $Y=4$. The grey curve is the tree-level distribution, for comparison.}\label{fig:punchline-plot}
\end{figure}

In the running coupling case, we present analytic results that are relevant for realistic values of $Y$ (not too large, typically smaller than $Y=5$). For such values, the effects of the non-linearity are mild and one can focus on the linear evolution. The saturation momentum reads in that case:
\begin{align}
\rho_s(Y)&=Y+2\sqrt{4b_0Y}+3\xi_1(4b_0Y)^{1/6}+\frac{1}{4}\ln(Y)+\kappa+\frac{7\xi_1^2}{180}\frac{1}{(4b_0Y)^{1/6}}\nonumber\\
&+\frac{5\xi_1}{108}\frac{1}{(4b_0Y)^{1/3}}-\left(\frac{9}{560}-\frac{1693\xi_1^3}{340200}-4b_0\rho_0\right)\frac{1}{(4b_0Y)^{1/2}}+\mathcal{O}\left(\frac{1}{Y^{2/3}}\right)\,.\label{eq:rhos-rc-summary}
\end{align}
The quenching parameters $\qhat(Y,x)$ which enters inside the TMB distribution is given by
\beq
&& \frac{\qhat(Y,x)L}{Q_s^2(L)}=\nn
&& \begin{dcases}
           \left[1+\left(\frac {\dot\rho_s-1}{\dot\rho_s}\right)x+\frac{1}{2}\left(\left(\frac{\dot\rho_s-1}{\dot\rho_s}\right)^2+\frac{\ddot\rho_s}{\dot\rho_s^3}-  \frac{\abar(\rho_s)}{\dot\rho_s}\right)x^2+\mathcal{O}\left(\frac{x^3}{Y^{1/3}}\right)\right]  & \textrm{ if } x\ge 0\label{eq:qhat-rc-summary}\nn
            \exp\left(\frac{\dot\rho_s-1}{\dot\rho_s}x+\frac{1}{2}\frac{\ddot\rho_s}{\dot\rho_s^3}x^2+\mathcal{O}\left(\frac{x^3}{Y^{1/3}}\right)\right) & \textrm{ if } x<0\,.               
               \end{dcases}\nn
\eeq
Note that compared to the fixed coupling result, we have not expanded in powers of $Y$ each coefficient of the polynomial functions in $x$ which appear in $\qhat(Y,x)$. The reason is that the convergence of the development of $\rho_s$ is slower (due to the $1/6$ power instead of $1/2$). Therefore, one must keep the entire $\rho_s$ development \eqref{eq:rhos-rc-summary} in order to reach realistic values of $Y$. With this expression, we are able to accurately estimate the $\kt$-distribution over 4 orders of magnitude around the saturation momentum, as shown Figure.~\ref{fig:punchline-plot}.

\section{Geometric scaling at fixed coupling }\label{sec:fc-analysis}
In the regime where the energy $E$ of the propagating parton is much larger than the characteristic frequency $\omega_c=\hat q L^2$, the broadening distribution does not depend on $E$. As a result, the physics will be governed by a single scale $Q_s(L)$ that is a function of system size and local medium properties such as the temperature $T$. Furthermore, the $k_\perp$-distribution will depend asymptotically on a single scaling variable $k_\perp/Q_s$ over a wide range of $k_\perp$'s.  The latter property will be refer to as geometric scaling. 

To study the asymptotics of this regime, we shall take the formal limit $E\to\infty$ to compute the asymptotic behavior of the quenching parameter $\qhat(L,\kt^2)$ and the saturation scale $Q_s(L)$ for large system sizes. Our calculation will be accurate up to terms of order $1/\ln(L)$. Solving exactly the non-linear evolutions for $\hat q$ is a difficult task. Therefore, in order to gain insight on the generic properties of solutions we shall first analyze the linearized version of the evolution equation for which exact analytic results can be obtain for arbitrary system sizes. This exercise will serve as an anchor point for the non-linear case where only asymptotic results will be derived. 

Furthermore, we will take advantage of a mathematical equivalence between the evolution of the quenching parameter and the formation of traveling wave-front in non-linear physics to discuss the onset of the extended geometric scaling. 

In section we focus on the fixed coupling case. The running coupling will lead to a different scaling which will be discuss in detail in Section~\ref{sec:rc-analysis}. 

\subsection{Asymptotics of the analytic solution to the linearized equation}\label{sec:fc-linear}

We remind the reader that the linearized evolution equation is obtained by replacing $Q_s$ in the lower bound of the  $k_\perp'$ integral in r.h.s. of Eq.\,\eqref{eq:qhat-DL} by its leading order value $\hat q_0 \tau$. This turns out to be a good approximation since non-linear effects do not alter the leading asymptotic behavior as we shall see. 

Thanks to the analytic solutions \eqref{eq:analytic-sol-1} and \eqref{eq:analytic-sol-2}, we want to show that the linearized evolution equation satisfies the following scaling property at large $Y$:
\begin{equation}
\qhat(Y,\rho)\underset{Y\to\infty}{\sim}\qhat_0 e^{\rho_s(Y)-Y}f(x=\rho-Y)\,,
\end{equation}
for some function $\rho_s(Y)$ to be determined. Using the asymptotic expansion of the Bessel functions, 
\begin{equation}
\mathrm{I}_n(z)\underset{z\gg1}{=}\frac{e^z}{\sqrt{2\pi z}}\left(1+\frac{1-4n^2}{8z}+\mathcal{O}\left(\frac{1}{z^2}\right)\right)\,,\label{eq:bessel-exp}
\end{equation}
Eq.\,\eqref{eq:analytic-sol-2} can be approximated for large values of  $Y$ by
\begin{equation}
\qhat(Y,\rho)=\qhat_0\frac{\rme^{2\sqrt{\abar Y\rho}}}{\sqrt{2\pi}(\abar Y\rho)^{1/4}}\left[1+\frac{1}{16\sqrt{\abar Y \rho}}-\frac{Y}{\rho}\left(1-\frac{15}{16\sqrt{\abar Y \rho}}\right)\right]\,.
\end{equation}
In order to get the correct scaling function when $Y\to\infty$, it is necessary to keep track of the sub-leading terms in the expansion \eqref{eq:bessel-exp}.
Replacing $\rho=x+Y$ and expanding for large $Y$, we find
\begin{equation}
\qhat(Y,\rho)=\qhat_0\,\rme^{\rho_s(Y)-Y}\,\left(1+\sqrt{\abar} x+\mathcal{O}\left(\frac{x}{Y}\right)\right)\,\exp\left(\sqrt{\abar} x-\frac{\sqrt{\abar} \tilde x^2}{4Y}\right)\,,\label{eq:qhat-linear-asymptote}
\end{equation}
with $\rho_s(Y)=(1+2\sqrt{\abar})Y-3/2\ln(Y)+{\rm const}$, which translates into the saturation scale 
\beq
Q^2_s(L) = \mu^2  \rme^{\rho_s(Y)} \propto  \frac{L^{^{(1+2\sqrt{\abar}) }}}{(\ln L)^{3/2}}\,.
\eeq 

Note that one obtains the same result starting from Eq.\,\eqref{eq:analytic-sol-2}, the difference between the two initial conditions is encoded in the constant term in $\rho_s(Y)$. This is an important observation as it illustrates the universality of the asymptotic solution for $\qhat(Y,\rho)$, namely, the fact that it loses sensitivity to the tree-level initial condition at large $Y=\ln L/\tau_0$. One can even infer from the exact analytic solutions for $Q_s^2(Y)$ the order of the non-universal coefficients. For a constant initial condition, we have (using Eq.\,\eqref{eq:def-Qs-linear}):
\begin{align}
\rho_s(Y)&=Y+\ln\left(\frac{\mathrm{I}_1(2\sqrt{\abar}Y)}{\sqrt{\abar} Y}\right)\,,\\
&=(1+2\sqrt{\abar})Y-\frac{3}{2}\ln(Y)-\frac{1}{2}\ln(4\pi\abar^{3/2})-\frac{3}{16}\frac{1}{\sqrt{\abar} Y}+\mathcal{O}\left(\frac{1}{Y^2}\right)\,,
\end{align}
while for the initial condition $\qhat^{(0)}(\rho)=\qhat_0\rho$, we get
\begin{align}
\rho_s(Y)&=Y+\ln\left(\frac{2\mathrm{I}_2(2\sqrt{\abar}Y)}{\abar Y}\right)\,,\\
&=(1+2\sqrt{\abar})Y-\frac{3}{2}\ln(Y)-\frac{1}{2}\ln(4\pi\abar^{5/2})-\frac{15}{16}\frac{1}{\sqrt{\abar} Y}+\mathcal{O}\left(\frac{1}{Y^2}\right)\,,
\end{align}
We conclude then that for the linearized evolution equation, only the first two terms are universal, whereas the constant and $\mathcal{O}(Y^{-1})$ terms depend on the initial condition. As we shall see, these statements remain correct in the non-linear case, with two important differences though: the coefficients of the $Y$ and $\ln(Y)$ terms are modified and another universal power $1/\sqrt{Y}$ appears in the development.

Finally, we emphasize that not only does Eq.\,\eqref{eq:qhat-linear-asymptote} account for the scaling limit, given by 
\begin{equation}
f(x)=\rme^{\sqrt{\abar} x}\left(1+\sqrt{\abar} x\right)\,,\label{eq:scaling-f-fc-lin}
\end{equation}
but it also encompasses the sub-asymptotic corrections via the $1/Y$ suppressed term inside the exponential and the prefactor. The latter informs us about the range of geometric scaling, that is, $ x \ll  Y\sim \rho_s$, or in terms of physical variables $\kt^2\ll Q_s^4/\mu^2$.

Or course, in this analysis the scaling is not exact since the scaling variable is $ x=\rho-Y$ which involves $Y$ rather than $\rho_s$. We shall see that in the non-linear equation $Y$ is replaced by $\rho_s(Y)$ resulting in a scaling with  $\rho_s$, i.e. $Q_s$, only .


\subsection{Asymptotic behaviour from non-linear wavefront formation}\label{sec:fc-geom}

In this section, we extend the previous results to the non-linear evolution equation for $\qhat(Y,\rho)$, exploiting a formal analogy with the physics of traveling waves propagation into unstable states \cite{2003,ARONSON197833,dee1983propagating,bramson1986microscopic,PhysRevA.39.6367,collet2014instabilities}. 
Such a mathematical connection turned out to be also fruitful for the study of the asymptotic behaviour of the saturation scale resulting from the non-linear BK evolution \cite{Munier:2003vc,Munier:2003sj,PhysRevD.70.077503,Beuf:2008mb,Beuf:2010aw}.

We will demonstrate that the double logarithmic evolution of the quenching parameter belongs to the same universality class as the Fisher-Kolmogorov-Petrovsky-Piscounoff (FKPP) equation \cite{fisher1937,10003528013}, a non-linear diffusion equation originally written to describe gene spreading in a population. This equation has extensively been studies, as a paragon for pulled front propagation \cite{ARONSON197833,collet2014instabilities,bramson1983convergence,van1987dynamical,van1988front,2000,Brunet:1997zz,brunet2015exactly,berestycki2017exact}.

The study of the linearized evolution equation suggests an asymptotic scaling solution of the form
\begin{equation}
\qhat(Y,\rho)\, e^{Y-\rho_s(Y)}\underset{Y\to\infty}{\sim}\qhat_0 \ f(\rho-\rho_s(Y))\,,\label{eq:qhat-scaling-fc-nl}
\end{equation}
where $ f$ is only a function of the scaling variable 
\begin{equation}
x=\rho-\rho_s(Y)\,.
\end{equation}

\begin{figure}[t] 
  \centering
     \includegraphics[page=1,width=0.55\textwidth]{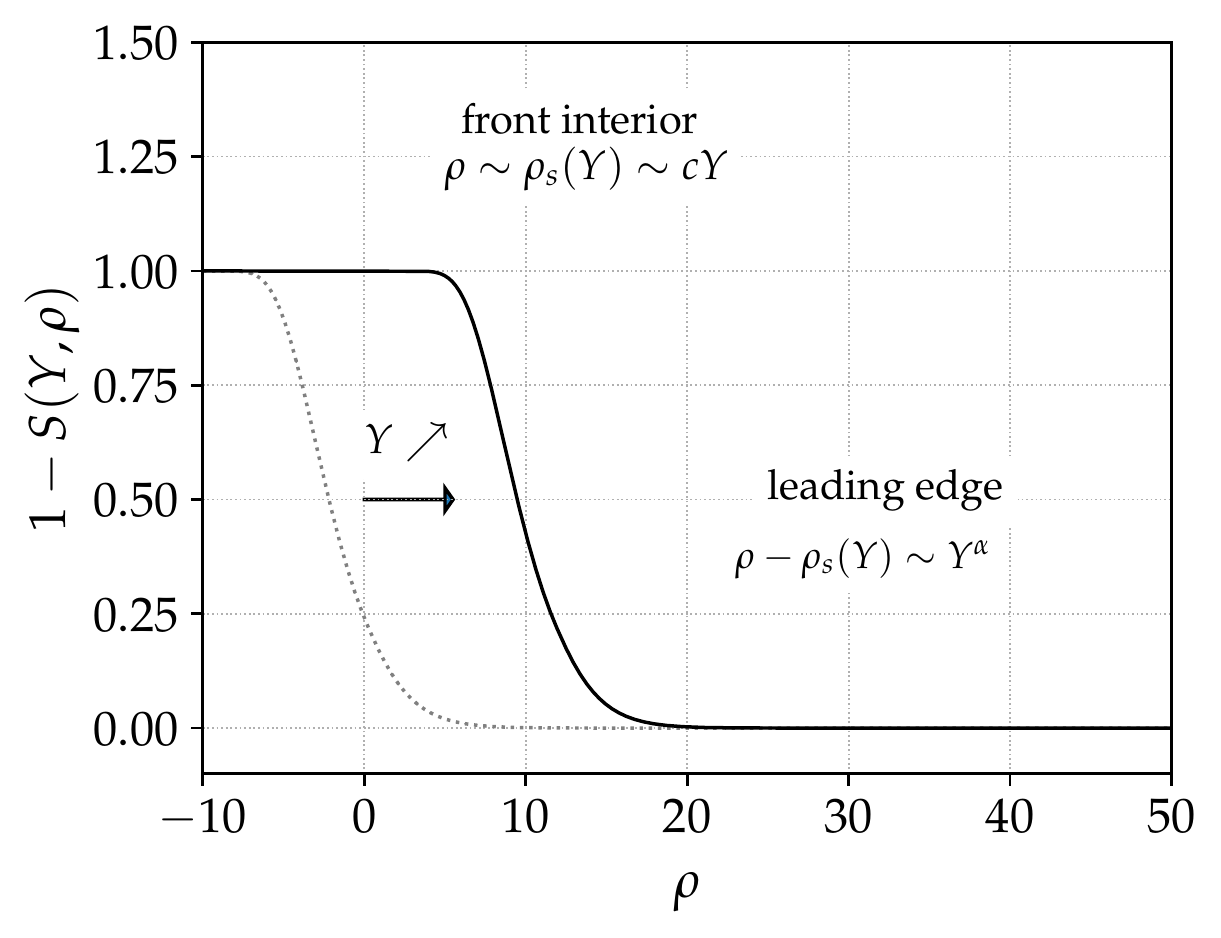} 
    \caption{Illustration of the traveling wave and its front propagation. The front propagates from the left to the right as $Y$ increases. The domains studied analytically in this paper are represented: the interior of the front, corresponding to $\rho-\rho_s(Y)\ll Y^\alpha$, where the non-linearities become important, and the leading edge regime with $\rho-\rho(Y)\sim Y^\alpha$. In the fixed coupling evolution, $\alpha=1/2$ while for the running coupling one, $\alpha=1/6$.}
    \label{fig:TW}
\end{figure}

The connection to traveling wave physics is made more transparent with the dipole $S$-matrix defined in Eq.\,\eqref{eq:Sxt-def}:
\begin{align}
1-S(\xt^2,L) &= 1-\exp\left(-\frac{1}{4}\frac{\qhat(Y,\rho)}{\qhat_0}e^{Y-\rho}\right)\,,\\
&\underset{Y\to\infty}{\sim}1-\exp\left(-\frac{1}{4}f(x)\,\rme^{-x}\right)\,.
\end{align}
where $\rho=-\ln(\xt^2\qhat_0\tau_0)$.
For $\rho \ll \rho_s$, $S$ saturates at $1$ owing to the unitarity constraint and drops to zero when $\rho \gg \rho_s$ as illustrated in Figure~\ref{fig:TW}. Hence, as $Y$ increases $\rho(Y)$ increases and the 
evolution of $S$ as function of $Y$  can be interpreted as the propagation of a front along the $\rho$ axis with speed $\dot \rho_s(Y)=\rm const.$ and $Y$ playing the role of the time. In the scaling regime, the wave propagates to the right while retaining its (universal) shape. The calculation of the deviation with respect to this uniformly translating profile at asymptotic times $Y$ can be done using techniques borrowed from non-linear physics of wavefront formation.

\subsubsection{Traveling wave solutions} 
\label{sub:TWsol}

We return now to the full non-linear problem for which \eqn{eq:evol-log-var} is replaced by 
\beq\label{eq:NL-evol-log-var}
 \hat q(Y,\rho) = \hat q_0(\rho)+\abar \int_0^Y \rmd Y' \int^\rho_{\rho_s(Y')} \rmd \rho' \hat q(Y',\rho')\,.
\eeq
The difference with \eqn{eq:evol-log-var} lies in the lower bound for the $\rho'$ integral, where we substituted $Y \to \rho_s(Y)$. 

In order to proof the existence of a family of front solution we insert the scaling ansatz \eqref{eq:qhat-scaling-fc-nl} into the fixed coupling evolution equation \eqn{eq:NL-evol-log-var} for $\qhat(Y,\rho)$. We obtain the following integro-differential equation  
\begin{align}
 \left[\dot\rho_s(Y)-1\right]f(x)-\dot\rho_s(Y)f'(x)&=\abar \int_0^{x}\rmd x'\,f(x')\,,\label{eq:diffeq-g-1}
\end{align}
which upon differentiation w.r.t.  $x$ reduces to a second order differential equation:
\begin{equation}
 -\dot\rho_s f''(x)+\left[\dot\rho_s-1\right]f'(x)-\abar  f(x)=0\,.
\end{equation}
Now, requiring that $\qhat$ satisfies the scaling relation \eqref{eq:qhat-scaling-fc-nl} implies that the $Y$ dependent coefficients must be constant: 
\begin{equation}
\dot\rho_s=c+\cO(1/Y)\,,
\end{equation}
for  large enough $Y$. Thus, we need to solve a the second order differential equation:
\begin{align}
-c f''(x)+(c-1)f'(x)-\abar f(x)&=0\,.
 \end{align}

The basis of solutions are given by exponential functions of the form $\exp(\beta x)$ with $\beta$ such that
\begin{equation}
c=\frac{\abar+\beta}{\beta(1-\beta)}\,.
\end{equation}
Rejecting imaginary values for $\beta$ (non-oscillatory solutions) provides the constraint $c\geqslant1+2\sqrt{\abar+\abar^2}+2\abar$ on the front velocity, which is equivalent to  $\beta \geqslant \beta_c=-\abar+\sqrt{\abar+\abar^2}$. The value of $c$ dynamically selected by the system depends on the initial condition at $Y=0$. It is straightforward to see that for asymptotic solutions of the form $\exp(\beta x)$, the function $1-S(0,\rho)$ behaves like $\exp((\beta-1)\rho)$ at large $\rho$ (note that $\beta-1<0$ for $\abar>0$). If the initial condition for $\qhat$ is such that $1-S(0,\rho)$ decays faster than $\exp((\beta_c-1)\rho)$, i.e.\  $\qhat(0,\rho)e^{-\beta_c \rho}\to0$ for large $\rho$, then the value of $c$ chosen by the system is the minimal one \cite{ARONSON197833,dee1983propagating,PhysRevA.39.6367,collet2014instabilities,bramson1983convergence,van1987dynamical,van1988front,PhysRevE.56.2597}:
\begin{equation}
c=1+2\sqrt{\abar+\abar^2}+2\abar\,,\qquad\beta=\beta_c=\frac{(c-1)}{2c}=\sqrt{\abar+\abar^2}-\abar\,.\label{eq:c0beta0}
\end{equation}
From now on we shall drop the subscript $c$ of $\beta_c$ since we will only consider physical initial conditions that satisfy this requirement.
For this critical value $c$, the solution $f$ is given by $f(x)=\rme^{\beta x}(a_1+a_2x)$
with $a_1$ and $a_2$ two integration constants. They are fixed by the very definition of $\rho_s$, which is given by the relation 
\beq 
\qhat(L,Q_s^2)L=Q_s^2\,,
\eeq
 that implies that $f (0)=1$ and $f '(0)=(c-1)/c$ on the saturation line $x=0$, i.e. $\rho=\rho_s(Y)$ yielding 
\begin{equation}
f(x)=\rme^{\beta x}(1+\beta x)\,.\label{eq:scaling-limit-fc}
\end{equation}

These results should be contrasted with those obtained for the linear evolution equation and reported in Eq.\,\eqref{eq:scaling-f-fc-lin}. First, the non-linear evolution imposes a single scale in the scaling limit, $Q_s$. Therefore, the function $f$ is a function of $x=\rho-\rho_s(Y)$ instead of $\rho-Y$. In terms of physical variables, the scaling variable is $\kt^2/Q_s^2$ in the non-linear case and $\kt^2/(\qhat_0L)$ in the linear case. Interestingly though, the shape of the scaling function $f$ and the sub-asymptotic corrections to $\qhat(Y,\rho)$ are the same within the double-logarithmic approximation in which $c\simeq 1+2\sqrt{\abar}$ and $\beta\simeq\sqrt{\abar}$ given that we work in the weak coupling limit, namely, $\abar \ll1$ . 

\subsubsection{Logarithmic shift of the front: study of the leading edge domain} 
\label{sub:log-shift}

The traveling wave solution $ f$ describes a uniformly translating front invading the large $\rho$ domain where $1-S$ vanishes. The corrections to the velocity $\dot\rho_s$ for non-asymptotic values of $Y$ are driven by the leading edge domain of the solution, that is the region where the front forms as a result of the growth of perturbations around the "unstable state" $1-S=0$. To obtain the behavior of the front on the leading edge, we look for a solution of the form \cite{2000,Munier:2003sj}
\begin{align}
 \qhat(Y,\rho)&=\qhat_0e^{\rho_s(Y)-Y}e^{\beta x} \ Y^\alpha G\left(\frac{x}{Y^\alpha}\right)\,,\\
 \dot\rho_s(Y)&=c+\dot\sigma_s(Y)\,.
\end{align}
This ansatz will allow us to describe the leading edge where $z=x/Y^\alpha\sim 1$. Assuming $\alpha$ to be positive and since $Y$ is large, it follows that  $x=\rho-\rho_s$ is large as well. Thus, we are indeed focusing on the front formation region. The power $\alpha$ describes a diffusion-like spreading of the traveling wave in the leading edge domain. Plugging this ansatz inside the differential equation satisfied by $\qhat$ and using the definitions of $c$ and $\beta$ given by Eq.\,\eqref{eq:c0beta0} in terms of $\abar$, one gets the following differential equation for $G_\alpha$:
 \begin{align}
&-\left(\alpha zY^{-1}+Y^{-\alpha}(c+\dot\sigma_s)\right)G''(z)-\left(\alpha\beta zY^{\alpha-1}+(2\beta -1)\dot\sigma_s\right)G'(z)\nonumber\\
&+\left(\alpha\beta Y^{\alpha-1}+\beta (1-\beta)\dot\sigma_sY^\alpha\right)G(z)=0\,.
\end{align}
The homogeneity constraint of this equation implies that $\alpha-1=-\alpha$ and thus $\alpha=1/2$ and $\dot\sigma_s= \delta_1Y^{-1}$ where $\delta_1$ is a constant. Neglecting the sub-leading powers $Y^{-1}$ and $Y^{-\alpha-1}$, we obtain
\begin{equation}
-c \ G''(z)-\frac{1}{2}\beta \ zG'(z)+\left(\frac{1}{2}\beta+\beta(1-\beta)\delta_1\right)G(z)=0\,.
\end{equation}
The general solution of this equation can be expressed in terms of confluent hypergeometric functions $\prescript{}{1}{\mathrm{F}}_1(a;b;x)$ and Hermite polynomials $\mathrm{H}(n;x)$:
\begin{align}
G(z)=\rme^{-\frac{\beta z^2}{4c}}&\left[a_1\mathrm{H}\left(2((\beta-1)\delta_1-1);\sqrt{\frac{\beta}{4c}}z\right)+a_2\prescript{}{1}{\mathrm{F}}_1\left(1-(\beta -1)\delta_1;\frac{1}{2};\frac{\beta}{4c}z^2\right)\right]\,.
\end{align}
The unknown constants $\delta_1$, $a_1$ and $a_2$ can be determined from the boundary conditions at $z=0$ and $z\to\infty$. 
$G$ has to decay faster than a power law at large $z$ \cite{2000}, meaning that the constant $a_2$ must vanish as it multiplies  the exponentially growing solution $\prescript{}{1}{\mathrm{F}}_1(a;1/2;x)\sim e^{x}x^{a-1/2}$.
The initial condition at $z=0$ is obtained by requiring that the solution $G$ matches with the scaling limit $f$ for large $x$, so that
\begin{equation}
G(z)\underset{z\to0}{\sim}\beta z\,.
\end{equation}
The small $z$ behavior of the Hermite function is
\begin{equation}
\mathrm{H}\left(2((\beta-1)\delta_1-1);\sqrt{\frac{\beta}{4c}}z\right)=\frac{4^{-1+(\beta-1)\delta_1}\sqrt{\pi}}{\Gamma\left(\frac{3}{2}-(\beta-1)\delta_1\right)}+\mathcal{O}(z)\,.
\end{equation}
In order for the constant term to vanish, the argument of the $\Gamma$ function must be a negative integer, constraining $\delta_1$ to be of the form $\delta_1=-\frac{3+2n}{1-\beta}$
with $n$ a non-negative integer. Finally, since $G$ must always be positive (no nodes) \cite{2000,PhysRevE.56.2597}, this fixes $n=0$ and therefore
\begin{equation}
\delta_1=-\frac{3c}{1+c}\,.
\end{equation}
This yields the following sub-asymptotic correction for the shape and position of the wave-front:
\begin{align}
G(z)&=\beta z \exp\left(-\frac{\beta}{4c}z^2\right)\,,\label{eq:G1/2-final}\\
\rho_s(Y)&=c Y-\frac{3c}{1+c}\ln(Y)+...\label{eq:rhos-final}
\end{align}
As for the case of the pulled front propagation into unstable states, our analysis of the leading edge domain shows that for non-asymptotic ``times" $Y$, the position of the wave-front undergoes a logarithmic shift, whose coefficient is given by the constant $\delta_1$. We observed the same logarithmic sub-asymptotic corrections in the linearized evolution. However, in the latter the coefficient was $-3/2$, whereas it is of order $-3/2(1+\sqrt{\abar})$ in the non-linear case. As argued in \cite{Caucal:2021lgf}, this is parametrically larger than single logarithmic corrections that we neglect in this paper. On the other hand, comparing Eq.\,\eqref{eq:G1/2-final} with Eq.\,\eqref{eq:qhat-linear-asymptote}, one notices that the sub-asymptotic corrections for the shape of the front in the leading edge domain are the same in the non-linear and linear case, within the double-logarithmic approximation where $\beta/(4c)\sim\sqrt{\abar}/4$ --- corrections of order $\alpha_s$ are relevant only at single logarithmic accuracy since $\alpha_s Y$ (or $\alpha_s \rho$) is of order $\sqrt{\alpha_s}\ll 1$ when $\sqrt{\alpha_s}\,Y=\mathcal{O}(1)$.

\subsubsection{Leading edge vs. front interior expansion} 

So far, we have obtained two distinct asymptotic solutions to the non-linear evolution equation for $\qhat(Y,\rho)$: $ f$ and $G$. The former is valid in the interior of the wave front corresponding to the saturation regime where $S\sim 1$ and $ 0 < x \lesssim 1$, whereas, the latter applies to the leading edge domain where $ x \sim Y^{1/2} \gg 1$ and $S\ll1$. This solution reads, in terms of $x$ and $Y$ (that is, in the front rest frame):
\begin{equation}
\qhat(Y,\rho)=\qhat_0\rme^{\rho_s(Y)-Y}\beta x\exp\left(\beta  x-\frac{\beta x^2}{4cY}\right)\,.
\end{equation}
Expanding for $x^2\ll Y$ (or equivalently $z\ll 1$), one finds
\begin{equation}
\qhat(Y,\rho)=\qhat_0\rme^{\rho_s(Y)-Y}\rme^{\beta  x}\left[\beta  x-\frac{\beta ^2x^3}{4cY}+\mathcal{O}\left(\frac{1}{Y^2}\right)\right]\,.\label{eq:G12-exp}
\end{equation}
This suggests the existence of an other expansion scheme of the form
\begin{equation}
\qhat(Y,\rho)=\qhat_0\rme^{\rho_s(Y)-Y}\rme^{\beta  x}\sum_{n\ge 0}\frac{1}{Y^{n/2}}f_n(x)\,,\label{eq:front-interior-exp}
\end{equation}
with $f_0(x)=\rme^{-\beta x} f(x)$.
This expansion is dubbed the ``front interior expansion" as it focuses on the behaviour of the solution in the rest frame of the front and near the saturation region ($z\ll 1$) \cite{2000}. In this region, it is not allowed to linearize the evolution equation. 

On the other hand, one can also study the corrections to the solution $G$ in the leading edge domain ($z\sim 1$). The natural expansion scheme there will be referred to as ``leading edge expansion" \cite{2000}, and it reads
\begin{equation}
 \qhat(Y,\rho)=\qhat_0e^{\rho_s(Y)-Y}e^{\beta  x}\left[Y^{1/2} G_{-1}(z)+G_0(z)+...+Y^{-n/2}G_{n}(z)+...\right]\,,
\end{equation}
with $G_{-1}\equiv G$. Each term in this series resums to all orders powers of $x/\sqrt{Y}$. By expanding in power series these analytic functions of $z$, one notices a relation between the leading edge and the front interior expansion. This is represented in Table~\ref{tab:expansion}: each function in the leading edge series accounts for a diagonal in the infinite triangular matrix of coefficients $f^j_n$. The function $G_{-1}$ resums the first red diagonal, the function $G_0$ resums the second diagonal and so forth.
There is a systematic way to compute each term in the leading edge, front interior and $\dot\rho_s$ expansion. As we shall see, these two expansions are related by matching conditions and the leading edge expansion constrains the coefficients of the asymptotic expansion of the saturation scale $\rho_s$.

We already know $f_0$ and $G_{-1}$. The function $f_1$ and $f_2$ can be obtained without difficulty by plugging the front interior expansion in the evolution equation for $\qhat$ and using $\dot\rho_s\simeq c+\delta_1/Y$. One gets the following differential equations:
\begin{align}
f_1''(x)&=0\,,\\
f_2''(x)&=\frac{\delta_1\beta }{4c^2}\left((c^2-1)x+2c(c+3)\right)\,.
\end{align}
The initial conditions for $f_1$ and $f_2$ are given by the definition of $Q_s$ which leads $f_1(0)=f_2(0)=0$, $f_1'(0)=0$ and $f_2'(0)=\delta_1/c^2$, such that
\begin{align}
f_1(x)&=0\,,\\
f_2(x)&=\frac{\delta_1x}{c^2}\left[1+\frac{(c-1)(3+c)}{8c}x+\frac{(c-1)^2(1+c)x^2}{48c^2}\right]\,.
\end{align}
Some features of this calculation are generic to all $f_n$ functions. First, they all satisfy a second order differential equation of the form $f_n''(x)=...$ whose  r.h.s. is determined by the $f_i(x)$ with $i<n$. The initial conditions are always provided by the definition of the saturation scale which yields $f_n(0)=0$ for $n\ge 1$. Therefore, all the $f_n(x)$ are polynomial functions of $x$, with degree at most $n+1$.

To better understand the interplay between the leading edge and front-interior expansion, we compute the functions $G_0$ and the next term in the development of $\dot\rho_s$. Inserting the leading edge expansion into the evolution equation for $\qhat$ and the development of $\dot\rho_s=c+\dot\sigma_s$, one finds a differential equation for $G_0$ whose homogeneity condition constrains the form of $\dot\sigma_s$ to be 
\begin{equation}
\dot\sigma_s=\frac{\delta_1}{Y}+\frac{\delta_2}{Y^{3/2}}+...
\end{equation}
The homogeneous part of the differential equation satisfied by $G_0$ is similar to the one satisfied by $G_{-1}$, while its inhomogeneous term depends on $G_{-1}$ and its derivatives:
\begin{align}
-c \ G_0''-\frac{\beta }{2} \ zG_0'-\frac{3\beta }{2} G_0&=\frac{1}{2}zG_{-1}''-\frac{\delta_1}{c}G_{-1}'-\beta (1+c)\frac{\delta_2}{2c}G_{-1}\,.
\end{align}
To solve this differential equation, it is convenient to perform the change of variable
\begin{equation}
G_0(z)=\rme^{-\frac{\beta }{4c}z^2}g_0\left(\frac{\beta }{4c}z^2\right)\,.
\end{equation}
Replacing the known values of $\beta $, $\delta_1$ and $G_{-1}$ in terms of $c$, one finds the following differential equation for $g_0(u)$:
\begin{align}
u \ g_0''(u)+\left(\frac{1}{2}-u\right)g_0'(u)+g_0(u)&=\kappa(u)\,,
\end{align}
with 
\begin{align}
\kappa(u)&=\left[-6+4C(1+C^2)\delta_2(2u)^{1/2}+24u-8u^2+C^4(2(3-2u)u+C\delta_2(2u)^{1/2})\right.\nonumber\\
&\left.-6C^2(1-5u+2u^2)\right]/\left(2(2+3C^2+C^4)\right)\,,
\end{align}
and $C=\sqrt{c-1}$.
The integration constants are fixed by matching the leading edge expansion with the front interior. More concretely, since $f_0(x)=1+\mathcal{O}(x)$ and $f_1(x)=0$, the development at small $u$ of $g_0$ must be $g_0(u)=1+\mathcal{O}(u)$ (terms of order $u^{1/2}$ are prohibited). The constant $\delta_2$ is fixed by demanding that at large $u$, $g_0(u)$ diverges no slower than $\rme^{u}u^{-3/2}$ \cite{2000}. We then obtain for $\delta_2$ the value 
\begin{equation}
\delta_2=\frac{3c\sqrt{2\pi(c-1)}}{(1+c)^2}\,,
\end{equation}
while the function $g_0$ can be expressed in terms of hypergeometric functions $\prescript{}{2}{F}_2$ and $\prescript{}{1}{F}_1$~\cite{2000,PhysRevD.70.077503}:
\begin{align}
&g_0(u)=2(1-2u)u \ \prescript{}{2}{F}_2(1,1;-1/2,2;-u)-\frac{3(3+c)}{1+c}u(1-2u) \ \prescript{}{2}{F}_2(1,1;1/2,2;-u)\nonumber\\
&+\frac{6(c-1)}{1+c}u(1-2u) \ \prescript{}{1}{F}_1(1/2;3/2;-u)\prescript{}{1}{F}_1(1/2;3/2;u)+\frac{6(c-1)u\rme^u}{1+c}\prescript{}{1}{F}_1(1/2;3/2;-u)\nonumber\\
&+\left(\frac{3\sqrt{\pi}(1-c)}{1+c} u^{1/2}(1-2u)+\frac{\rme^{-u}}{1+c}2u(1-2u)(3-4u-c(3+4u))\right)\prescript{}{1}{F}_1(1/2;3/2;u)\nonumber\\
&-\frac{3(c-1)\sqrt{\pi}}{1+c}u^{1/2}\rme^{u}+\frac{1}{1+c}\left(6(c-1)\sqrt{\pi}u^{1/2}-u(-5+6u+c(7+6u))\right)+1\,.
\end{align}
It is enlightening to expand the function $G_0(z)$ in powers of $z$. From the expression above, one finds
\begin{equation}
G_0(z)=1+\frac{\delta_1(c-1)(3+c)}{8c^3}z^2+\mathcal{O}(z^3)\,.
\end{equation}
Combined with Eq.\,\eqref{eq:G12-exp}, one observes that the function $G_{1/2}$ and $G_0$ resums respectively the leading and sub-leading powers of the polynomial functions $f_n(x)$ for all values of $n$, as illustrated in Table~\ref{tab:expansion}.

\begin{table}
\centering
\begin{tabular}{ |c | c c c c c c c c c|}
\hline
\backslashbox{$f_n(x)$}{$x^j$} & $x^0$ & $x^1$ & $x^2$ & $x^3$ & ... & $x^j$ & $x^{j+1}$ & $x^{j+2}$ &  ... \\\hline
 $ f_0(x)$ & \textcolor{blue}{1}                    & \textcolor{red}{$\beta $}                             & /                                      & /                                      & /   & / & /   & /   & /   \\ 
 $f_1(x)$        & 0                    & \textcolor{blue}{0}                                    & \textcolor{red}{0}                               & /                                      & /   & / & /  & /   & /  \\  
 $f_2(x)$        & 0  &  $\frac{\delta_1}{c^2}$  & \textcolor{blue}{$\frac{\delta_1(c-1)(c+3)}{8c^3}$} & \textcolor{red}{$\frac{\delta_1(c-1)^2(c+1)}{48c^4}$}    & /   & /   & /      & /   & /  \\
 ...        & ...                  & ...                                   & ...                                    & ...   & ... & ...  & /     & / & / \\
 $f_n(x)$        & 0                & $f_n^1$                                 & $f_n^2$                                  & $f_n^3$ & ... & \textcolor{blue}{$f_n^{n}$ }& \textcolor{red}{$f_n^{n+1}$} & /   & / \\
  ...        & ...                  & ...                                   & ...                                    & ...   & ... & ...  & ...     & ... & / \\ \hline
\end{tabular}
  \caption{Relation between the front interior and leading edge expansion. The function $G_{-1}$ resums the red powers in $x$ to all orders, which correspond to leading powers of the $f_n$ polynomials, the function $G_0$ resums the blue terms, the next-to-leading powers, and so forth.  }\label{tab:expansion}
\end{table}

\begin{figure}[t] 
  \centering
  \begin{subfigure}[t]{0.48\textwidth}
     \includegraphics[page=1,width=\textwidth]{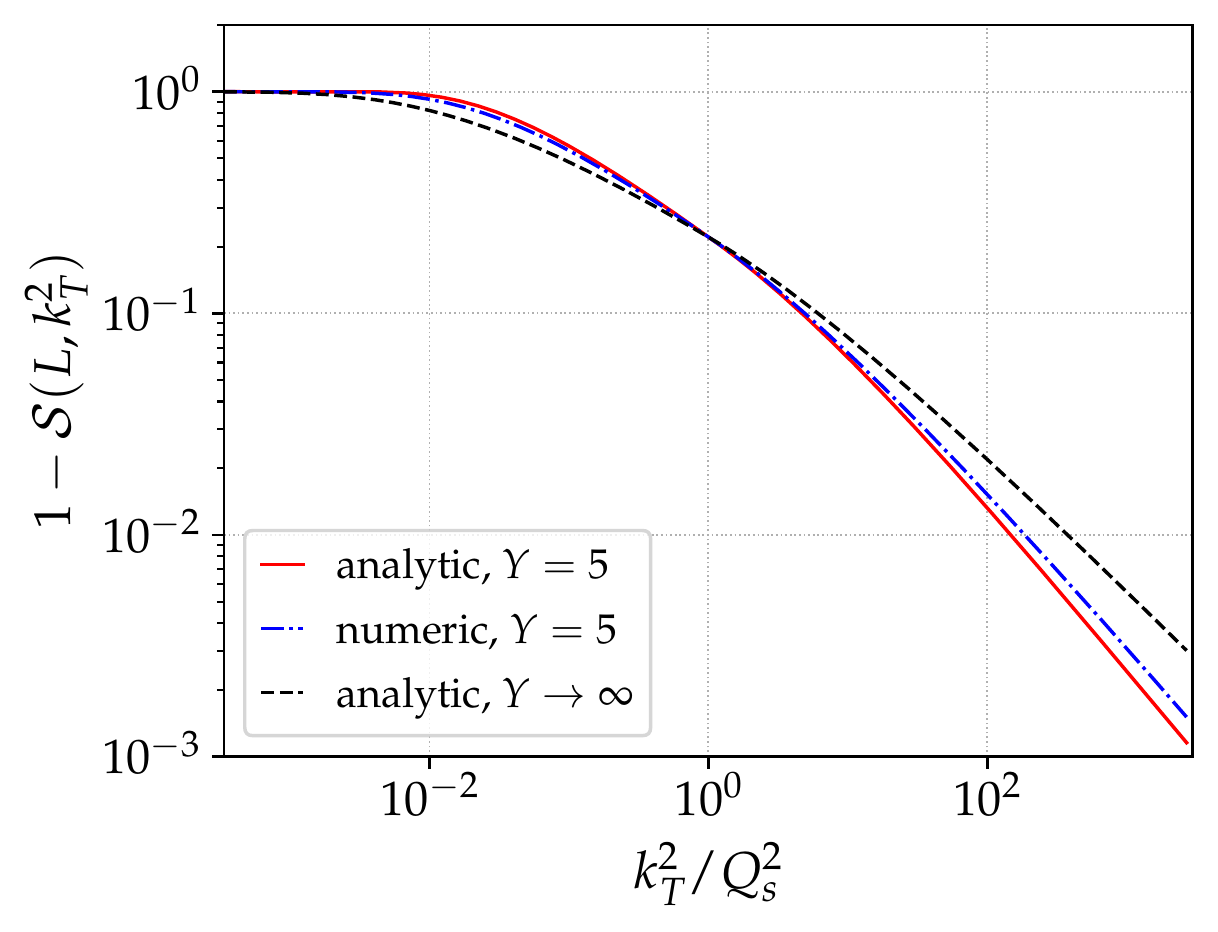} 
    \caption{\small }\label{fig:wave-front-fc}
  \end{subfigure}
  \hfill
  \begin{subfigure}[t]{0.48\textwidth}
     \includegraphics[page=2,width=\textwidth]{fc-analytic.pdf}    
     \caption{\small }\label{fig:rhos-fc}
  \end{subfigure}
    \caption{(Left) The function $1-S(Y,\rho)$ in the rest frame of the front. (Right) The velocity of the front as function of time $Y$.}
    \label{fig:fc-summary-plot}
\end{figure}

In Fig.\,\ref{fig:fc-summary-plot}, our analytic results are compared to numerical simulations of the non-linear system \eqref{eq:NL-evol-log-var}. On the left figure~\ref{fig:wave-front-fc}, the dashed blue line corresponds to the dipole $S$-matrix in the rest frame of the front, i.e.\, as a function of $\kt^2/Q_s^2(L)$, given by the analytic expression of $\qhat$ including the first two terms in the leading edge development, while the red curve is the numerical solution. Even for this rather realistic value of $Y=5$, the leading edge development shows a rapid convergence (the analytic curve could be further improved around the transition at $x\approx 0$ by including more terms in the front interior expansion, see the discussion in Sec.\,\ref{sec:numerics}). The black dotted line is the asymptotic scaling limit given by the function $f$. The right plot \ref{fig:rhos-fc} displays the velocity of the front $\dot\rho_s$ as a function of time $Y$, as given by our asymptotic expansion
\begin{equation}
\dot\rho_s(Y)= c \ Y-\frac{3c}{1+c}\frac{1}{Y}+\frac{3c\sqrt{2\pi(c-1)}}{(1+c)^2}\frac{1}{Y^{3/2}}+\mathcal{O}\left(\frac{1}{Y^2}\right)\,,
\end{equation}
and its truncation up to order $Y$ and $1/Y$, compared to the numerical solution. Once again, the convergence of the development is very good, down to small values of $Y\approx 2\div 5$.

\section{Modified geometric scaling with running coupling}\label{sec:rc-analysis}

In this section, we consider the double logarithmic evolution of the quenching parameter including the running of the strong coupling constant. One would naively expect that running coupling corrections modify the fixed coupling asymptotic results for $\rho_s(Y)$ via terms of order $\alpha_s Y$ (single logarithmic corrections instead of double logarithmic ones). However, it is not the case as the power structure of the asymptotic development of $\rho_s$ at large $Y$ is dramatically modified compared to the fixed coupling scenario. This is mainly a consequence of the evolution equation with running coupling belonging to a different universality class than the FKPP equation of the fixed coupling evolution.

\subsection{Proof of the Iancu-Triantafyllopoulos's conjecture}\label{sec:rc-linear}

In this subsection, we derive the scaling limit and the sub-asymptotic corrections of the solution to the linearized evolution equation with running coupling and thus, provide the proof of the expansion for $\rho_s(Y)$ conjectured in \cite{Iancu:2014sha} based on a numerical analysis.
The equation we aim at solving is as follows
\beq\label{eq:rc-qhat-evol}
\frac{\del }{\del Y}  \hat q(Y,\rho) = \int_Y^\rho \rmd \rho' \abar(\rho' ) \hat q(Y,\rho') \simeq b_0 \int_Y^\rho\frac{\rmd \rho'}{\rho'}  \hat q(Y,\rho') \,.
\eeq
where we have approximated the running coupling by
\beq
\abar(\rho)=\frac{b_0}{\rho+\rho_0} \approx \frac{b_0}{\rho}\,.
\eeq
Although neglecting $\rho_0$ is justified asymptotically by the fact that for large $Y$, $\rho > \rho_s \gg \rho_0$ the following analysis can be extended easily by simply shifting  $\rho \to \rho+\rho_0$ and $Y \to Y+\rho_0$. 
Following \cite{Iancu:2014sha}, we introduce a new variable $u$ such that 
\beq 
u=\ln (\rho/Y)\quad \text{and} \quad  f(Y,u) = \hat q(Y,\rho) \,.
\eeq
Near the saturation line $\rho=Y$, the variable $u$ behaves like $(\rho-Y)/Y= x/Y$. We now look for a solution of the form 
\beq\label{eq:polynom-g}
f(Y,u) = \qhat_0\sum_{n=0}^\infty (b_0Y)^n \, g_n(u) \,.
\eeq
We will see that the $g_n$ functions are polynomial functions of $z$ of degree $n$. After the change of variable $\qhat(Y,\rho)\to f(Y,u)$, \eqn{eq:rc-qhat-evol} becomes
\beq\label{eq:rc-f-evol}
\left(\frac{\del }{\del Y} -\frac{1}{Y}\frac{\del }{\del u}\right)  f(Y,u) = b_0 \int_0^u \rmd u'\,  f(Y,u') \,.
\eeq
Plugging \eqn{eq:polynom-g} in \eqn{eq:rc-f-evol} and differentiating with respect to $u$, we find
\beq\label{eq:ddg-recurrence}
g''_{n+1}(u)  -(1+n) g'_{n+1}(u) = - g_n(u)\,.
\eeq
Since $g_0(u)=1$ is a polynomial, it is straightforward to show by recursion that the $g_n(u)$ functions are polynomial in $u$, assuming that they do not diverge exponentially at large $u$. They are also positive for $u\ge 0$.
On the tree-level saturation line, that is, $u=0$, the differential equation satisfied by $g_n$ implies
\beq
g_{n+1}(0)  = \frac{1}{n+1} \left[\int_0^{\infty} \rmd u \ \rme^{-(n+1) u } g_n(u)  \right]\,.\label{eq:laplace-gn0}
\eeq
This equation is very interesting: it shows that the $n+1$ coefficient of the saturation line is related to a Laplace transform of the polynomial that multiplies the previous power $Y^n$\,. Hence, if we can obtain the asymptotic form of $g_n(u)$  when $n\to \infty$ we then may use the steepest descent method to integrate over $u$. This is the method that we shall follow to derive the scaling limit of $\qhat(Y,\rho)$.

\begin{figure}[t] 
  \centering
  \begin{subfigure}[t]{0.48\textwidth}
     \includegraphics[page=1,width=\textwidth]{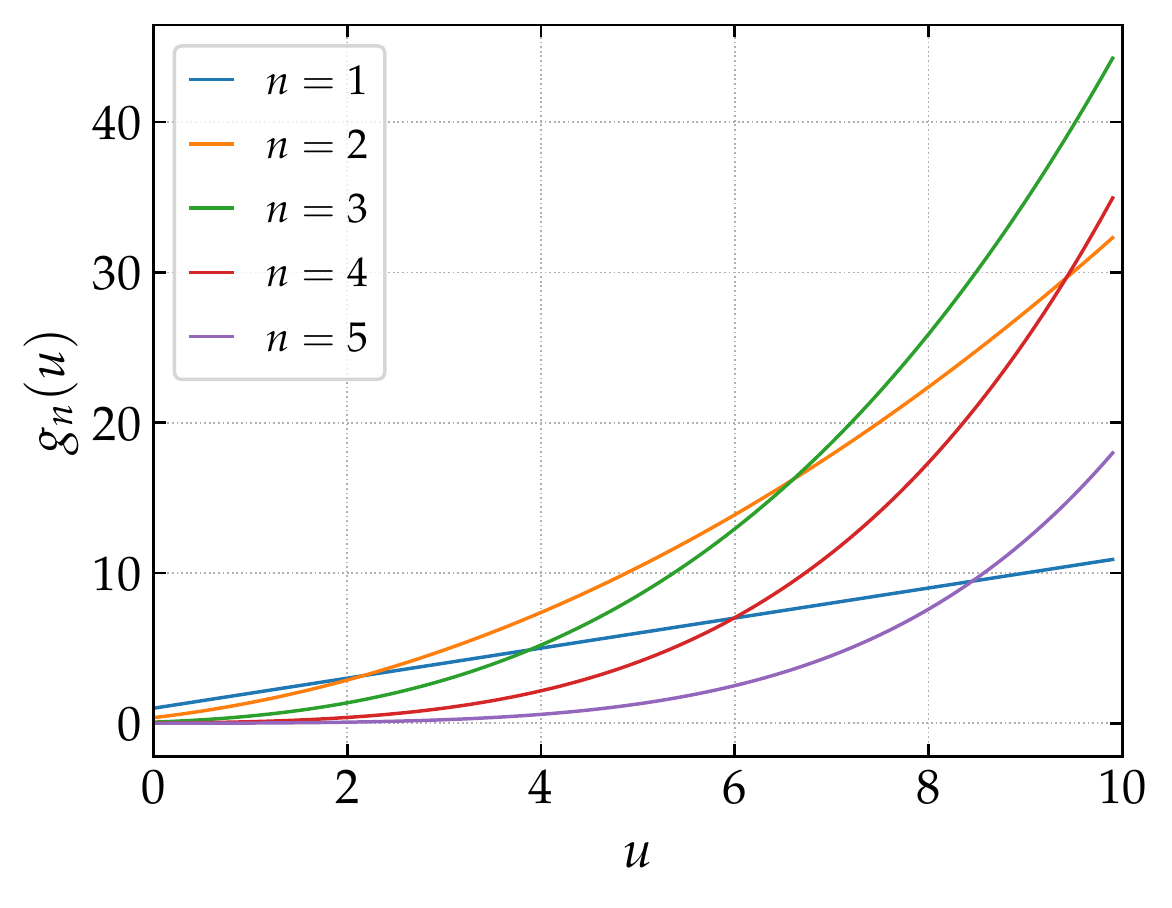} 
    \caption{\small }\label{fig:gn-unscaled}
  \end{subfigure}
  \hfill
  \begin{subfigure}[t]{0.48\textwidth}
     \includegraphics[page=2,width=\textwidth]{gn_polynoms.pdf}    
     \caption{\small }\label{fig:gn-sclaed}
  \end{subfigure}
    \caption{(Left) The first five polynomial functions $g_n(u)$. (Right) Scaling behavior of the polynomial functions $g_n(u)$ for large $n$ (see text for details).}
    \label{fig:polynoms-gn}
\end{figure}

\subsubsection{Scaling solution}
First, we would like to have a visual insight on the shape of the polynomial functions $g_n(u)$. The functions $g_n(u)$ for $0\le n\le 5$ are shown in Fig.\,\ref{fig:gn-unscaled}. They do not seem to follow any organizing principle. However, once one considers the rescaling $g_n(u)\to \rme^{-u}g_n(u/n)/g_n(0)$, inspired by Eq.\,\eqref{eq:laplace-gn0}, we observe, as shown in Fig.\,\ref{fig:gn-sclaed}, that the corresponding curves all tend to lie on the same universal scaling limit as $n$ becomes large. It is then tempting to conjecture that there exists a scaling behavior of these polynomial functions at large $n$.

Thus, inserting the following ansatz:
\beq\label{eq:scal-form}
g_n(u) =  g_n(0)h(n u )\,,
\eeq
in  \eqn{eq:scal-form} in \eqn{eq:ddg-recurrence}  we obtain 
\beq
g_{n+1} (0) (n+1)^2 \left[ h''(\nu)-h'(\nu)\right] = -g_n(0) h\left(\frac{n\nu}{n+1}\right) \simeq -g_n(0) h(\nu) \,,
\eeq
with the new variable $\nu\equiv nu$. 

Assuming a separation of variables we can write:
\beq
 g_{n+1} (0)  = \frac{c}{(n+1)^2 } g_n(0)\,,
\eeq
and 
\beq
 h''(\nu)-h'(\nu) +\frac{1}{c} h(\nu) =0  \,,
\eeq
where $ c$ is a constant de be determined. 
Comparing with \cite{Iancu:2014sha}, we should expect $c=4$. The solution that satisfies $h(0)=h'(0)=1$ reads
\beq\label{eq:h-fct}
h(\nu) = \left(1+ \frac{\nu}{2}\right)\, \rme^{\frac{\nu}{2}} \,.
\eeq
In fact, $c=4$ appears to be a critical value (it is a double root)  that separates oscillatory solutions from exponentially decaying ones. Like with fixed coupling evolution, for realistic initial conditions, the value of $c$ chosen by the system at large time is the smallest among all physical values (those which do no lead to an oscillatory behaviour). We shall see in the next subsection that the traveling wave interpretation of the running coupling evolution allows us to understand this criterion in the same way as at fixed coupling. Hence, at large $n$ we readily find
\beq\label{eq:gn-fct}
  g_{n} (0) = \frac{4^n}{(n!)^2}\,.
\eeq

We are now equipped to  derive the leading asymptotic behavior of the quenching parameter from the expansion \eqref{eq:polynom-g}. Using \eqn{eq:h-fct} and \eqn{eq:gn-fct} in \eqn{eq:polynom-g}, we have
\beq
\hat q (Y,\rho)=  \qhat_0\sum_{n=0}^\infty  \frac{4^n}{(n!)^2}  \left(1+\frac{ n u}{2}  \right) \left(b_0 Y\rme^{\frac{u}{2}}\right)^n\,.\label{eq:qhat-linear-sumexact}
\eeq
Let us first consider the first term in the parentheses. The treatment of the $nu/2$ term is similar.
At large $n$ we can turn the sum into an integral 
\beq
 \qhat(Y,\rho)=\qhat_0\int_0^\infty  \rmd n \, \exp\left[ n \ln (b_0 Y \rme^{\frac{u}{2}}) -2 \ln n! + n\ln 4\right] \,,\,
\eeq
which can be solve using the steepest descent method. It follows from
\beq 
\frac{\rmd }{\rmd n}( n \ln (b_0Y \rme^{\frac{u}{2}}) -2 \ln n! + n\ln 4)=0\,,
\eeq
and using the Stirling formula, that at the saddle-point the value of $n$ is
\beq
n=   \sqrt{4 b_0 Y \rme^{\frac{u}{2}}}\,.
\eeq
This justifies that the variable $\nu=nu$ scales like $ x/\sqrt{Y}=(\rho-Y)/\sqrt{Y}$ near the saturation line.
Finally, we obtain for the leading behavior of the quenching parameter
\beq
\qhat(Y,\rho) \approx  \qhat_0\exp\left[ n \ln (4b_0Y \rme^{\frac{u}{2}}) -2  n(\ln n -1) \right]  =\qhat_0\exp\left(4 \sqrt{ b_0Y \rme^{\frac{u}{2}}} \right) \,.
\eeq
Expressing the above result in terms of the variable $ x=\rho-Y$ by approximating  $z = \ln (\rho/Y)\simeq  x/Y \ll 1$, we find 
\beq
 \qhat(Y,\rho) \approx \qhat_0 \exp\left[4 \sqrt{ b_0 Y \left(1+\frac{x}{2Y}\right) } \right]\approx \qhat_0 \exp\left[4 \sqrt{ b_0Y } +\sqrt{\frac{b_0}{Y}}\, x\right]\,.
\eeq
We recognize in the first factor the saturation scale 
\beq
  \hat q (Y,Y)\sim  \qhat_0\rme^{4 \sqrt{b_0 Y } } \,.
\eeq
Turning now to the exact result for $\qhat(Y,\rho)$, one obtains
\begin{equation}
\qhat(Y,\rho)=\qhat_0\rme^{4 \sqrt{ b_0Y } +\sqrt{\frac{b_0}{Y}}\, x}\left(1+\sqrt{\frac{b_0}{Y}} x\right)\,,\label{eq:rc-lin-scaling-limit}
\end{equation}
where the second term inside the parenthesis comes from the $nu/2$ term in Eq.\,\eqref{eq:qhat-linear-sumexact}.
Not surprisingly, we recover the fixed coupling exponent $\beta\sim \sqrt{\bar \alpha_s}\sim\sqrt{b_0/Y}$. However, contrary to the fixed coupling case, the scaling variable is no longer $x$, but $ x/\sqrt{Y}$. The leading behavior of the saturation scale, which appears as a pre-factor in \eqref{eq:rc-lin-scaling-limit} is 
\begin{equation}
\rho_s(Y)=Y+4\sqrt{b_0}Y+...
\end{equation}

\subsubsection{Sub-asymptotic corrections }

In order to draw a complete picture of the solution we need to address the sub-asymptotic terms. Guided by the scaling analysis of the previous section, we may look for a solution of the form
\beq
g_n(u)  = g_0(0) \, h(\nu=nu,n) \, \rme^{\frac{nu}{2}}\,,
\eeq
which we insert in \eqn{eq:ddg-recurrence} and obtain 
\beq\label{eq:ddf-recurrence}
 \frac{g_n(0)n^2}{4g_{n-1}(0)} \left[4h''(\nu,n) -h(\nu,n)\right] = - \rme^{\frac{u}{2}} \, h\left(\nu-\frac{\nu}{n},n-1\right) \,.
\eeq
The pre-factor in the r.h.s. encodes the pre-asymptotic correction to $g_n(0)$. Now, we define the function $\dot a(n)$ as
\beq
 \frac{g_n(0)n^2}{4g_{n-1}(0)}  &\equiv \rme^{\dot a}\,.\label{eq:gn-ratio}
\eeq
Using the dotted notation for the discrete derivative with respect to $n$, we have
\beq
\dot  a(n) =\dot L_n(0) +\ln n^2 -\ln 4 \,.
\eeq 
with $L_n(0) =\ln g_n(0)$. In the scaling limit, we have $\dot a=0$ by construction, therefore $\dot a$ quantifies the deviation of the ratio \eqref{eq:gn-ratio} w.r.t. the scaling behavior.
Expanding \eqn{eq:ddf-recurrence} for large $n$ we have up to subleading terms 
\beq
 \left(4h'' - h\right)   \rme^{\dot a} = -h +\dot h +\frac{\nu}{2n} h+ \frac{\nu}{n}h'\,.
\eeq
Similarly to the fixed coupling pre-asymptotic analysis, we may assume a diffusion-like Ansatz 
\beq
h(\nu,n) = n^\alpha G\left(y=\frac{\nu}{n^\alpha} \right)\,.\label{eq:G-ansatz}
\eeq
It follows that
\beq 
\dot h(\nu,n)= \alpha n^{\alpha-1} G(y)-\frac{y}{n}G'(y)\simeq  \alpha n^{\alpha-1} G(y)\,,
\eeq
where the second term can be neglected at large $n$ since we expect $\alpha>0$. A similar argument allows us to neglect other $G'$ and $G$ terms in what follows.  Hence, $G$ obeys the equation 
\beq
4 n^{-\alpha}\rme^{\dot a } G''(y)  +\left[n^\alpha \left(1- \rme^{\dot a}\right)-\frac{1}{2}n^{-1+2\alpha} y \right]G(y) =0\,.
\eeq
The homogeneity condition implies that  
\beq 
\alpha = 1/3, \quad \quad \rme^{\dot a} \simeq 1+\dot a = 1+ \beta n^{-2/3}\,,
\eeq
and 
\beq
G''(y)  = \frac{1}{4}\left(\frac{1}{2} y-\beta \right) G(y) \,.\label{eq:G-airyeq}
\eeq
The exponent of the diffusive ansatz \eqref{eq:G-ansatz} is interesting as it differs from the fixed coupling evolution equation. Indeed, in terms of the scaling variable $\nu$ and $Y$, the function $G$ is a function of $\nu/n^\alpha\sim \nu/Y^{1/6}$. This $1/6$ power, smaller the power $1/2$ at fixed coupling, changes the structure of the asymptotic expansion of $\rho_s$ quite dramatically.
Turning back to the equation \eqref{eq:G-airyeq}, the latter is the Airy equation with a regular solution at large $y$ that reads 
\beq
 G(y) = { \rm const. }\, {\rm Ai}\left(\frac{1}{2}y - \beta\right)\,,
\eeq
In addition, in order to match onto the scaling solution  we must impose that $G(y) \sim y $ at large $n$. This condition fixes the value of $\beta$ to be
\beq
\beta= |\xi_1|\,, \,\,
\eeq
where $\xi_1= -2.338...$  is the rightmost zero of the Airy function. 

In terms of the original variables we finally obtain at large $n$ 
\beq\label{eq:pre-asymptotic}
g_n(z) \propto \, g_n(0) \, n^{1/3}\, {\rm Ai}\left(\xi_1 +  \frac{1}{2}n^{2/3} u \right) \, ,
\eeq
with
\beq
\frac{\rmd }{\rmd n }\ln  g_n(0)  = -2\ln n + \ln 4  +   |\xi_1| n^{-2/3}\,,
\eeq
or upon integration 
\beq
\ln  g_n(0)  = -2n(\ln n-1) + n \ln 4   + 3   |\xi_1| n^{1/3}+{\rm const.}\,,
\eeq
which is the form conjectured by the authors of \cite{Iancu:2014sha} to fit their numerical results. Using again the steepest descent method as in the previous subsection, one obtains the sub-asymptotic corrections to $\rho_s(Y)$:
\begin{equation}
\rho_s(Y)=Y+4\sqrt{b_0 Y}+3\xi_1(4b_0 Y)^{1/6}+\cO\left(1\right)\,.
\end{equation}
The analysis of the large $n$ behaviour of the polynomial functions $g_n(u)$ enables us to obtain the first sub-asymptotic correction to the saturation scale, in agreement with the numerical findings in \cite{Iancu:2014sha}. One could follow the same method to obtain the next corrections in the development of $\rho_s$. Nevertheless, the traveling wave method detailed in following subsection turns out to be more efficient, since it offers a systematic way to compute the sub-asymptotic corrections and it can be extended to the non-linear case in which the saturation boundary in the evolution equation is fixed at $\rho_s(Y)$ instead of $Y$.

\subsection{Asymptotic analysis of the non-linear equation }\label{sec:fc-nonlinear}

We now consider the non-linear evolution equation with running coupling
\begin{equation}
\frac{\partial \qhat(Y,\rho)}{\partial Y}=\int_{\rho_s(Y)}^\rho\der \rho'\,\bar\alpha_s(\rho')\qhat(Y,\rho')\,,\qquad\bar\alpha_s(\rho)=\frac{b_0}{(\rho+\rho_0)}\,.
 \label{eq:rc-evol}
\end{equation}
Even though the running coupling evolution of the quenching parameter does not fall into the same universality class as the fixed coupling equation (or the FKPP equation), we will follow the same strategy as in section \ref{sec:fc-geom} in order to get the scaling limit and the sub-asymptotic corrections.
 
\subsubsection{Scaling limit and diffusive deviations}

\paragraph*{Modified scaling variable.} We start by re-deriving the scaling limit given by Eq.\,\eqref{eq:rc-lin-scaling-limit} for the linear evolution equation. Inspired by the form of the scaling variable $(\rho-Y)/\sqrt{Y}$ in that case, we try the following scaling form
\begin{equation}
\qhat(Y,\rho)\underset{Y\to\infty}{\sim}\,\qhat_0\rme^{\rho_s(Y)-Y}f(\chi)\,,
\end{equation}
with
\begin{equation}
\chi=\frac{\rho-\rho_s(Y)}{\sqrt{Y}}\,.
\end{equation}
This scaling variable $\chi$ should be contrasted with the corresponding variable $x=\rho-\rho_s(Y)$ in the fixed coupling evolution.
Inserting this ansatz into the non-linear evolution equation for $\qhat(Y,\rho)$ with running coupling, one finds that the function $f$ satisfies
\begin{equation}
-\left(\frac{\chi}{2Y}+\frac{\dot\rho_s}{Y^{1/2}}\right)f'(\chi)+\left(-1+\dot\rho_s\right) f(\chi)=Y^{1/2}\int_0^\chi\der \chi' \ \frac{b_0}{\chi'Y^{1/2}+\rho_s+\rho_0} f(\chi')\,.
\end{equation}
Setting $\chi=0$, and using the definition of the saturation momentum $ f(0)=1$, the velocity of the front behaves as
\begin{equation}
\dot\rho_s=1+\frac{f'(0)}{Y^{1/2}}+...
\end{equation}
Differentiating once more w.r.t. $\chi$ and using the asymptotic behaviour of $\dot\rho_s$ given above yields
\begin{equation}
-\left(\frac{2c+\chi}{2Y}+\frac{1}{Y^{1/2}}\right)f''(\chi)+\left(\frac{c}{Y^{1/2}}-\frac{1}{2Y}\right)f'(\chi)=\frac{b_0Y^{1/2}}{\chi Y^{1/2}+\rho_s+\rho_0}f(\chi)\,,
\end{equation}
with $c=f'(0)$. Expanding in powers of $Y$ on both side, and extracting the leading $1/Y^{1/2}$ terms, one gets
\begin{equation}
-f''(\chi)+cf'(\chi)-b_0  f(\chi)=0\,.
\end{equation}
As in the fixed coupling case, the value of $c$ is fixed by requiring the discriminant of this differential equation to vanish, so that
\begin{equation}
c = 2\sqrt{b_0}\,.
\end{equation}
We have then recovered the results \eqref{eq:rc-lin-scaling-limit}, namely, 
\begin{align}
f(\chi)&=\rme^{\sqrt{b_0}\chi}\left(1+\sqrt{b_0}\chi\right)\,,\\
\rho_s(Y)&=Y+4\sqrt{b_0Y}+o\left(Y^{1/2}\right)\,.
\end{align}

\paragraph*{Diffusive ansatz around the scaling limit.} Following the method detailed in Sec.\,\ref{sec:fc-geom}, we need to compute the perturbations in the leading edge domain of the front in order to get the next term in the development of $\rho_s$ at large $Y$. We consider the following diffusive ansatz, similar to the fixed coupling case:
\begin{align}
\qhat(Y,\rho)&=\qhat_0 e^{\rho_s(Y)-Y}\rme^{\sqrt{b_0}\chi} \ Y^\alpha G\left(\frac{\chi}{Y^\alpha}\right)\\
\dot\rho_s(Y)&=1+\frac{2\sqrt{b_0}}{Y^{1/2}}+\dot\sigma_s(Y)
\end{align}
Plugging this ansatz inside Eq.\,\eqref{eq:rc-evol} and differentiating twice with respect to $\chi$, we find
\begin{align}
&-\frac{Y^{-3/2-\alpha}}{2}\left((1+2\alpha)\zeta Y^\alpha+2Y^{1/2}\dot\rho_s\right)G''\nonumber\\
&-\frac{Y^{-3/2}}{2}\left(1+2Y+2(1+\alpha)\sqrt{b_0}\zeta Y^\alpha+(4\sqrt{b_0}Y^{1/2}-2Y)\dot\rho_s\right)G'\nonumber\\
&-\frac{\sqrt{b_0}Y^{-3/2+\alpha}}{2}\left(1-2\alpha+2Y+\sqrt{b_0}\zeta Y^\alpha+2(\sqrt{b_0}Y^{1/2}-Y)\dot\rho_s\right)G\nonumber\\
&=\frac{b_0Y^\alpha}{Y}\left(1-\frac{\zeta Y^\alpha}{Y^{1/2}}-\frac{\rho_s(Y)-Y}{Y}+...\right)G\,,\label{eq:diffeq-Galpha}
\end{align}
with $\zeta=\chi/Y^{\alpha}$. To ease the counting of $Y$ powers, we have expanded the right hand side, equal to $\bar\alpha_s(\zeta Y^{1/2+\alpha}+\rho_s)Y^\alpha G$ for large $Y$. 

The strategy is to find the leading power in $Y$ in Eq.\,\eqref{eq:diffeq-Galpha} in order to get a simpler second order differential equation for $G$. The r.h.s.\ being proportional to $\abar$, it must contribute to this differential equation, and therefore it must contain this leading power. Since the first term $b_0Y^{\alpha-1} G$ cancels against a similar term in the l.h.s., the leading power is associated with the second term, proportional to $Y^{2\alpha-3/2}$. As we aim at finding a second order differential equation, the leading power in the coefficient of $G''$ must also be proportional to $Y^{2\alpha-3/2}$. This homogeneity condition $2\alpha-3/2=-1-\alpha$ yields $\alpha=1/6$, implying that the leading power in Eq.\,\eqref{eq:diffeq-Galpha} is $Y^{-7/6}$. As noted in the previous section, the diffusion power $\alpha=1/6$ is not the same as the $1/2$ found in the fixed coupling evolution, showing that in the running coupling case, the evolution equation does not belong to the same universality class as the FKPP equation. 

For consistency, one has to include a term of order $Y^{-5/6}$ in the development of $\dot\rho_s$, namely
\begin{equation}
\dot\sigma_s(Y)=\frac{\delta_1}{Y^{5/6}}+\mathcal{O}\left(\frac{1}{Y}\right)\,,
\end{equation}
with an unknown constant $\delta_1$ to be determined. Indeed, such term generates contributions of the same order as $Y^{-7/6}$. Gathering all terms proportional to $Y^{-7/6}$ provides the differential equation
\begin{equation}
-G''(\zeta)+\left(\sqrt{b_0}\delta_1-\frac{b_0}{2}\zeta\right)G(\zeta)=-b_0\zeta G(\zeta)\,.\label{eq:diffeq-G16}
\end{equation}
Notice that the $\zeta$ dependence of the right hand side of this equation comes from the $\rho$ argument in the QCD coupling. If we had replaced $\abar(\rho)$ by $\abar(\rho_s(Y))$ in the evolution equation, this term would not be there, changing the value of the constant $\delta_1$ that we now determine.

The only solution of the Airy-type differential equation \eqref{eq:diffeq-G16} which satisfies $G(\zeta)=\sqrt{b_0}\zeta+\mathcal{O}(\zeta^2)$ in order to recover the leading behaviour of the scaling solution $f$ near the interior of the front and with vanishing boundary conditions at $\zeta=+\infty$ is 
\begin{equation}
G(\zeta)=\frac{2^{1/3}b_0^{1/6}}{\mathrm{Ai}'(\xi_1)}\mathrm{Ai}\left[\xi_1+2^{-1/3}b_0^{1/3}\zeta\right]\,.\label{eq:G16-final}
\end{equation}
The constant $\delta_1$ is then fixed by these boundaries conditions:
\begin{equation}
\delta_1=2^{-2/3}b_0^{1/6}\xi_1\,.
\end{equation} 
The study of the diffusion around the scaling limit in the leading edge domain enables us to recover the asymptotic expansion of $\rho_s(Y)$:
\begin{equation}
\rho_s(Y)=Y+4\sqrt{b_0Y}+3\xi_1(4b_0Y)^{1/6}+o\left(Y^{1/6}\right)\,,
\end{equation} 
and provides the first correction $G$ to the scaling limit. Compared to the fixed coupling case, one notices that the first two terms in the development of $\rho_s$ does not depend on the linearization of the evolution equation.

\subsubsection{Corrections in the interior and on the edge of the wavefront}

So far, we have essentially recovered the results obtained in section \ref{sec:rc-linear} for a linear saturation boundary.
The main advantage of the mathematical approach of wave front formation is that it provides a systematic way of calculating the higher orders in the asymptotic development of $\rho_s(Y)$ and the sub-asymptotic corrections of $\qhat(Y,\rho)$ even in the presence of a non-linear saturation boundary.

\paragraph*{Front interior expansion.} 
The series expansion of the diffusive solution \eqref{eq:G16-final} is instructive. One finds 
\begin{align}
G(\zeta)&=\sqrt{b_0}\zeta+\frac{2^{1/3}b_0^{7/6}\xi_1}{12}\zeta^3+\frac{b^{3/2}}{24}\zeta^4+...\label{eq:G16-expansion}
\end{align}
This expansion enables one to infer the form of the front interior expansion: at fixed $\chi$, the term in $\zeta^3$ brings a power $Y^{-1/3}$, the term in $\zeta^4$ brings a power $Y^{-1/2}$ and so on. We therefore define the front interior expansion as
\begin{equation}
\qhat(Y,\rho)=\rme^{\rho_s(Y)-Y}\rme^{\sqrt{b_0}\chi}\sum_{n\ge 0}(4b_0Y)^{-n/6}f_{n}\left(\sqrt{b_0}\chi\right)\,.\label{eq:Gform}
\end{equation}
The rescaling $Y\to4 b_0Y$ and $\chi\to \sqrt{b_0}\chi$ is purely conventional and simplifies the expressions of the functions $f_n$.
The function $f_0$ has already been computed. From the expression of the scaling limit $f(\chi)$, one gets $ f_0(X)=1+X$.

Using the evolution equation and the following development of $\dot{\rho}_s$
\begin{equation}
\dot{\rho}_s=1+\frac{c}{Y^{1/2}}+\frac{\delta_1}{Y^{5/6}}+\frac{\delta_2}{Y}+\frac{\delta_3}{Y^{7/6}}+\frac{\delta_4}{Y^{4/3}}+...\label{eq:dotrhos-ansatz}
\end{equation}
where the presence of the corrections in $1/Y$, $1/Y^{7/6}$ and $1/Y^{4/3}$ will be justified {\it a posteriori} from our calculation of the next terms in the leading edge expansion, we find that the functions $f_{n}$ follow an infinite hierarchy of second order differential equation of the form $f_n''(X)=...$. As in the fixed coupling calculation, this hierarchy can be solved iteratively since the system is triangular (the right hand side depends only on $f_i(X)$ with $i<n$).
Concerning the initial conditions, the definition of the saturation boundary yields $f_n(0)=0$ for all $n\ge 1$. The conditions on the first derivative is obtained thanks to the differential equation in integral form. 
The first two terms read
\begin{align}
f_1(X)&=0\,\\
f_{2}(X)&=\xi_1\left[X+X^2+\frac{1}{6}X^3\right]\,.
\end{align}
One observes that the leading power in the polynomial functions $f_{0}$, $f_1$ and $f_{2}$ are included in the series expansion of $G$ displayed in Eq.\,\eqref{eq:G16-expansion}. In the front interior expansion \eqref{eq:Gform}, the first six terms (up to $n=5$) are universal. By universal, we mean that they do not depend on the initial conditions for the evolution nor on the constant term in the development of $\rho_s(Y)$ at large $Y$. All the functions $f_n(X)$ for $0\le n\le 5$ are provided in appendix \ref{app:G0-rc}.

\paragraph*{Leading edge expansion.} The front interior expansion does not enable to fix the value of the coefficients $\delta_i$ in the asymptotic expansion of $\dot\rho_s$. These coefficients are determined by matching the leading edge expansion with the front interior one.
In the running coupling evolution, the leading edge expansion takes the form
\begin{equation}
\qhat(Y,\rho)=\rme^{\rho_s-Y}\rme^{\sqrt{b_0}\chi}\left[(4b_0Y)^{1/6}G_{-1}(\zeta)+G_0(\zeta)+...+(4b_0Y)^{-n/6}G_n(\zeta)+...\right]\,,
\end{equation}
with $G_{-1}(\zeta)\equiv (4b_0)^{-1/6}G(\zeta)$.
After a rather tedious calculation similar to the one leading to the solution $G_{-1}$, one gets the following differential equation for $G_0$:
\begin{equation}
-G_0''+\left(\sqrt{b_0}\delta_1+\frac{b_0}{2}\zeta\right)G_0=\left(-\delta_1+\frac{7}{6}\sqrt{b_0} \zeta\right)G'-\frac{1}{3}\sqrt{b_0}(-1+6b_0+3\delta_2)G\,,
\end{equation}
by looking at the coefficient in front of the $Y^{-4/3}$ power on both side of the equation. The right hand side depends on the coefficient $\delta_2$ in front of the $Y^{-1}$ power in the asymptotic development of $\rho_s$, it is then a consistency requirement to include such a term, and justify a posteriori the form of \eqref{eq:dotrhos-ansatz}. 

Contrary to the calculation of section \ref{sec:fc-geom}, the homogeneous equation is the same as the one satisfied by $G$. Since the inhomogeneous right hand side is known, this equation can be solved by the method of variational parameters.
The initial conditions for $G_0$ are $G_0(0)=1$ and $G_0'(0)=0$ in order to match with the front interior expansion which has no term of order $\chi Y^{-1/6}$. These two conditions fix the two constants of integration. The constant $\delta_2$ is then determined by demanding the solution $G_0$ to decay exponentially at large $\zeta$. 
We find that the coefficient $\delta_2$ in front of the $\ln(Y)$ term reads 
\begin{equation}
\delta_2=\frac{1}{4}-2b_0\,.
\end{equation}
The coefficient in front of the $\ln(Y)$ term is different from the coefficient $1/4$ found in \cite{Iancu:2014sha} by numerically solving the linearized evolution equation for $\qhat$\footnote{We have also checked that our analytic approach enables to recover this $1/4$ coefficient for the linearized evolution equation, cf.\ section \ref{sub:lin-vs-nlin}.}. As in the fixed coupling case, the non-linearity of the saturation boundary brings a sizeable correction in front the logarithmic term in the asymptotic development of the saturation scale. Parametrically, this correction is of order $b_0\ln(Y)\sim \alpha_s(\rho_s)Y\ln(Y)$ and is then larger than pure single log corrections, or order $\alpha_s Y$.

The sub-asymptotic correction $G_0(\zeta)$ can be expressed in terms of the Airy function and its derivative,
\begin{align}
G_0(\zeta)=\frac{1}{\mathrm{Ai}'(\xi_1)}\left[\mathrm{Ai}'(s(\zeta))+\left(-\frac{7}{12}s^2(\zeta)+\frac{5\xi_1}{3}s(\zeta)-\frac{13\xi_1^2}{12}\right)\mathrm{Ai}(s(\zeta))\right]\,,
\end{align}
with $s(\zeta)=\xi_1+2^{-1/3}b_0^{1/3}\zeta$. It is quite remarkable that the dependence upon the QCD constant $b_0$ enters only through this shift function $s(\zeta)$. This feature does not persist in the higher orders of the leading edge development.
One can also verify that this function admits the series expansion
\begin{equation}
G_0(\zeta)=1+\sqrt{b_0}\delta_1 \zeta^2+\frac{b_0}{6}(-1+2b_0+\delta_2)\zeta^3+...
\end{equation}
and therefore accounts for the sub-leading powers of the front interior expansion.

\paragraph*{Universality.} Finally, one may wonder which terms in the asymptotic expansion of 
\begin{equation}
\rho_s(Y)= Y+4\sqrt{b_0Y}+6\delta_1Y^{1/6}+\delta_2\ln(Y)+\kappa+...
\end{equation}
are universal. By universal, we mean independent of the initial condition for $\qhat$. In particular, the integration constant $\kappa$ is not determined by the leading edge regime and depends on the initial condition. Therefore, the terms in the development of $\rho_s$ or $\qhat(Y,\rho)$ at large Y which depends on $\kappa$ are not universal. Expanding $\alpha_s(\rho=\zeta Y^{2/3}+\rho_s)$ in powers of $Y$, one gets
\begin{align}
\frac{b_0}{\zeta Y^{2/3}+\rho_s(Y)+\rho_0}&=\frac{b_0}{Y}-\frac{b_0\zeta}{Y^{4/3}}-\frac{4b_0^{3/2}}{Y^{3/2}}+\frac{b_0\zeta^2}{Y^{5/3}}+\frac{2b_0(-3\delta_1+4\sqrt{b_0}\zeta)}{Y^{11/6}}\nonumber\\
&-\frac{b_0\delta_2\ln(Y)}{Y^2}+\frac{b_0(16b_0-\kappa-\rho_0-\zeta^3)}{Y^2}+o\left(\frac{1}{Y^2}\right)\,.\label{eq:asrho-dev}
\end{align}
When this development is multiplied by the leading edge expansion, the smallest power of $Y$ which involves the coefficient $\kappa$ is $Y^{-2+1/6}=Y^{-11/6}$. This is the power which determines the differential equation satisfied by $G_3$. Therefore, the function $G_3$ is not universal while $G_1$ and $G_2$ are. These functions can be determined by following the same method as in the computation of $G_0$. The coefficients $\delta_3$ and $\delta_4$ in the development \eqref{eq:dotrhos-ansatz} of $\dot\rho_s$ are obtained from the  matching of $G_1$ and $G_2$ with the front interior expansion and from the boundary conditions in $\zeta=\infty$:
\begin{equation}
\delta_3=-\frac{7\xi_1^2}{1080}\frac{1}{(4b_0)^{1/6}}\,,\qquad \delta_4=-\xi_1\left(\frac{5}{324}+6b_0\right)\frac{1}{(4b_0)^{1/3}}\,.
\end{equation}
The two functions $G_1$ and $G_2$ are provided in appendix \ref{app:G0-rc}. Last, one notices the presence of a term of order $\ln(Y)/Y^2$ in \eqref{eq:asrho-dev}. In principle, such term would spoil the shape of the leading edge development, introducing a contribution of the form $Y^{-1/2}\ln(Y) \tilde G_3(\zeta)$. Yet, it is not the case\footnote{Beyond this order $n=3$, it is possible that the leading edge expansion contains terms of order $Y^{-n/6}\ln(Y)\tilde G_n(\zeta)$ for $n\ge 4$.} as one can prove that the only solution $\tilde G_3$ consistent with the front interior expansion is $\tilde G_3=0$, provided that the asymptotic expansion of $\dot \rho_s$ reads
\begin{equation}
\dot{\rho}_s=1+\frac{c}{Y^{1/2}}+\frac{\delta_1}{Y^{5/6}}+\frac{\delta_2}{Y}+\frac{\delta_3}{Y^{7/6}}+\frac{\delta_4}{Y^{4/3}}+\frac{\delta_{5a}\ln(Y)}{Y^{3/2}}+\frac{\delta_{5b}}{Y^{3/2}}+...\label{eq:dotrhos-ansatz2}
\end{equation}
with $\delta_{5a}=-\delta_2\sqrt{b_0}$. The coefficient $\delta_{5b}$ in Eq.\,\eqref{eq:dotrhos-ansatz2} reads 
\begin{equation}
\delta_{5b}=\frac{9}{2240\sqrt{b_0}}-\frac{1693\xi_1^3}{1360800\sqrt{b_0}}-\sqrt{b_0}(\kappa+\rho_0)+\sqrt{b_0}\left(-\frac{1}{4}+5b_0\right)\,,
\end{equation}
and is indeed not universal since it is $\kappa$ and $\rho_0$ dependent.
For $\rho_s$, our final result including all universal terms (except for the unknown integration constant $\kappa$) in the large $Y$ expansion is
\begin{align}
\rho_s(Y)&=Y+2\sqrt{4b_0Y}+3\xi_1(4b_0Y)^{1/6}+\left(\frac{1}{4}-2b_0\right)\ln(Y)+\kappa+\frac{7\xi_1^2}{180}\frac{1}{(4b_0Y)^{1/6}}\nonumber\\
&+\xi_1\left(\frac{5}{108}+18b_0\right)\frac{1}{(4b_0Y)^{1/3}}+b_0\left(1-8b_0\right)\frac{\ln(Y)}{\sqrt{4b_0Y}}+\mathcal{O}\left(Y^{-1/2}\right)\,.\label{eq:rhos-exp}
\end{align}
One notices that the $b_0$ dependence cannot be absorbed into a redefinition of the variable $Y\to b_0 Y$ because of the $b_0$ dependent terms inside the parenthesis of the $\ln(Y)$ and $Y^{1/3}$ terms.

\subsection{Comparison with the linearized evolution equation}
\label{sub:lin-vs-nlin}

We now discuss the asymptotic expansion of $\qhat(Y,\rho)$ and $\rho_s(Y)$ for the linearized evolution equation. In this scenario, one can also apply the analytic techniques of front propagation into unstable states to derive the large $Y$ development of these quantities. The only differences are the change of scaling variable
\begin{equation}
\chi=\frac{\rho-\rho_s(Y)}{\sqrt{Y}}\quad \to\quad \chi\equiv\frac{\rho-Y}{\sqrt{Y}}\,,
\end{equation}
and the definition of the saturation scale which is not defined by an implicit relation but rather as
\begin{equation}
\rho_s(Y)=Y+\ln\left(\frac{\qhat(Y,Y)}{\qhat_0}\right)\,.
\end{equation}
One can then compute the front interior, the leading edge and the $\rho_s$ asymptotic expansions as in the non-linear case. The formula for the front interior and leading edge functions are provided in appendix \ref{app:G0-rc}. They display interesting scaling properties in terms of the variable $b_0 Y$ and $s$ respectively. This feature persists for the asymptotic expansion of $\rho_s$ which reads
\begin{align}
\rho_s(Y)&=Y+2\sqrt{4b_0Y}+3\xi_1(4b_0Y)^{1/6}+\frac{1}{4}\ln(Y)+\kappa\nonumber\\
&+\frac{7\xi_1^2}{180}\frac{1}{(4b_0Y)^{1/6}}+\frac{5\xi_1}{108}\frac{1}{(4b_0Y)^{1/3}}+\mathcal{O}\left(\frac{1}{Y^{1/2}}\right)\,.\label{eq:rhos-exp-linear}
\end{align}
Thus, the presence of the terms linear in $b_0$ in the coefficient of the asymptotic series in Eq.\,\eqref{eq:rhos-exp} is a consequence of the back-reaction of the quantum evolution of $\qhat$ on the saturation boundary.

An important remark concerns the absence of term of order $\ln(Y)/\sqrt{Y}$ in this development. The reason is that for the linear evolution, the right hand side of the differential equation satisfied by the functions of the leading edge development is given by the expansion of $\alpha_s(\rho=\zeta Y^{2/3}+Y)$
\begin{equation}
\frac{b_0}{\zeta Y^{2/3}+Y+\rho_0}=\frac{b_0}{Y}-\frac{b_0\zeta}{Y^{4/3}}+\frac{b_0\zeta^2}{Y^{5/3}}-\frac{b_0(\rho_0+\zeta^3)}{Y^2}+\mathcal{O}\left(Y^{-7/3}\right)\,.
\end{equation}
In other words, the scale $\rho_s(Y)$ is simply replaced by $Y$ due to the absence of back-reaction, so that there is no spurious $\ln(Y)$ terms in the power development above. For the same reason, the non-universal contribution of order $Y^{1/2}$ in Eq.\,\eqref{eq:rhos-exp-linear} does not depend on $\kappa$ and simply reads
\begin{equation}
-2\left(\frac{9}{2240\sqrt{b_0}}-\frac{1693\xi_1^3}{1360800\sqrt{b_0}}-\sqrt{b_0}\rho_0\right)\frac{1}{Y^{1/2}}\,.\label{eq:d5coef-linear}
\end{equation}
In spite of being non-universal, this contribution enables to accurately describe the function $\rho_s(Y)$ down to small $Y$ values ($Y\sim 1\div 2$).
\subsection{Numerical study: convergence of the asymptotic espansion}

\begin{figure}[t] 
  \centering
  \begin{subfigure}[t]{0.48\textwidth}
     \includegraphics[page=1,width=\textwidth]{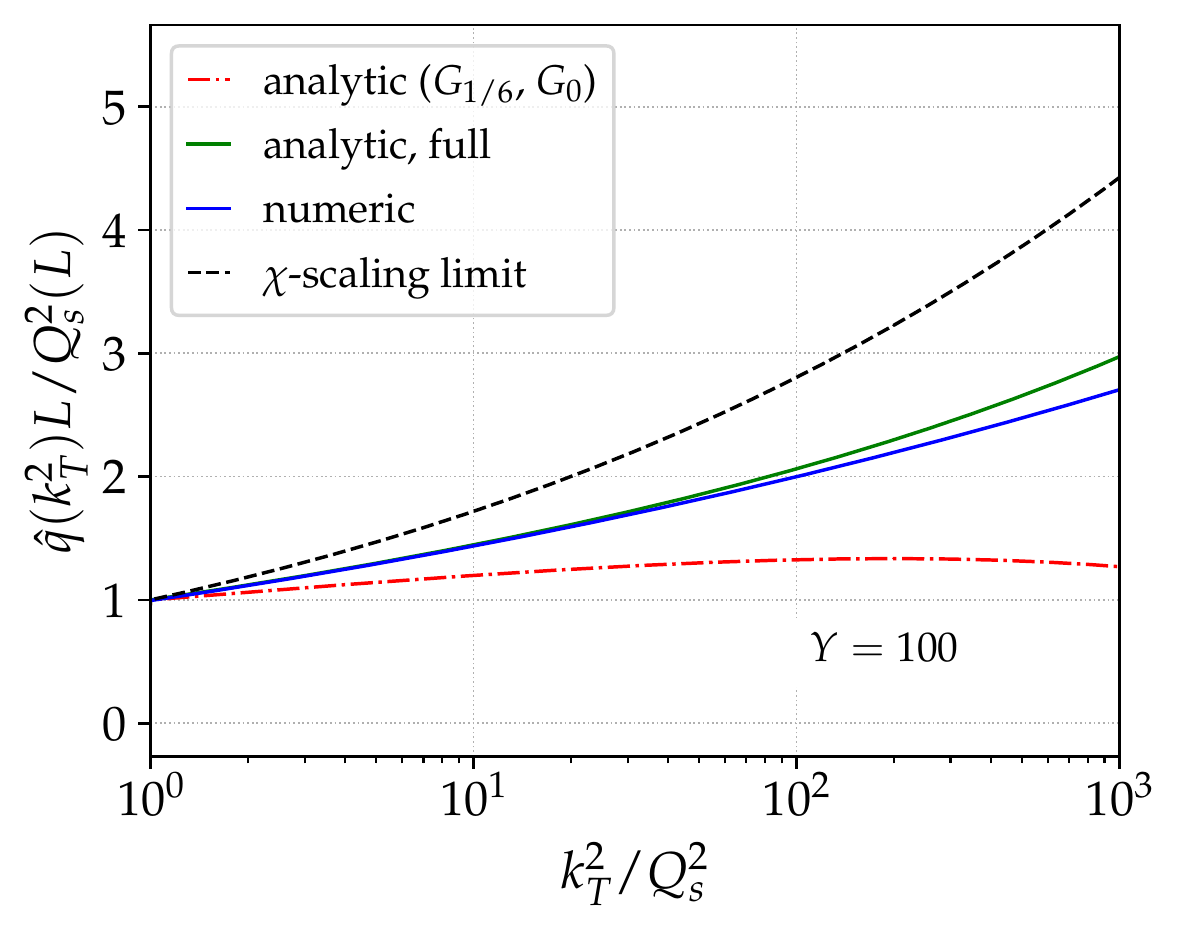} 
    \caption{\small }\label{fig:wave-front-rc}
  \end{subfigure}
  \hfill
  \begin{subfigure}[t]{0.48\textwidth}
     \includegraphics[page=2,width=\textwidth]{rc-analytic.pdf}    
     \caption{\small }\label{fig:rhos-rc}
  \end{subfigure}
    \caption{(Left) The dimensionless parameter $\qhat(L,k_T^2)L/Q_s^2(L)$ as a function of the variable $k_T^2/Q_s^2$ including the first three terms in the leading edge expansion. (Right) Comparison between the numerical calculation of $\dot\rho_s(Y)$ and its asymptotic expansion.}
    \label{fig:rc-summary-plot}
\end{figure}

The expansion \eqref{eq:rhos-exp} and its truncations up to the orders $\mathcal{O}(Y^{-1/2})$ and $\mathcal{O}(Y^{-1})$ are compared to numerical computation of $\rho_s(Y)$ in Fig.\,\ref{fig:rhos-rc}. Contrary to the fixed coupling case, the convergence of the this asymptotic expansion at moderate values of $Y$ is much slower. This is essentially a consequence of the weaker power $Y^{-1/6}$ of the scaling deviation compared to the fixed coupling case $Y^{-1/2}$, due to the fact that the two evolution equations do not belong to the same universality class. Because of the large $b_0$ dependent contribution in front of the $\ln(Y)$ and $Y^{-1/3}$ terms, the coefficients of the power expansion in $Y$ of $\dot\rho_s$  behave like the coefficients of an asymptotic series, and therefore there is no convergence at small or moderate values of $Y$. 

The convergence of the leading edge expansion is shown in Fig.\,\ref{fig:wave-front-rc}. The red curve includes the first two terms $G_{-1}$ and $G_0$, while the green curve includes all the universal terms in both the leading and front interior expansion. For $Y=100$, a rapid convergence of the series is observed. Unfortunately, for the same reason as for $\rho_s(Y)$, at low and moderate values of $Y$, the truncated series has a pathological behavior due to both the slow convergence of the series and the linear terms in $b_0$ which behave like a divergent asymptotic series.

In the next section we will address this short coming at moderate values of $Y$.  We abandon, in particular,  the leading edge expansion for a Taylor series in the vicinity of the saturation line, $x=0$ , i.e.,  $\rho \sim \rho_s(Y)$.  

\section{Phenomenology of transverse momentum broadening}\label{sec:numerics}

In this section, we study the transverse momentum distribution after resummation of the leading radiative corrections, including running coupling effects. We aim at providing analytic expressions for the $\kt$ distribution, that may be used for the phenomenology of transverse momentum broadening in heavy-ion collisions or as initial conditions for the small-$x$ evolution of gluon distributions in large nuclei.
\subsection{Taylor expansion about $\rho_s$}

Although the leading edge expansion for $\hat q$ is essential to determine systematically  the sub-asymptotic corrections to $\rho_s$, we have shown that for moderate values of $Y$ it fails to converge, and therefore cannot be used at small values of $Y=1\div 5$, the typical values relevant for phenomenology. This limitation is only apparent. Indeed, the leading edge expansion applies to large $x$ values away from the saturation regime. Hence, if one is mostly interested in finite region near the saturation scale then a Taylor expansion around $x=0$ should be enough to achieve a desirable accuracy:
\beq
\hat q(Y,x) = \hat q_0\,\rme^{\rho_s(Y)-Y}\, \mathcal{F}(x,Y)\,,
\eeq
where
\beq \label{eq:F-fct}
 \mathcal{F}(x,Y)= \, 1 +   \mathcal{F}'(0,Y) \,x +  \frac{1}{2!}\mathcal{F}''(0,Y) \,x^2+ O(x^3)\,,
\eeq
with $x=\rho-\rho_s(Y)$ as usual.
This is not to say that the leading edge expansion is not useful, quite the contrary. Indeed, we will show that the coefficients of this Taylor expansion depends only on the function $\rho_s(Y)$ which itself is determined by the leading edge expansion.

Therefore, if we have a good analytic knowledge of this function at small $Y$, we may expect that the resulting TMB distribution will be close to the exact numerical result. It turns out that at small $Y$, the effect of the non-linearities in the evolution equation are mild, so that the function $\rho_s(Y)$ in the non-linear case is in fact  close to $\rho_s(Y)$ in the linear case. Since we have a very good analytic control of $\rho_s(Y)$ in the linear case, we will use its form as our analytic input inside the Taylor expansion that we now detail.
As reported in Sec.\,\ref{sub:DLAresum}, when using the functions $\qhat(Y=\ln(L/\tau_0),\rho)$ and $\rho_s(Y)$ to evaluate the transverse momentum distribution, one must distinguish between the dense regime $\rho<\rho_s$ ($x<0$) and the dilute one $\rho>\rho_s$ ($x>0$).
First, we want to determine the Taylor expansion of $\mathcal{F}(x,Y)=\mathcal{F}_>(x,Y)$, to the right of the saturation line, up to second order in $x$ around $x=0$ ($\rho=\rho_s(Y)$). As a matter of fact, there is a one-to-one correspondence between $\rho_s(Y)$ and the Taylor coefficients of $\mathcal{F}_>(x,Y)$ at $x=0$. Indeed, plugging the definition of $\mathcal{F}_>$ inside the differential equation satisfied by $\qhat(Y,\rho)$ one gets
\begin{equation}\label{eq:evol-eq-f}
 (\dot\rho_s-1) \mathcal{F}_>(x,Y) + \dot{\mathcal{F}}_>(x,Y) - \dot\rho_s \mathcal{F}'_>(x,Y) =\int_0^x \der x'  \abar(x+\rho_s) \mathcal{F}_>(x',Y)\,.
\end{equation}
Using the fact that $ \mathcal{F}_>(0,Y)  =1 $, from which we also deduce that $\dot{\mathcal{F}}(0,Y) =0$, \eqn{eq:evol-eq-f} is readily solved yielding 
\begin{equation}\label{eq:evol-eq-df}
 \mathcal{F}_>'(0,Y)  = \frac {\dot\rho_s(Y)-1}{\dot\rho_s(Y)} \,.
\end{equation}
To obtain the second derivative we need to differentiate \eqn{eq:evol-eq-f} w.r.t. $x$
\begin{equation}\label{eq:evol-eq-df}
 (\dot\rho_s-1)  \mathcal{F}'_> + \dot{\mathcal{F}}'_>- \dot\rho_s \mathcal{F}''_>= \abar(x+\rho_s)  \mathcal{F}_>\,.
\end{equation}
Evaluating the latter at $x=0$ and using \eqn{eq:evol-eq-df}, we obtain
\begin{equation}
 \mathcal{F}_>''(0,Y) = \left(\frac{\dot\rho_s-1}{\dot\rho_s}\right)^2+\frac{\ddot\rho_s}{\dot\rho_s^3}-  \frac{\abar(\rho_s)}{\dot\rho_s} \,.
\end{equation}
As a result we obtain 
\beq\label{eq:taylor>}
\hat q_>(Y,x) = \hat q_0\,\rme^{\rho_s(Y)-Y}\, \left[1+ \frac {\dot\rho_s-1}{\dot\rho_s} x + \frac{1}{2}\left(\left(\frac{\dot\rho_s-1}{\dot\rho_s}\right)^2+\frac{\ddot\rho_s}{\dot\rho_s^3}-  \frac{\abar(\rho_s)}{\dot\rho_s}\right)x^2 +\mathcal{O}(x^2)\right]\,,\nn
\eeq
with $\rho_s$ and its derivatives evaluated at $Y=\ln(L/\tau_0)$.
Higher order derivatives, $\mathcal{F}^{(n)}_>(0,Y)$, can be computed in a similar fashion by iteration.

Let us turn now to $\qhat(\rho)=\qhat_<(\rho)$ to the left of the saturation line. By definition we have 
\begin{equation}
\qhat_<(\rho)=\qhat_0\, \rme^{\rho-Y_s(\rho)}\,.
\end{equation}
Writing $Y_s(\rho)=Y_s(\rho_s(Y)+x)$, and Taylor expanding with respect to $x$, one finds that
\begin{equation}
Y_s(\rho)=Y+\frac{1}{\dot\rho_s}x-\frac{1}{2}\frac{\ddot\rho_s}{\dot\rho_s^3}x^2+\mathcal{O}(x^3)\,.
\end{equation}
In the end, one can approximate the function $\qhat_<$ using
\begin{equation}
\hat q_<(Y,x) = \hat q_0\,\rme^{\rho_s(Y)-Y}\,\exp\left(\frac{\dot\rho_s-1}{\dot\rho_s}x+\frac{1}{2}\frac{\ddot\rho_s}{\dot\rho_s^3}x^2+\mathcal{O}(x^3)\right)\,.\label{eq:taylor<}
\end{equation}
If one expands again this result in powers of $x$, one notices that the coefficients of $x^0$ and $x$ are equal to those of $\qhat_>$, but not the coefficient of $x^2$. It means that the function $\qhat(\rho)$ we will employ in our analytic results is continuous and derivable, but not twice differentiable.

\subsection{Numerical results}

\begin{figure}[t] 
  \centering
  \begin{subfigure}[t]{0.48\textwidth}
     \includegraphics[page=1,width=\textwidth]{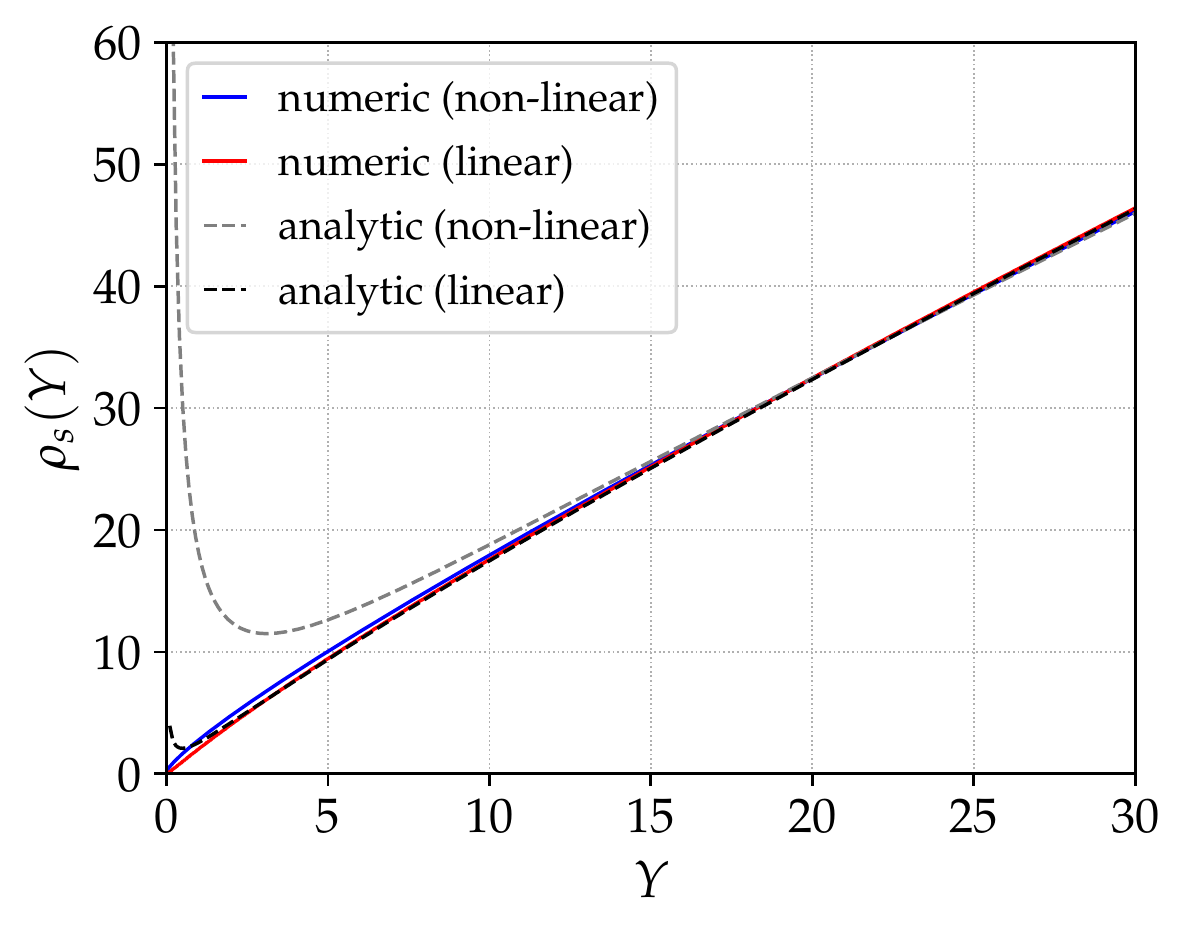} 
    \caption{\small }
    \label{fig:rhosY-linear-vs-non-linear}
    \end{subfigure}
  \begin{subfigure}[t]{0.48\textwidth}
     \includegraphics[page=1,width=\textwidth]{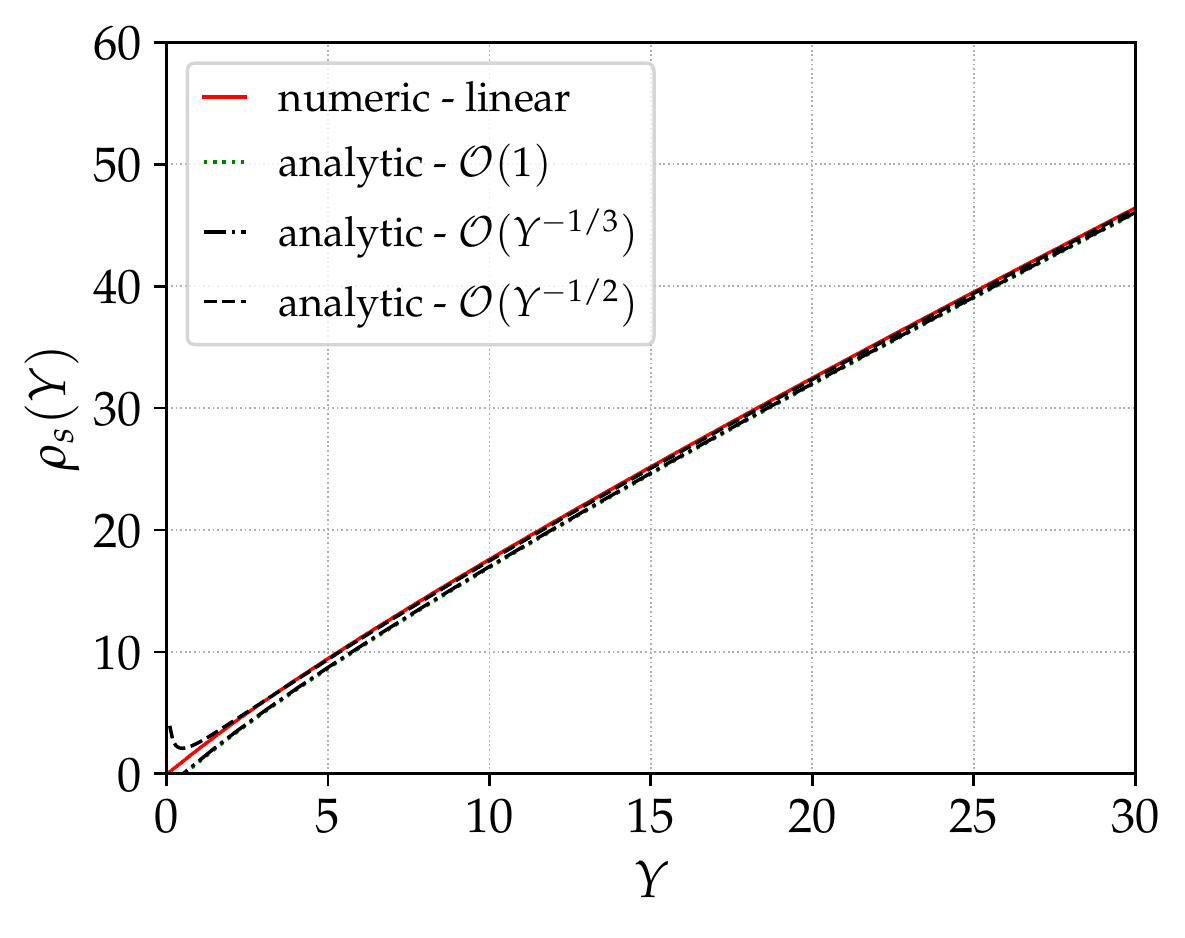} 
    \caption{\small}
    \label{fig:rhosY-convergence}
    \end{subfigure}    
        \caption{The saturation scale $\rho_s(Y)$ obtained numerically for the non-linear and linear scenario. The dotted black curve is the analytic result given by Eq.\,\eqref{eq:rhos-exp-linear} and \eqref{eq:d5coef-linear}.}\label{fig:rhosY}
\end{figure}

The formula for $\qhat(Y,\rho)$ obtained  from the Taylor expansion approach depends on $\rho_s(Y)$  only. If one aims at quantifying the effects of the quantum corrections for realistic values of $Y$, say $Y=2\div 5$, one needs an analytic expression for $\rho_s(Y)$ that correctly describes this range of values. In Fig.\,\ref{fig:rhosY-linear-vs-non-linear}, we observe that at moderate $Y$, the non-linear (blue curve) and the linear (red curve) evolution of the saturation scale are very close. On the other hand, even though the effects of the non-linearities are mild, we are already in the universal regime which is not driven by the initial condition.

One can then take advantage of this fact, since contrary to the non-linear evolution, the asymptotic expansion of $\rho_s(Y)$ is convergent in the linear case, even at small values of $Y$. The convergence of the asymptotic development \eqref{eq:rhos-exp-linear} is demonstrated in Fig.\,\ref{fig:rhosY-convergence}, where the dashed curve labeled $\mathcal{O}(Y^{-1/2})$ also includes the term given by Eq.\,\eqref{eq:d5coef-linear}.  In these analytic curves, the value of the unknown integration constant $\kappa$ is determined by fitting the large $Y$ tail of the numerical data. In the non-linear case, as we have shown in the previous section, the development is divergent and does not describe the numerical data for $Y$ smaller than 10. This is clear from the grey curve in Fig.\ref{fig:rhosY-linear-vs-non-linear}, when compared to the numerical result in blue. 

In the linear case, the asymptotic expansion of $\rho_s(Y)$ at $\mathcal{O}(Y^{-1/2})$ accuracy provides a very good approximation of both the linearized and "exact" $\rho_s(Y)$: in Fig.\,\ref{fig:rhosY}, the dotted black curve overlaps with the red curve even at $Y$ of order 1 and is also close to the blue curve. The mathematical reason is that the linear development at large $Y$ is related to the non-linear one by dropping the problematic terms (proportional to $b_0$, induced by the back-reaction of the quantum evolution to the saturation boundary) which makes the series divergent as $Y$ becomes smaller.

\begin{figure}[t] 
  \centering
  \begin{subfigure}[t]{0.48\textwidth}
     \includegraphics[page=1,width=\textwidth]{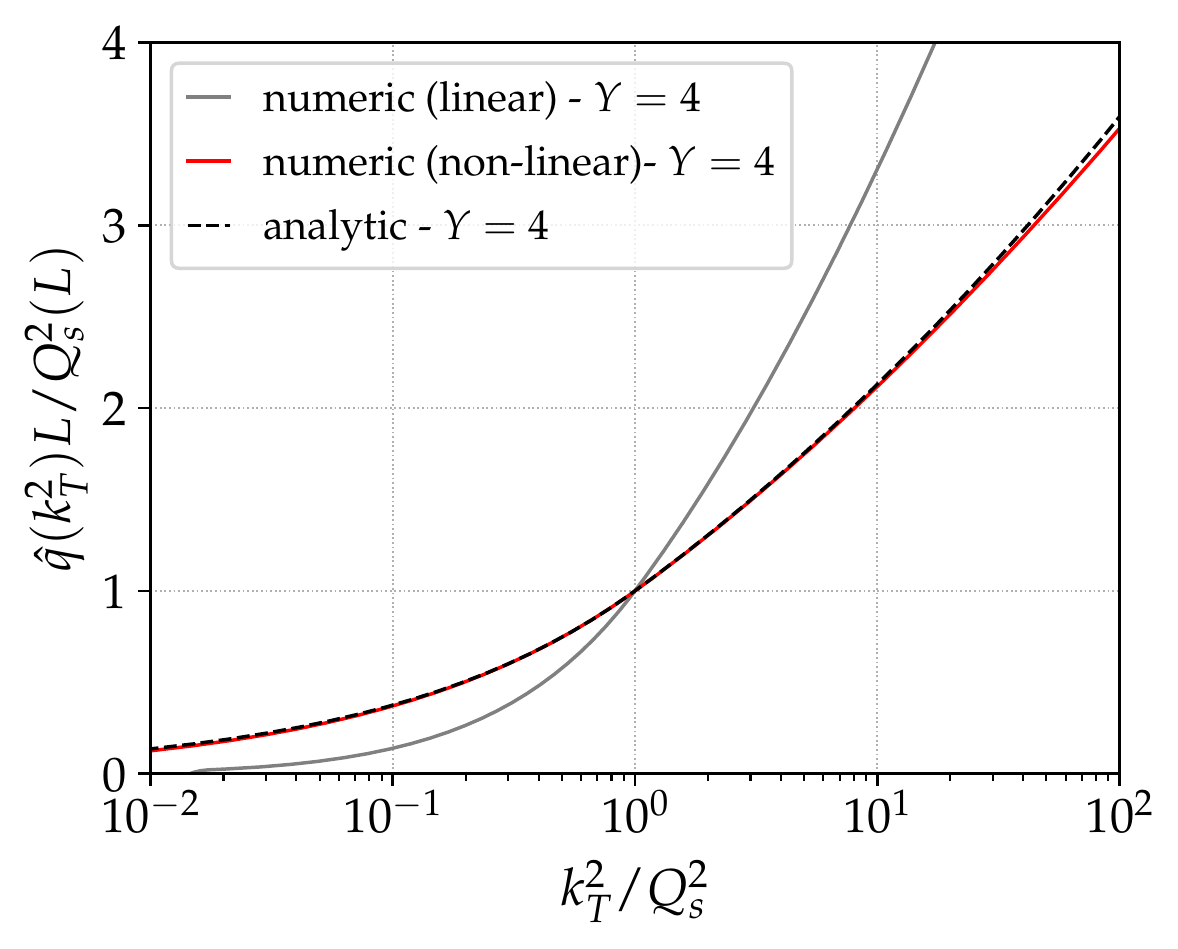} 
    \caption{\small }\label{fig:rc-pheno}
  \end{subfigure}
  \hfill
  \begin{subfigure}[t]{0.48\textwidth}
     \includegraphics[page=1,width=\textwidth]{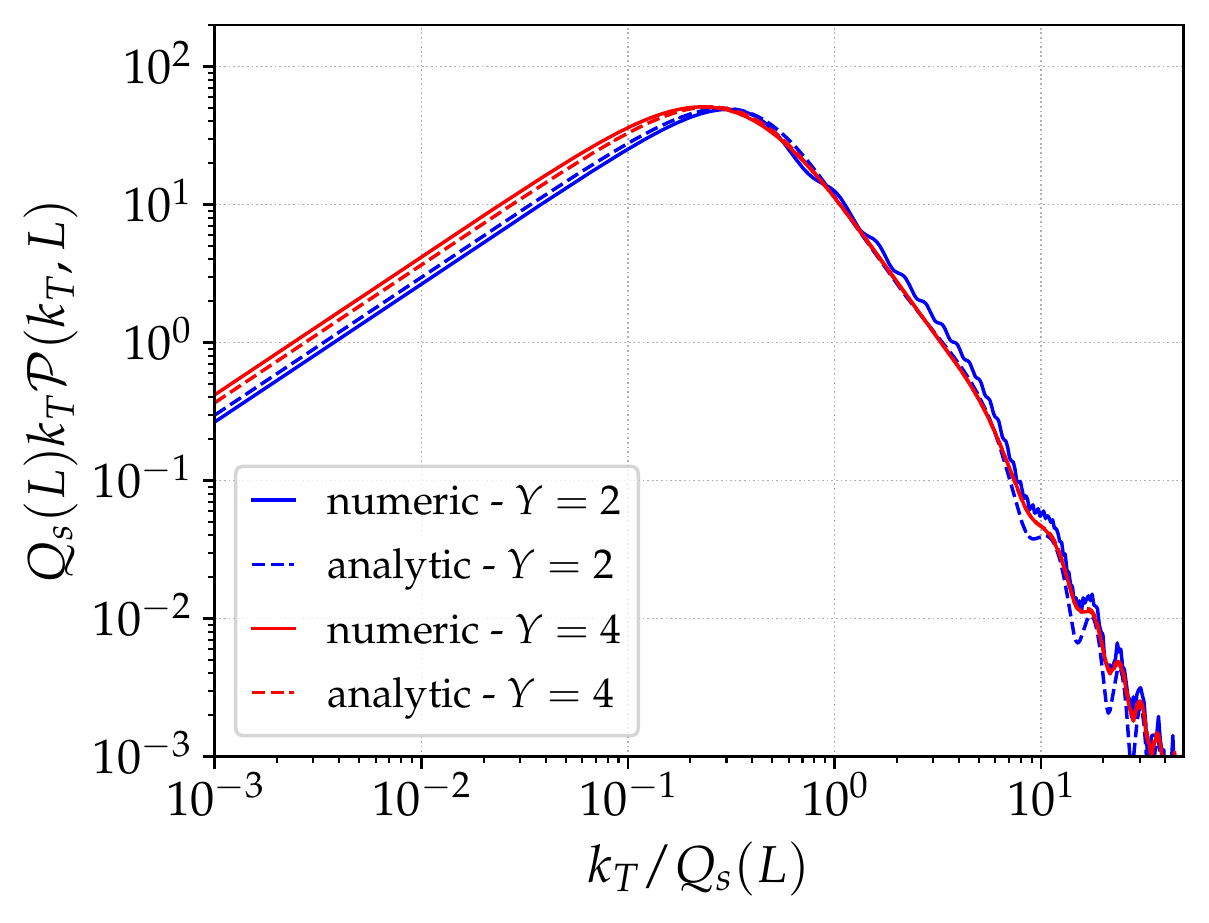}    
     \caption{\small }\label{fig:kt-distrib}
  \end{subfigure}
    \caption{(Left) The dimensionless parameter $\qhat(L,k_T^2)L/Q_s^2(L)$ as a function of the variable $k_T^2/Q_s^2$ compared to a truncation of the divergent series. (Right) Transverse momentum broadening distribution after quantum evolution with running coupling, compared to two phenomenological ansatz.}
    \label{fig:kt-plot}
\end{figure}

Since at moderate $Y$ values, the effect of the non-linearities on the saturation scale $\rho_s(Y)$ are mild, it is legitimate to use the expressions \eqref{eq:rhos-exp-linear} and \eqref{eq:d5coef-linear} as our analytic input for $\rho_s$ in the Taylor expansions given by Eq.\,\eqref{eq:taylor>}-\eqref{eq:taylor<}. 
We have now all the ingredients to compare the function $\qhat(Y,\rho)$ obtained by solving the non-linear evolution equation numerically with our analytic expressions \eqref{eq:taylor>}-\eqref{eq:taylor<}. This is shown Fig.\,\ref{fig:rc-pheno}: the red curve is our numerical result for $\qhat(Y,\rho)$ as a function of $x=\rho-\rho_s$, while the dashed black curve correspond to Eq.\,\eqref{eq:taylor>}-\eqref{eq:taylor<}. This analytic approach is very conclusive and can be systematically improved by including more terms in the Taylor expansion in order to describe the large $x$ domain. 

In Fig.\,\ref{fig:rc-pheno}, the grey curve is the numerical result for $\qhat(Y,\rho)$ from the linearized evolution equation. In contrast with what we observed for  $\rho_s(Y)$, we notice that the non-linearity has an important effect on $\qhat$: it slows down the evolution. The difference comes from the coefficients of the Taylor expansion that we have established in the previous sub-section: for the linear evolution equation, the dependence upon $\rho_s$ of these coefficients is not the same. In particular, the first derivative in $x=0$ is $\dot\rho_s-1$ which is significantly larger than $1-1/\dot\rho_s$ at moderate values of $Y$. Hence, even though we observed that the asymptotic development of the linearized $\rho_s$ is a good proxy for the "exact" $\rho_s$ for $Y$ of order $1$, the linearization turns out to be a bad approximation for the function $\qhat(Y,\rho)$ and consequently, for the TMB distribution itself.

The resulting TMB distribution is shown in Fig.\,\ref{fig:kt-distrib} for two values of $Y$, $Y=2$ and $4$. The dashed curves correspond to the analytic expressions after Fourier transform of the dipole $S$-matrix. For $Y=4$, the agreement is excellent, and even for $Y=2$, our formulas correctly captures the general trend of the distribution. We point out that the oscillatory behaviour at large $\kt$ is a consequence of the discontinuity of the second derivative of $\qhat(L,\kt^2)$ with respect to $\kt^2$ in $Q_s^2$.

\section{Physics discussion}
\label{sec:physics}

We now detail the physical interpretation of the asymptotic limit of the quenching parameter when the system size $L$ goes to infinity. We rely mainly on the equation established in section \ref{sub:TWsol}. The discussion will be divided into two subsections: one related to the behaviour of the asymptotic TMB distribution around its peak, and an other about its large $\kt$ tail.

\subsection{Anomalous diffusion and L\'{e}vy flights}

Using Eq.\,\eqref{eq:scaling-limit-fc} and $\rho_s(Y)= cY$ at large $Y$, one can obtain the scaling limit of $\qhat$ in the dense regime that controls the peak of the $\kt$ distribution. Indeed, we have
\begin{equation}
\qhat(Y=\ln(\tau/\tau_0),\rho)\approx \qhat_0\rme^{\rho_s(Y)-Y}f(\rho-\rho_s(Y))\,.
\end{equation}
For $\kt^2\le Q_s^2(L)$, we need to evaluate $Y$ along the saturation line as shown by Eq.\eqref{eq:qhat-geoscal}. Slightly abusing the notation, since we now name $Y=\ln(L/\tau_0)$, we have
\begin{align}
\qhat_<(Y,\rho)&=\qhat_0\rme^{Y-Y_s(\rho)}f(0)\,,\\
&=\qhat_0\rme^{2\beta x}\,,
\end{align}
where we have used $f(0)=1$ and $Y_s(\rho)\approx \rho/c$ in the scaling limit (we recall that $\beta=(c-1)/(2c)$ and $c=1+2\sqrt{\abar+\abar^2}+2\abar$ is the velocity of the traveling wave front). Plugging this expression inside the dipole S-matrix and transverse momentum distribution, one obtains the two dimensional Fourier transform of a ``stretched exponential":
\begin{equation}
\mathcal{P}(\kt,L)=\int\der^2\xt \rme^{-i\kt\cdot\xt}\exp\left(-\frac{1}{4}(|\xt|Q_s(L))^{2-4\beta}\right)\,.\label{eq:Levydistrib}
\end{equation}
Therefore, the leading effect of the radiative corrections for large system size is a modification of the power exponent of the effective dipole size $|\xt|$. In the running coupling case, one would obtain similar results, with the value of $\beta\approx\sqrt{\abar}$ replaced by $\sqrt{b_0/Y}$. $\rho_s(Y)\sim Y$ is indeed the natural choice of the sliding scale in $\alpha_s$. 

Instead of having a Gaussian like distribution, the TMB probability distribution is of L\'{e}vy type. Such probability distribution naturally arises in the context of random walks called L\'{e}vy flights, which constitutes a generalization of Brownian motion. For the problem at hand, the emergence of a L\'{e}vy flight at asymptotically large times can be understood as a consequence of a scale invariant dynamics, due to the self similarity of multiple gluon fluctuations over all time scales, as shown Fig.\,\ref{fig:cartoon-resum}. 

Such random walks lead to an anomalous scaling of the moments with time. For instance, the median $\mathcal{M}$ of the distribution typically behaves like the saturation momentum as a function of the system size, namely
\begin{equation}
\mathcal{M}\propto L^{c/2}\approx L^{1/2+\sqrt{\abar}}\,.
\end{equation}
The positive deviation with respect to the $1/2$ power of standard diffusion betrays the onset of a super diffusive regime at large time. The sub-leading corrections with respect to the simple power law scaling have been obtained in section \ref{sub:log-shift}.

On top of this anomalous scaling of the ``moments" of the $\kt$ distribution, we emphasize that its whole shape around the peak differs significantly from the tree-level result. This also applies even at moderate values of the system size and not only in the rather formal asymptotic $L\to\infty$ limit, as shown in Fig.\,\ref{fig:punchline-plot}, where one observes that the resummed distribution is significantly broader than the tree-level one.

\subsection{Large-$\kt$ tail of the TMB distribution} 
\label{sub:kt-tail}
In the previous section we have discussed the behavior of the distribution around its peak that is controlled by the saturation scale $Q_s(L)$. At large $\k$, away from the saturation line where the physics is dominated by a single hard scattering there is an interesting relation between the large $\kt$ tail of the TMB distribution and the medium gluon distribution associated with the correspondence \eqref{eq:qhat-gpdf}. 
To extract the large $\kt$ behavior of the TMB distribution, we expand the exponential in the definition of the dipole $S$-matrix
\beq\label{eq:P-UV}
\mathcal{P}(\kt) &\simeq & -  \frac{1}{4}  \int \dif^2\xt \,  \qhat(L,1/\xt^2)L\,\xt^2\,   \,  \rme^{-i\xt\cdot \kt }\,,\\
&=& \frac{\pi}{2} \vec{\nabla}^2_{\kt}   \frac{1}{\kt^2}\int_0^{+\infty } \dif z\,z  \,   \qhat(L,1/z^2 \kt^2)L\,  \,  \mathrm{J}_0(z)\,.
\eeq
In the limit $k_T \to + \infty$, the $z$ integral is dominated by $z \sim 1 $, and we thus have  $\ln 1/z^2 \ll \ln \kt^2$. It allows us to expand $\qhat$
as 
\begin{equation}
\qhat(L,1/z^2 \kt^2)=\qhat(L,\kt^2)+\ln\left(\frac{1}{z^2}\right)\frac{\partial \qhat(L,\kt^2)}{\partial\ln \kt^2}+\mathcal{O}\left(\ln^2(z)\right)\,,
\end{equation}
Using 
 \beq
  \int_0^{+\infty } \dif z\,z   \,  \mathrm{J}_0(z)=0  \,,\qquad   \int_0^{+\infty } \dif z\,z  \,   \ln\left(\frac{1}{z}\right) \mathrm{J}_0(z)=1\,,
 \eeq
 we obtain the formula \cite{Caucal:2021lgf}
\begin{align}
 \mathcal{P}(\kt) &\simeq \vec{\nabla}_{\kt}^2\frac{\pi}{\kt^2}\frac{\dif \qhat (\kt^2)L }{\dif \ln \kt^2 }\,,\label{eq:P1-largekt}\\
 &\propto \frac{1}{\kt^4}\left.\frac{\partial xG_N(x,Q^2) }{\partial \ln Q^2 }\right|_{Q^2=\kt^2}\,,
\end{align}
where $xG_N$ is the gluon distribution function defined in Sec.\,\ref{eq:gluon-pdf} and $x = \max(\tau_0/L, \k^2/2 P^+T)$ (the value of $x$ used in this expression will be commented in the next section). 
In the second line, we have assumed that $\kt^2$ dependence of $\dif \qhat/\dif\ln(\kt^2)$ is relatively slow compared to the power law behavior due to the $1/\kt^2$ factor. To measure the qualitative effect of the $\kt^2$ dependence of $\qhat$ as a result of quantum corrections, it is convenient to parametrize the deviation to the Rutherford behavior as follows:
\begin{equation}
 \mathcal{P}(\kt) \propto\frac{1}{\kt^{4-\mathcal{D}(\kt^2)}}\,,\qquad\mathcal{D}(\kt^2)\equiv\frac{2}{\rho}\ln\left(\frac{\partial \qhat(Y,\rho)}{\partial\rho}\right)\,.\label{eq:Dpower}
\end{equation}

\begin{figure}[t] 
  \centering
     \includegraphics[width=10cm]{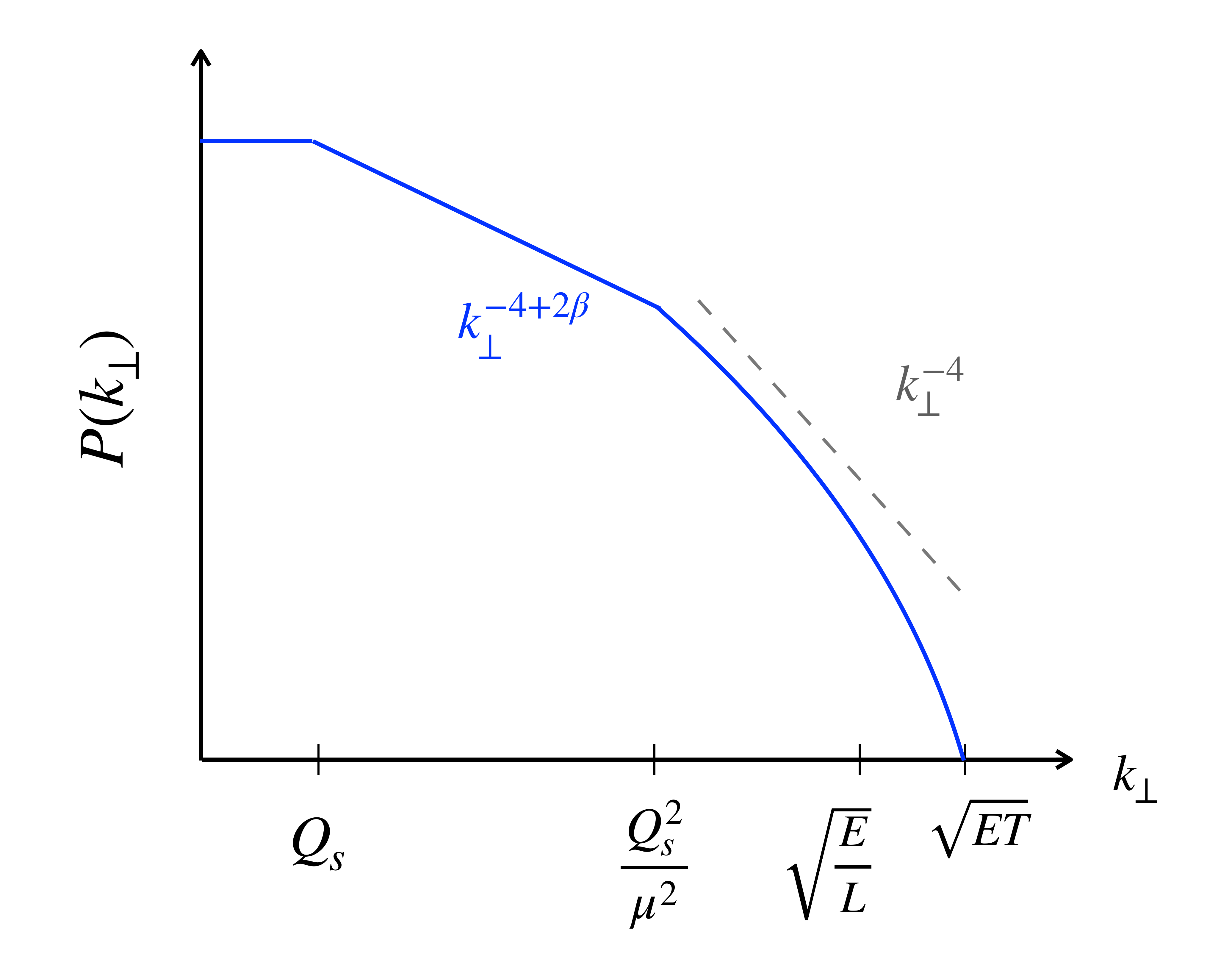} 
    \caption{An illustration of the various regimes of the transverse momentum broadening distribution $P(k_\perp)$. In the non-linear regime $k_\perp \lesssim Qs = \sqrt{\hat q L}$ the distribution is dominated by multiple soft scattering and the distribution exhibits geometric scaling, i.e., it is only a function of the scaling variable $k_\perp/Q_s$. The regime $Q_s \lesssim k_\perp \lesssim Q^2_s/\mu$ corresponds the extended geometric scaling window that is characterized by a heavy tailed distribution.  For  $k_\perp > Q^2_s/\mu$  the dynamics is linear and thus, not sensitive to saturation physics. $k_\perp=\sqrt{E/L}$ scale marks the transition from $k^-\equiv xP^-\sim 1/L$ to $k^-\equiv xP^-\sim k_\perp^2/2E$. The latter relates to the standard Feynman $x$ of the medium PDF.  }\label{fig:kt-dist-regimes}
\end{figure}

When $E\gg \omega_c$, one must distinguish several different regimes at large $k_\perp$, as shown Figure~\ref{fig:kt-dist-regimes}. In the domain $Q_s\ll k_\perp\ll Q_s^2/\mu$, we are in the extended geometric scaling window, whose behaviour is driven by the asymptotic limit ($Y\to\infty$) of the quenching parameter $\qhat(Y,\rho)$. This asymptotic limit satisfies geometric scaling, in so far as $\qhat(Y,\rho)$ is a function of $x=\rho-\rho_s=\ln(\kt^2/Q_s^2)$ only. Indeed using Eq.\,\eqref{eq:qhat-DLA-approx}, one finds that for $x\simeq \rho-Y\ll Y$, or equivalently, $k_\perp\ll Q_s^2/\mu$,
\begin{equation}
\qhat(Y,\rho)\simeq \qhat_0\rme^{2\sqrt{\abar}Y}\rme^{\sqrt{\abar}x}\,.
\end{equation}
From Eq.\,\eqref{eq:Dpower}, such a behavior leads to a heavy-tail with
\begin{align}
\mathcal{D}(\kt^2)\simeq\begin{cases}
              2\sqrt{\abar} & \textrm{ fixed coupling}\,,\label{eq:heavy-tail}\\
              2\sqrt{\frac{b_0}{Y}} & \textrm{ running coupling}\,.                
               \end{cases}
\end{align}
In the extended geometric scaling window, the TMB distribution behaves like a power law, with slower decay than the Rutherford one. This heavy tail is also characteristic of Levy flight processes. At fixed coupling at least, the analogy with L\'{e}vy flights enables to understand both the anomalous behavior of the diffusion process, since $Q_s\propto L^{\frac{1}{2}+\sqrt{\abar}}$ instead of $L^{1/2}$ and the heavy-tailed distribution. One could argue that even at tree-level, the TMB distribution displays a heavy tail with a Rutherford like decay. However, we emphasize that the heavy tail highlighted in this paper is entirely a consequence of the radiative corrections. Indeed, even in the case of a constant initial condition $\qhat^{(0)}(\rho)=\qhat_0$ (the so-called "harmonic approximation") for which the tree-level distribution is gaussian and therefore decays exponentially at large $k_\perp$, the resummation of self-similar gluon fluctuations leads to a heavy, power law tail given by Eq.\,\eqref{eq:heavy-tail}.

Beyond the extended geometric scaling window, for $k_\perp\gg Q_s^2/\mu$, but for $k_\perp\le \sqrt{E/L}$, the $\kt$-distribution becomes sensitive to the double log of the DGLAP evolution, and the exponent $\mathcal{D}(\kt^2)$ reads
\begin{align}
\mathcal{D}(\kt^2)\simeq\begin{cases}
              4\sqrt{\frac{\abar Y}{\rho}} & \textrm{ fixed coupling}\,,\label{eq:dglap-tail}\\
              4\sqrt{\frac{b_0 Y \ln(\rho)}{\rho^2}} & \textrm{ running coupling}\,.                
               \end{cases}
\end{align}
The function $\mathcal{D}$ is therefore slowly tending to $0$ at large $\kt$, meaning that one recovers asymptotically the Rutherford $1/\kt^4$ power law, as illustrated Fig.\,\ref{fig:kt-dist-regimes} between the scales $Q_s^2/\mu$ and $\sqrt{E/L}$.

\section{Energy dependence of the quenching parameter}
\label{sec:Edep-qhat}

So far, we have considered the propagation of a highly energetic parton, with $E=P^+\gg\omega_c$. In this section, we comment on the opposite regime $E\lesssim\omega_c$. In this limit, the characteristic quantum diffusion time $2E/\kt^2$ of the incoming parton wave-function can become smaller than the system size $L$ even in the saturation regime where $\kt^2\sim Q_s^2\approx\qhat_0L$. In order to understand what the dominant radiative corrections are in this case, we make a short detour by the operator definition of $\qhat$ and its link with parton distribution functions (PDFs). Then, we address the effect of the leading radiative corrections when $E\lesssim\omega_c$, and derive the typical energy dependence of the quenching parameter in this regime.

\subsection{On the $x$-dependence of $\hat q$ }\label{sub:xdep-qhat}

\paragraph*{$\qhat$ and gluon PDF.} The quenching parameter is to some extend related to the gluon parton distribution function. However, the connection is not straightforward and certainly the two quantities are not equivalent across the full $k_\perp$ spectrum. To begin with, PDF's are defined in the dilute regime of weakly interacting partons while the quenching parameter $\hat q$ is sensitive to non-linear or saturation effects. Bearing in mind these differences, it is instructive to explore the identification for cold nuclear matter 
\beq
\hat q(Q^2)  \to 4\pi^2 \alpha_s \,n  xG_N (x,Q^2)\,, \label{eq:qhat-gpdf}
\eeq
The function $xG_N(x,Q^2)$ is the the gluon PDF of a nucleon target, which reads in the parton model \cite{Collins:1989gx}
\begin{equation}
xG_N(x)=\frac{1}{2\pi P^-}\int_0^{\infty} \rmd x^+ \rme^{i x P^- x^+}\langle P|  F^{i-}(x^+) [x^+,0^+] F^{i-}(0^+)|P\rangle\,,\label{eq:gluon-pdf} 
\end{equation}
where $F^{i-}(x^+) \equiv F^{i-}(x^+,0^-,\0)$ is the field strength tensor and $[x^+,0^+]$ is a gauge link connecting the points $(x^+,0^-,\0)$ and $(0^+,0^-,\0)$, in light-cone variables. 
In doing so we recover the Glauber-Mueller formula for the forward dipole scattering amplitude \cite{Baier:1996sk,Liou:2013qya}. 

However, there is an ambiguity in the Feynman $x$-dependence of the gluon distribution in the idenfication \eqref{eq:qhat-gpdf}. 
The latter can be resolved by computing quantum corrections. At sufficiently low  $k_\perp$  the one loop correction is dominated by soft and collinear double logarithm 
\beq\label{eq:DL}
\abar\int_{\tau_0}^{\tau_{\rm max}} \frac{\rmd \tau  }{\tau  }\int^{Q^2}_{\mu^2} \frac{\rmd \kt'^2 }{\kt'^2} =\abar \ln \frac{\tau_{\rm max}}{\tau_0} \ln \frac{Q^2}{\mu^2}\,,
\eeq
where $\tau \sim 1/k^-$ is the formation time of the soft gluon radiation and $\kt'$ its transverse momentum. Including these radiative corrections through the renormalization of the gluon PDF makes \eqref{eq:gluon-pdf} $Q^2$-dependent. Similarly, $\qhat$ acquires a $Q^2$ dependence through the resummation of double logarithmic radiative corrections like \eqref{eq:DL}. In the case of $\qhat$, this $Q^2$ dependence should not be confused with the ``tree level" $Q^2$-dependence coming from the upper limit of the $\kt^2$ integration when $\qhat$ is defined as $\langle \kt^2\rangle/L$, as in \cite{Baier:1996sk,Benzke:2012sz}. In that case, the $Q^2$ dependence is a higher twist effect resulting from the Coulomb tail of the single gluon exchange in the $t$ channel which makes the second moment of the TMB distribution divergent.

In a similar fashion, the $x$-dependence of $\qhat$ can be understood from the $\tau_{\rm max}$ dependence induced by higher order corrections. To set the typical value of $\tau_{\rm max}$, one must distinguish two regimes. Indeed, from the on-shellness condition of the radiated gluon, one gets the relation $k^-=\kt^2/(2k^+)$. Since $k^+<P^+$, the latter equality gives the kinematic constraint $\tau_{\rm max}\sim 2P^+/\kt^2$.\footnote{This constraint also applies for virtual fluctuations due to probability conservation, see e.g.\ \cite{Iancu:2016vyg}.} In the TMB process, the two regimes are then either $\tau_{\rm max}> L$ or $\tau_{\rm max}<L$.

In the case $\tau_{\rm max}> L$, the largest formation time is actually set by the medium size $L$, namely $1/k^-\sim 1/(xP^-)<L$, and one recovers after resummation the double logarithmic limit \eqref{eq:asymp-exp} (with $Y=\ln(L/\tau_0)$) of the evolution considered in this paper. Indeed, fluctuations larger than $L$ are strongly suppressed due to the LPM effect owing to the fact that these long-lived gluons do not resolve the medium from the hard scattering that originates the jet. Notice that this constraint does not apply in the case of an asymptotic quark scattering off a shock-wave where the so called small-$x$ gluons can stretch beyond the extent of the target up to $\tau_{\rm max} \sim 2 P^+/k_\perp^2 > L$. 
Effectively, the relation $\tau_{\rm max}\sim L$ translates into the upper bound $L$ for the $x^+$ integration in the operator definition of $\hat q$,
\beq\label{eq:qhat-operator-1}
\hat q =\frac{
4\pi^2 \alpha_s n}{P^-}\int_0^{\sim L} \frac{\rmd x^+}{2\pi}  \langle P|  F^{i-}(x^+) [x^+,0^+] F^{i-}(0^+)|P\rangle \,.
\eeq

On the other hand, when $\tau_{\rm max} \sim 2 P^+/k_\perp^2 \ll L$ the gluon fluctuation is not sensitive to the size of the system and we must recover the standard gluon PDF 
 \beq \label{eq:qhat-operator-2}
\hat q = \frac{4\pi^2 \alpha_s \,n}{P^-}\int_0^{\infty} \frac{\rmd x^+}{2\pi}  \rme^{i x P^- x^+}\langle P|  F^{i-}(x^+) [x^+,0^+] F^{i-}(0^+)|P\rangle\,,
\eeq
where $x P^-\equiv \kt^2/2P^+$ and $P^- \sim T$.

The small and large $x$ regimes of $\hat q$ that are encompassed by \eqn{eq:qhat-operator-1} and \eqn{eq:qhat-operator-2}, respectively, can be combined in 
 \beq \label{eq:qhat-operator-3}
\hat q \equiv \frac{4\pi^2 \alpha_s n}{P^-}\int_0^{\infty} \frac{\rmd x^+}{2\pi}   \,\rme^{i x P^- x^+} \langle P|  F^{i-}(x^+) [x^+,0^+] F^{i-}(0^+)|P\rangle \,\Theta(x^+<L)\,.
\eeq
This definition differs from other definitions encountered in the literature \cite{Casalderrey-Solana:2007xns,Idilbi:2008vm,Majumder:2012sh}, in which $\hat q $ is defined as the second moment of the TMB distribution at leading twist. We believe that our calculation is more natural given that $\qhat$ appears in the unintegrated distribution and thus depends locally on $\k$. To sum up this discussion, our main findings are that the $x$-dependence, or equivalently the $\tau$-dependence of the quenching parameter should be given by
\begin{equation}
\tau= \mathrm{min}\left(\frac{2P^+}{\kt^2},L\right)\,,\label{eq:x-dep-final}
\end{equation}
or in terms of $x$ 
\beq
x = \mathrm{max}\left(\frac{\kt^2}{2P^+T},\frac{1}{LT}\right)\,.
\eeq

\paragraph*{Differences between the evolution of $\qhat$ and DGLAP/BFKL.} Based on the above insight on the relation between $\qhat$ and the gluon PDF, let us now analyze more closely the double logarithmic structure in \eqn{eq:DL} in order to identify the major differences with the two main regimes of QCD evolution, namely, DGLAP and BFKL. First, DGLAP evolution equations resum powers of the single logarithm $\ln Q^2/\mu^2 $ which in our context would be $\ln \k^2 /\mu^2$, while assuming that $x$ is of order 1 such that there is no need for resumming $\ln 1/x$ powers. This criterion is obviously met at large enough $k_\perp$ at the end of the power tail as shown in section \ref{sub:kt-tail} (cf.\ also figure~\ref{fig:kt-dist-regimes} and the shape of the $\kt$ distribution between $Q_s^2/\mu$ and $\sqrt{E/L}$). Now when $x \ll 1$ in \eqn{eq:qhat-operator-3} but at the same time we have $xP^- L \gg 1$, i.e, $k_\perp > \sqrt{P^+/L}$, we would have at next to leading order 
\beq
\frac{\hat q_{\rm NLO}}{\hat q_{\rm LO}}\approx  \abar \ln  \frac{\k^2}{\mu^2}\ln \frac{1}{x} =\abar \ln \frac{\k^2}{\mu^2}\ln \frac{P^+}{\k^2 L}\,. 
\eeq
After resummation, using 
\beq
\qhat\sim \qhat_0\exp\left(2\sqrt{\abar\ln \frac{\k^2}{\mu^2}\ln \frac{P^+}{\k^2 L}}\right)\,,
\eeq
one finds that the deviation $\mathcal{D}$ to the Rutherford behaviour defined in Sec.\,\ref{sub:kt-tail} is given by $\mathcal{D}(\kt^2)\simeq 4\sqrt{(\rho_E-\rho)/\rho}$. This is illustrated in figure~\ref{fig:kt-dist-regimes} in the domain $k_\perp>\sqrt{E/L}$ (when $k_\perp$ gets close to $\sqrt{ET}$, one approaches the kinematical limit beyond which the eikonal approximation is no longer valid).

For  $k_\perp <  \sqrt{P^+/L}$ the quantum phase in \eqn{eq:qhat-operator-3} is no longer relevant and the integral over $x^+$ must be cut-off at $L$. This yields
\beq
\frac{\hat q_{\rm NLO}}{\hat q_{\rm LO}}\approx  \abar \ln  \frac{\k^2}{\mu^2}\ln \frac{L}{\tau_0}\,.
\eeq
In this regime the collinear logarithm $ \ln  \frac{k_\perp^2}{\mu^2}$  is always larger that the soft logarithm  $\ln  \frac{L}{\tau_0}$ and we may wonder whether the relative importance of these logs would be reversed, in which case the Regge kinematics where single logs of $1/x$ are resummed would be more appropriate. However, saturation effects are relevant precisely in the regime where both logs are equally important. 

Indeed, multiple scattering responsible for the unitarization of the dipole amplitude are effective when $k^2_\perp = Q_s^2=\hat q L$. As a result, we obtain the double log structure $ \abar   \ln^2 \frac{L}{\tau_0}$.  Unless we are interested in the deep saturation regime $k_\perp \ll Q_s$ a single log approach {\it \`a la} BFKL is not necessary.  

On the other hand, in the case of DIS in the standard Regge kinematics the saturation scale scales as $Q_s^2(x) \sim x^{\lambda}$ where the anomalous dimension $\lambda \propto \alpha_s \ll 1$ \cite{kovchegov_levin_2012},  we thus have $\ln  Q_s^2(x) =   \alpha_s \ln 1/x \ll \ln 1/x$, where we readily see that the collinear log is suppressed compared to the soft log by the coupling constant.

In contrast,  in the case of momentum broadening the leading order comes with a factor $Q_s^2 \sim L \sim 1/x $  (where we used that $xP^-\sim {1/L}$) and quantum evolution will only slightly depart from this behavior. As we have seen, we have in particular $Q_s^2 \sim L^{1+2\sqrt{\abar}}$ at leading logarithmic accuracy. In other words, the problem of transverse momentum broadening near the saturation line is of double logarithmic nature and hence, does not favor neither DGLAP nor BFKL evolution equations. This justifies a posteriori why the dilute regime is sufficient for the qualitative analysis of the phase space.

\subsection{$E \ll \hat q L^2$ and energy dependence of the saturation scale }\label{sub:Edep-Qs}

We now return to discussing the qualitative consequences of the relation \eqref{eq:x-dep-final} on the energy dependence of $Q_s$.
In DLA, the asymptotic behavior of $\hat q$ reads
\beq\label{eq:qhat-DLA-approx} 
\hat q(\tau,\kt^2) \approx \hat q_0 \exp\left(2\sqrt{\abar \ln \frac{\kt^2}{\mu^2}\ln\frac{\tau}{\tau_0}}\right)\,,
\eeq
as can be shown from Eq.\,\eqref{eq:analytic-sol-1}\footnote{We have dropped sub-leading power prefactors.}. In this expression, the value of $\tau$ is bounded by the minimum between the jet path length $L$ and the quantum diffusion time given by $2E/\kt^2$. In the high energy limit, i.e.\ $P^+ = E \gg \omega_c\equiv\qhat L^2$, which is the main focus in this paper there is a potentially large phase space in which 
\begin{equation}
\textrm{min}\left(L,\frac{2E}{\kt^2}\right)=L\,.
\end{equation}
This phase space is given by $Q_s^2\le\kt^2\le \sqrt{2E/L}$. In this regime, using the defining equation for $Q_s$ and Eq.\,\eqref{eq:qhat-DLA-approx}, one readily finds
\begin{equation}
\rho_s(Y)=c\,Y\,,\qquad c= 1+2\sqrt{\abar +\abar^2}+2\abar\,,\label{eq:rhos-smallY}
\end{equation}
corresponding to the superdiffusive scaling.

\begin{figure}[t] 
  \centering
  \begin{subfigure}[t]{0.48\textwidth}
     \includegraphics[page=1,width=\textwidth]{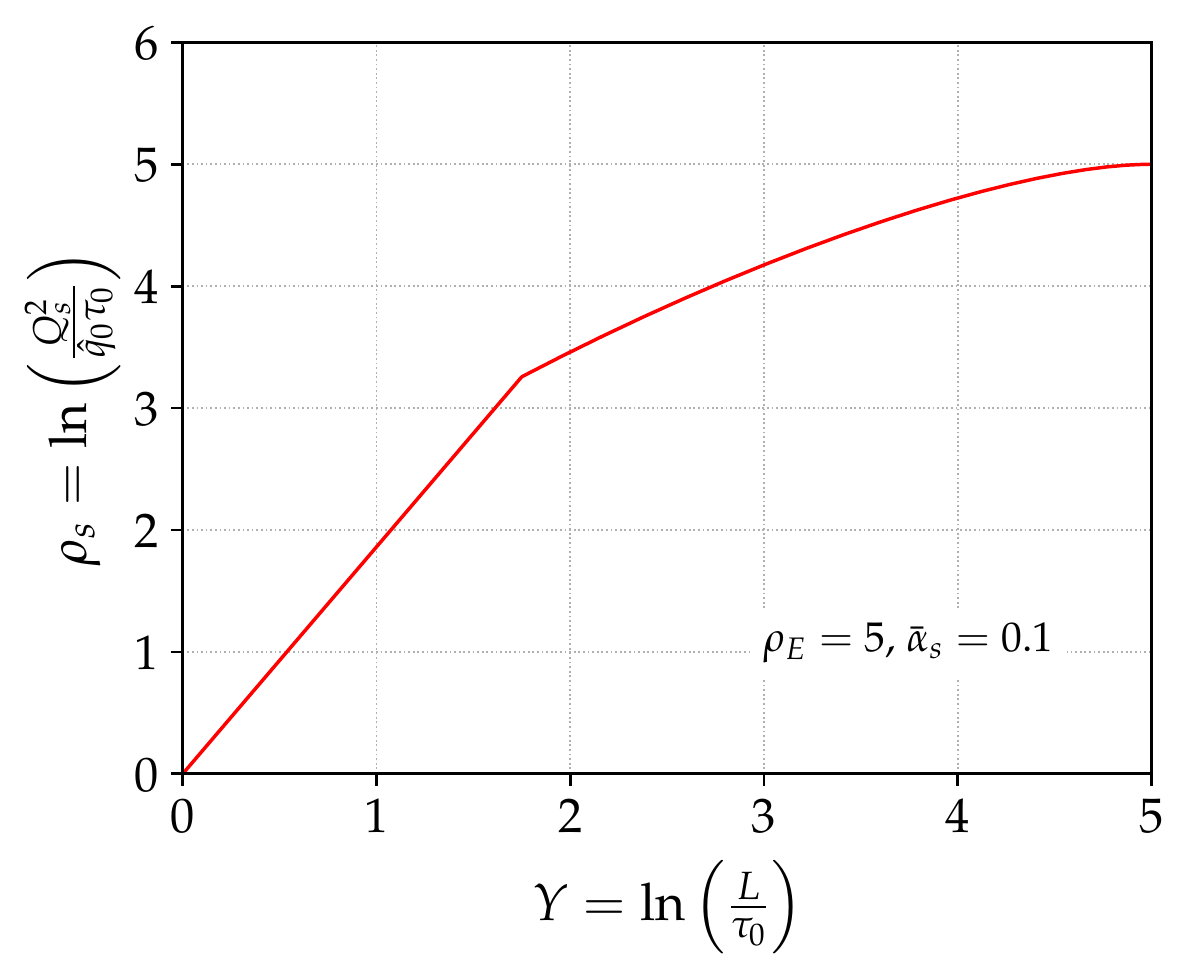} 
    \caption{\small }\label{fig:QsL}
  \end{subfigure}
  \hfill
  \begin{subfigure}[t]{0.48\textwidth}
     \includegraphics[page=2,width=\textwidth]{rhos-DLA.pdf}    
     \caption{\small }\label{fig:QsE}
  \end{subfigure}
    \caption{System size (left) and energy (right) dependence of the saturation momentum from the DLA asymptotic result \eqref{eq:qhat-DLA-approx}.}
    \label{fig:polynoms-gn}
\end{figure}

Let us now explore  the opposite limit, which corresponds to $E\ll\omega_c$. In this case we have $1/\tau= k_\perp^2/2E \gg 1/L$ not only at high $k_\perp$ but also deep inside the saturation regime. We can determine the behavior of the saturation momentum from its defining equation and the asymptotic form \eqref{eq:qhat-DLA-approx}:
\beq
 \hat q(\tau(Q_s),Q_s) L= Q_s^2=\hat q_0 L \exp\left(2\sqrt{\abar \ln \frac{Q_s^2}{\mu^2}\ln\frac{2E}{Q_s^2 \tau_0}}\right)\,.
\eeq
Defining the variable $\rho_E=\ln(2E/(\qhat_0\tau_0^2))\sim\ln 2ET/\mu^2$, the above equation can be rewritten as
\beq
( \rho_s-Y)^2 =4 \abar \rho_s (\rho_E-\rho_s)\,.
\eeq
Solving the quadratic equation for $\rho_s$ and retaining the larger solution we obtain 
\beq 
\rho_s&=& \frac{(Y+ 2\abar \rho_E) + \sqrt{ (Y + 2\abar \rho_E)^2-(1+4 \abar)Y^2}}{(1+4 \abar)}\,.\label{eq:rhos-largeY}
\eeq
In the limit $\abar\ll 1$ and $\abar \rho_E\ll Y$, we may simplify further
\beq 
\rho_s
&\simeq& Y+2\sqrt{\abar Y(\rho_E-Y)}+\mathcal{O}(\abar)\,,
\eeq
that is 
\beq
Q_s^2\simeq \qhat_0 L\,.
\eeq 
This is the typical leading order result, which does not display the anomalous behavious. As a function of $L$ for a fixed parton energy $E$, the saturation momentum recovers its linear scaling with the system size for $L$ large enough, $L\gg \sqrt{2E/\qhat_0}$. The leading behaviour of the saturation scale as a function of the system size $Y=\ln(L/\tau_0)$ is shown  in Fig.\,\ref{fig:QsL}, where one observed the two regimes \eqref{eq:rhos-smallY} and \eqref{eq:rhos-largeY} with a transition around the scale $\sqrt{2E/\qhat_0}$.

Conversely, for a fixed system size $L$, $Q_s$ exhibits an energy dependence below the scale $\omega_c$. This dependence, given by Eqs.\,\eqref{eq:rhos-smallY}-\eqref{eq:rhos-largeY}, is represented on Fig.\,\ref{fig:QsE} where $\rho_s$ is plotted as a function of $\ln(2E/(\qhat_0\tau_0))\sim\ln(E/T)$.

\section{Summary }\label{sec:conclusion}
In this article, we investigate quantum corrections to transverse momentum broadening in the double-logarithmic approximation by solving analytically and numerically the non-linear evolution equation for the quenching parameter $\hat q$ recently put forward.  

The effects of quantum evolution are three-fold: i) They cause a heavy tailed distribution, akin to L\`{e}vy random walks, to form at large $Y=\ln L/\tau_0$.  i) The TMB distribution loses sensitivity to initial condition given by the tree-level and tends to a universal distribution that can be computed analytically.  iii)  The TMB distribution obeys a geometric scaling in an extended region above the saturation scale, i.e., $ Q_s \lesssim k_\perp \ll Q_s^2/m_D$. 

Our analytic approach is based on an asymptotic expansion of the solution for large $Y$ in both fixed and running coupling using techniques that we borrowed from traveling waves analyses and gluon saturation physics.  In the regime where the fast parton energy  tends to infinity, we show that the transverse momentum distribution quickly reaches a universal regimes where it is approximately a function of a single scaling variable  $k_\perp/Q_s(L)$ with mild scaling violations that can be computed systematically. We have in particular computed the first six terms in the asymptotic expansion of $Q_s(Y)$. 

Furthermore, we have derived new results for the running coupling case that differs substantially from the fixed coupling case discussed in a previous work. We have in particular provided a formal proof for the asymptotic formula for $Q_s(Y)$ conjectured in Ref.~\cite{Iancu:2014sha} and computed the corrections to their results caused by the non-linearity of the saturation boundary. The running coupling exhibits a weak scaling which can be understood approximately from the fixed coupling analysis as an additional $Y$ dependence through the coupling constant that enters the exponent of the power spectrum. 

Our approach consists in searching for diffusion-like solutions near the exact scaling solutions away from saturation line. This region at the edge of the wave front leads the propagation of the front and in turn, is responsible for setting the overall speed of the wave given by $\dot \rho(Y)$.  
 
In the fixed coupling case, where geometric scaling is asymptotically exact,  we were able to obtain a very good agreement between our universal asymptotic expansion with the numerical computation for $Q_s(Y)$ down to small values of $Y\sim 2-5$. On the other hand, the asymptotic expansion in the running coupling case, when the non-linear effects are accounted for,  does not converge at low values of $Y$. We nevertheless observed that the expansion in the linearized equation provides a good approximation. 

The phenomenological interest of the analytic formulas derived in this paper is twofold. First, they can be used to estimate at low numerical cost the TMB of an energetic jet propagating through the quark-gluon plasma formed in heavy-ion collisions. In the case of the dijet azimuthal asymmetry, computed in the Sudakov formalism in \cite{Mueller:2016gko,Chen:2016vem}, TMB enters in the calculation as a single parameter, the average transverse momentum squared $\langle \kt^2\rangle$, or the saturation scale $Q_s$, which is fitted to experimental data. The relation between this parameter and the medium physical properties can then be obtained from our expressions for $Q_s(L)$. It would also be interesting to experimentally observe the super-diffusive regime by scanning the saturation scale over various system sizes (e.g.\ various centrality classes or nuclei).

Another possible application of our results pertains to QCD at small-$x$. At high energy and fixed $Q^2$, the building block of the fully inclusive DIS cross-section is the dipole S-matrix \eqref{eq:Sxt-def} whose evolution with Bjorken-$x$ is governed by the non-linear BK equation. In phenomenological studies, the BK equation is usually solved using the McLerran-Venugopalan model \cite{McLerran:1993ni,McLerran:1993ka} as the initial condition at moderate values of $x=x_0\sim 0.01$. The MV model is analogous to the tree-level form of $S(\xt,L)$ that gives the tree-level TMB distribution. This paper shows that including the leading radiative corrections enhanced by the nucleus size, of order $\alpha_s^n\ln^{2n}(A^{1/3})$ for all $n\ge 0$ leads to a universal distribution which becomes insensitive to the detail of the tree-level physics. This motivates the use of our analytic expression as a new initial condition for the BK equation. This approach would differ from other attempts to go beyond the MV model such as \cite{Dumitru:2011zz, Dumitru:2011ax} which focuses on power of $1/A$ suppressed corrections in the MV effective action or \cite{Dumitru:2020gla,Dumitru:2021tvw} which considers only a single gluon emission from a valence quark in the computation of the two-point correlator.

We conclude this summary with a brief perspective on future studies. The asymptotic expansions calculated in this paper are valid in the double logarithmic limit. We expect the single logarithmic corrections to change some of the coefficients in this series, as shown in \cite{Beuf:2010aw} in the context of the BK or BFKL evolution. For the jet quenching parameter problem, the resummation of the single logarithmic corrections raises additional questions related to the proper kernel to be used in the evolution or the exponentiation of the single log terms. Nevertheless, it is crucial for precision phenomenology to go beyond the present results. In order to draw a complete picture of the transverse momentum broadening in a dense QCD medium over all transverse momenta, such single logarithmic resummation should be matched with NLO results with exact kinematics \cite{Zakharov:2018rst}, NLO corrections to the collision kernel \cite{Arnold:2008vd,Caron-Huot:2008zna,Ghiglieri:2018ltw} and a proper determination of the non-perturbative small $k_\perp$ domain (e.g from lattice calculations \cite{Moore:2020wvy,Moore:2021jwe}). Regarding the phenomenological applications to heavy-ion collisions, one could also investigate the relative importance between higher order corrections and the effects of inhomogeneities in the plasma as computed in \cite{Barata:2022krd}.

\section*{Acknowledgements} 
 This work is supported by the U.S. Department of Energy, Office of Science, Office of Nuclear Physics, under contract No. DE- SC0012704.  Y. M.-T. acknowledges support from the RHIC Physics Fellow Program of the RIKEN BNL Research Center. Jaxodraw  was used to generate Feynman diagrams \cite{BINOSI200476}. 


\appendix 

\section{Universal leading edge and front interior expansion with running coupling}
\label{app:G0-rc}
In this appendix, we provide the analytic expressions of the \textit{universal} front interior and leading edge functions for the non-linear $\qhat$ evolution with running coupling. We also explain how to obtain the corresponding expansions in the linearized evolution.
\subsection*{The universal front interior functions}

The front interior expansion can be expressed as
\begin{equation}
\qhat(Y,\rho)=\qhat_0\rme^{\rho_s(Y)-Y}\rme^{\sqrt{b_0}\chi}\sum_{n\ge 0}(4b_0Y)^{-n/6}f_n(\sqrt{b_0}\chi)\,,\label{eq:fi-expansion}
\end{equation}
with $\chi=(\rho-\rho_s(Y))/Y^{1/2}$. Only the first five terms in this sum are universal:
\begin{align}
f_0(X)&=1+X\,,\\
f_1(X)&=0\,,\\
f_{2}(X)&=\xi_1\left[X+X^2+\frac{1}{6}X^3\right]\,,\\
f_{3}(X)&=\left(\frac{1}{2}-12b_0\right)X-\left(\frac{1}{2}+6b_0\right)X^2-\frac{1}{4}X^3+\frac{1}{12}X^4\,,\\
f_4(X)&=\xi_1^2\left[\frac{-7}{540}X+\left(\frac{-7}{540}+\frac{1}{2}\right)X^2+\left(\frac{-7}{3240}+\frac{1}{2}\right)X^3+\frac{1}{8}X^4+\frac{1}{120}X^5\right]\,,\\
f_5(X)&=-\xi_1\left[\left(\frac{5}{162}+20b_0\right)X+\left(\frac{59}{162}+25b_0\right)X^2+\left(\frac{1301}{972}+9b_0\right)X^3\right.\nonumber\\
&\left.+\left(\frac{7}{12}+b_0\right)X^4+\frac{1}{180}X^5-\frac{1}{120}X^6\right]\,.
\end{align}
One notices that all these $f_n$ functions are functions of $X=\sqrt{b_0}\chi$ and $b_0$, $f_n\to f_n(X,b_0)$. One can show that in the linear case, the functions $f_n$ are given by setting $b_0=0$ in the second argument, i.e. $f_n=f_n(X,0)$ (the variable $\chi$ becomes also $\chi=(\rho-Y)/\sqrt{Y}$). Therefore, the $b_0$ scaling associated with Eq.\,\eqref{eq:fi-expansion} is exact for the linear evolution equation.

\subsection*{The universal leading edge functions}

The leading edge expansion resums all the leading powers in the front interior expansion, as explained in the main text. In mathematical terms, the leading edge series reads
\begin{equation}
\qhat(Y,\rho)=\qhat_0\rme^{\rho_s(Y)-Y}\rme^{\sqrt{b_0}\chi}\sum_{n\ge -1}(4b_0 Y)^{-n/6}G_{n}(\zeta)\,,
\end{equation}
with $\zeta=\chi/Y^{1/6}=x/Y^{2/3}$.
The universal functions $G_{n}(\zeta)$ (with $n\le 2$) can be expressed as combinations of the Airy function and its derivative, with polynominal coefficients in $\zeta$:
\begin{equation}
G_{n}(\zeta)=\frac{1}{\mathrm{Ai}'(\xi_1)}\left[P_n(s)\mathrm{Ai}(s)+Q_n(s)\mathrm{Ai}'(s)\right]\,,
\end{equation}
with $s=\xi_1+2^{-1/3}b_0^{1/3}\zeta$. The polynomial functions $P_n(s)$ and $Q_n(s)$ read
\begin{align}
P_{-1}(s)&=1\,,\\
P_0(s)&=-\frac{13}{12}\xi_1^2+\frac{5}{3}\xi_1s-\frac{7}{12}s^2\,,\\
P_1(s)&=\left(-\frac{247\xi_1}{540}+\frac{169\xi_1^4}{288}+6\xi_1  b_0\right)+\left(\frac{151}{180}-\frac{65\xi_1^3}{36}-6b_0\right)s+\frac{97\xi_1^2}{48}s^2\nonumber\\
&-\frac{35\xi_1}{36}s^3+\frac{49}{288}s^4\,,\\
P_2(s)&=\left(\frac{73}{1620}-\frac{3419\xi_1^3}{6480}-\frac{2197\xi_1^6}{10368}-\frac{13\xi_1^3}{2}b_0\right)+\left(\frac{923\xi_1^2}{6480}+\frac{845\xi_1^5}{864}+\frac{33\xi_1^2}{2}b_0\right)s\nonumber\\
&+\left(\frac{6449\xi_1}{6480}-\frac{6383\xi_1^4}{3456}-\frac{27\xi_1}{2}b_0\right)s^2+\left(-\frac{3953}{6480}+\frac{2365\xi_1^3}{1296}+\frac{7b_0}{2}\right)s^3\nonumber\\
&-\frac{3437\xi_1^2}{3456}s^4+\frac{245\xi_1}{864}s^5-\frac{343}{10368}s^6\,.
\end{align}
and
\begin{align}
Q_{-1}(s)&=0\,,\\
Q_0(s)&=1\,,\\
Q_1(s)&=-\frac{1057}{540}\xi_1^2+\frac{89\xi_1}{27}s-\frac{241}{180}s^2\,,\\
Q_2(s)&=\left(-\frac{737\xi_1}{1620}+\frac{19877\xi_1^4}{12960}+12\xi_1b_0\right)+\left(\frac{737}{1620}-\frac{181\xi_1^3}{36}-12b_0\right)s+\frac{4367\xi_1^2}{720}s^2\nonumber\\
&-\frac{1031\xi_1}{324}s^3+\frac{2639}{4320}s^4\,.
\end{align}
The functions $P_n$ and $Q_n$ are functions of the variable $s$ and $b_0$ through the coefficients of these polynomial functions. As for the front interior expansion, the functions $P_n$ and $Q_n$ for the linearized evolution equation are obtained by setting $b_0=0$ in all these coefficients.

\bibliographystyle{utcaps}
\bibliography{biblio.bib}

\end{document}